\documentclass[12pt]{article}
\usepackage{algorithm}
\usepackage{algpseudocode}
\usepackage{amsmath}
\usepackage{graphicx,psfrag,epsf}
\usepackage{enumerate}
\usepackage{natbib} 
\usepackage{url} 
\usepackage[utf8]{inputenc} 
\usepackage[T1]{fontenc}    
\usepackage{hyperref}       
\usepackage{url}            
\usepackage{booktabs}       
\usepackage{amsfonts}       
\usepackage{nicefrac}       
\usepackage{microtype}      
\usepackage{xcolor}         
\usepackage{mathtools}
\usepackage{bm}
\usepackage{soul}
\usepackage{amsmath}
\usepackage{float} 
\usepackage{subcaption}
\usepackage{multicol}
\usepackage{multirow}
\usepackage{arydshln}
\usepackage{siunitx}
\usepackage{booktabs}
\usepackage[version=4]{mhchem}
\usepackage{collcell}

\DeclareMathOperator*{\argmax}{arg\,max}

\usepackage[toc,page,titletoc]{appendix}

\newcommand{\blind}{1}

\newcommand{\beginsupplement}{%
        \setcounter{table}{0}
        \renewcommand{\thetable}{S\arabic{table}}%
        \setcounter{figure}{0}
        \renewcommand{\thefigure}{S\arabic{figure}}%
     }

\begin{document}

\def\spacingset#1{\renewcommand{\baselinestretch}%
{#1}\small\normalsize} \spacingset{1}


\if1\blind
{
  \title{\bf Manifold-constrained Gaussian process inference for time-varying parameters in dynamic systems}
  \author{Yan Sun\hspace{.2cm}\\
    and \\
    Shihao Yang\thanks{
    Corresponding Author. shihao.yang@isye.gatech.edu} \\
    H. Milton Stewart School of Industrial and Systems Engineering \\
    Georgia Institute of Technology}
  \maketitle
} \fi

\if0\blind
{
  \bigskip
  \bigskip
  \bigskip
  \begin{center}
    {\LARGE\bf Manifold-constrained Gaussian process inference for time-varying parameters in dynamic systems}
\end{center}
  \medskip
} \fi

\bigskip
\begin{abstract}
Identification of parameters in ordinary differential equations (ODEs) is an important and challenging task when modeling dynamic systems in biomedical research and other scientific areas, especially with the presence of time-varying parameters. This article proposes a fast and accurate method, TVMAGI (Time-Varying MAnifold-constrained Gaussian process Inference), to estimate both time-constant and time-varying parameters in the ODE using noisy and sparse observation data. TVMAGI imposes a Gaussian process model over the time series of system components as well as time-varying parameters, and restricts the derivative process to satisfy ODE conditions. Consequently, TVMAGI completely bypasses numerical integration and achieves substantial savings in computation time. By incorporating the ODE structures through manifold constraints, TVMAGI enjoys a principled statistical construct under the Bayesian paradigm, which further enables it to handle systems with missing data or unobserved components. The Gaussian process prior also alleviates the identifiability issue often associated with the time-varying parameters in ODE. Unlike existing approaches, TVMAGI can be applied to general nonlinear systems without specific structural assumptions. Three simulation examples, including an infectious disease compartmental model, are provided to illustrate the robustness and efficiency of our method compared with numerical integration and Bayesian filtering methods.
\end{abstract}

\noindent%
{\it Keywords:}  ordinary differential equations, inverse problem, time-varying parameter estimation, Gaussian process, Bayesian inference
\vfill

\newpage
\spacingset{1.45} 

\section{Introduction}
\label{sec:intro}

Ordinary Differential Equations (ODEs) are often used to analyze the behavior of dynamic systems, such as the spread of infectious diseases (\cite{seir}),  interactions between species (\cite{LVmodel}), and viral dynamics (\cite{perelson1996hiv}). This paper studies a general formulation of ODE equations, where some of the parameters are allowed to be time-varying:

\begin{equation}
\bm{\dot x}(t) \equiv \frac{d\bm{x}(t)}{dt}=\mathbf{f}(\bm{x}(t),\bm{\theta}(t), \bm{\psi}, t), t \in [0,T]    
\end{equation}

Here, $\bm{x}(t)$ is the series of system outputs from time $0$ to $T$, $\bm{\bm{\psi}}$ denotes time-constant parameters, $\bm{\bm{\theta}}(t)$ denotes time-varying parameters, and $\mathbf{f}$ is a set of general functions that characterize the derivative process. When $\mathbf{f}$ is non-linear, the system outputs $\bm{x}(t)$ typically do not have analytic solutions. To solve $\bm{x}(t)$ given initial conditions $\bm{x}(0)$ and parameters $\bm{\bm{\theta}}(t)$ and $\bm{\bm{\psi}}$, numerical integration methods are often required, such as Euler's Method or Runge-Kutta Method (\cite{lapidus1971numerical}).

This paper focuses on the inverse problem that, given the observations, how to efficiently draw inference on the ODE parameters. Our goal is to estimate time-constant parameters $\bm{\bm{\psi}}$ and time-varying parameters $\bm{\bm{\theta}}(t)$ inside the ODE from data. In real world, the observation data of system components $\bm{x}$ are often obtained at discrete time points and are subject to measurement errors. We thus assume that we observe $\bm{y}(\bm{\tau})=\bm{x}(\bm{\tau})+\bm{\epsilon}(\bm{\tau})$, where $\bm{\tau}$ denotes the observation time points while error $\bm{\epsilon}(\bm{\tau})$ denotes Gaussian noise. We focus on the inference of $\bm{\theta}(t)$ and $\bm{\psi}$ given $\bm{y}(\bm{\tau})$, with emphasis on nonlinear structure $\mathbf{f}$.

The time-varying parameter $\bm{\theta}(t)$ in the ODE is often important yet challenging to recover from real-world data. For example, during a pandemic, the time-varying disease reproduction number is critical for public health policy decisions. However, its estimation can still be crude despite the best effort (\cite{abbott2020estimating}). The time-varying $\bm{\theta}(t)$ provides too much degree of freedom to the ODE system, and two different $\bm{\theta}(t)$ can both give $\bm{x}(t)$ that fits the observation data, resulting in identifiability issues (\cite{miao2011identifiability}).

\subsection*{Review of related literature}

Most existing methods for ODE inference only accommodate time-constant parameters (\cite{dondelinger2013ode, Yange2020397118, wenk2019gpode}). For time-varying parameters inference in the ODE system, existing methods all have their deficiencies (\cite{wu2005}). For example, \cite{Li2002} relied on time-consuming numerical integration; \cite{huang2006} proposed a Bayesian parametric approach to model time-varying coefficients in the HIV-1 dynamic model, sacrificing some flexibility; \cite{HLW} and \cite{cao2012penalized} developed an efficient two-stage local polynomial estimation method that circumvents numerical integration for a non-parametric time-varying parameters, but required ODE system to have linear dependency on the time-varying parameters (see Eq.\eqref{simplified_hiv}). Bayesian filtering methods are also explored in the time-varying ODE parameter inferences, although lacking some statistical rigor. For example, \cite{pei2020initial} and \cite{shaman2012forecasting} apply Ensemble Adjustment Kalman Filter (EAKF) algorithm to estimate parameters in a metapopulation SEIR model. \cite{schmidt2021probabilistic} proposed an extended Kalman Filter approach based on Gauss-Markov process that can infer time-varying parameter but cannot accommodate time-constant parameters any more. To the best of our knowledge, there is no existing Bayesian inference method that eliminates numerical integration for a general ODE system with both time-constant and time-varying parameters.

\subsection*{Our contribution}
We propose a fast and statistically principled method to infer time-varying $\bm{\bm{\theta}}(t)$ and  time-constant $\bm{\psi}$ from noisy observations of ODE. The key idea is to use Bayesian approach and place Gaussian process (GP) prior on $\bm{x}(t)$ and time-varying parameters $\bm{\theta}(t)$, and thus the identifiability issue is mitigated using the informative prior that favors smoother parameter curves. Our method is built upon the prior work of MAnifold-constrained Gaussian process Inference (\cite{Yange2020397118}) where the Gaussian process $\bm{x}(t)$ is restricted on a manifold that satisfies the ODE system. Placing a Gaussian process on $\bm{x(t)}$ facilitates a fast inference on $\bm{\theta}(t)$, as it completely bypasses numerical integration. Our approach also adheres to the classical Bayesian paradigm with rigorous theoretical derivation. Through a Gaussian process model on $\bm{\bm{\theta}}(t)$, we are able to generalize the MAnifold-constrained Gaussian process Inference to the situation where time-varying and time-constant parameters co-exist. We name our method TVMAGI (Time-Varying MAnifold-constrained Gaussian process Inference), emphasizing its capability in handling time-varying parameters. We demonstrate the effectiveness of TVMAGI through three realistic simulation examples, where TVMAGI works well even when some of the system components $\bm{x}(t)$ are partially observed. Through these simulation examples, we also show that TVMAGI can outperform benchmark methods including a numerical integration approach, a Bayesian filtering approach, and a two-stage approach. Thanks to the computational savings of bypassing numerical integration, TVMAGI has great potential to be generalized in high-dimensional and large-scale systems. TVMAGI achieves a substantial leap from the previously-proposed time-constant parameter inference methods \cite{wenk2019gpode, Yange2020397118} by investigating a much more complicated problem with functional estimate of time-varying parameters. The change from time-constant parameter to time-varying parameter also creates a notable difference in the scientific context, as parameters to be inferred in most real-world phenomena are non-stationary or changing over time.

\section{Method of TVMAGI}\label{sec:overview}

\paragraph{The prior.} Following standard Bayesian notation, the $D$-dimensional dynamic system $\bm{x}(t)$ is a realization of stochastic process $\bm{X}(t)=(X_1(t),...,X_D(t))$, and the $P$-dimensional time-varying parameters $\bm{\theta}(t)$ is a realization of stochastic process $\bm{\Theta}(t)=(\Theta_1(t),...,\Theta_P(t))$. We assume that $\bm{\Theta}(t)$ is continuous and differentiable in $t$ during time period $[0, T]$, which helps to prevent overfit and to alleviate identifiability issue, but can be relaxed later. The prior distribution  of $\bm{X}$ and $\bm{\Theta}$ in each dimension is independent Gaussian process.  That is, 
\begin{equation}\label{eq:tvmagi-theta-prior}
    \Theta_p(t) \sim \mathcal{GP}(\mu^\Theta_p, \mathcal{K}^\Theta_p), t \in [0,T], p \in \{1,\ldots, P\}
\end{equation}
\begin{equation}\label{eq:gp_prior}
X_d(t) \sim \mathcal{GP}(\mu_d^X, \mathcal{K}^X_d), t \in [0,T], d \in \{1,\ldots, D\}
\end{equation}
where $\mathcal{K}^X_d$ and $\mathcal{K}^\Theta_p$: $\mathbb{R} \times \mathbb{R} \rightarrow \mathbb{R}$ are positive definite covariance kernels for GP, while $\mu_d^X$ and $\mu_d^\Theta$: $\mathbb{R} \rightarrow \mathbb{R}$ denote mean functions. 

\paragraph{The likelihood.} The observations are denoted as $\bm{y}(\bm{\tau})=(y_1(\bm{\tau}_1),..., y_D(\bm{\tau}_D))$, where $\bm{\tau}=(\bm{\tau}_1, \bm{\tau}_2,...,\bm{\tau}_D)$ is the collection of observation time points across all components. Each component $\bm{X}_d(t)$ can have its own set of observation time points $\bm\tau_d = (\tau_{d,1}, \ldots, \tau_{d, N_d})$, where $N_d$ is the number of observations of the $d$-th component. If the $d$-th component is not observed, then $N_d = 0$, and $\bm\tau_d = \emptyset$. The observation likelihood is thus assumed to be
\begin{equation}\label{eq:obs_likelihood}
\bm{Y}_{d}(\bm{\tau}_{d}) = \bm{X}_{d}(\bm{\tau}_{d}) + \bm{\epsilon}(\bm{\tau}_d),\quad \bm{\epsilon}(\bm{\tau}_d) \sim \mathcal{N}(0, \sigma_{d}^2 )
\end{equation}
In this paper, notation $t$ shall refer to time generically, and $\bm{\tau}$ shall denote specifically the observation time points. 

\paragraph{The manifold constraint.} We introduce a variable $W$ to quantify the difference in the derivative process $\bm{\dot X}(t)$ between Gaussian process and ODE:
\begin{small}
\begin{equation}\label{eq:norm_inf}
W=\mathop{\sup}\limits_{t \in [0,T],d \in \{1,...,D\}}| \dot X_d(t)-\mathbf{f}(\bm{X}(t),\bm{\Theta}(t), \bm{\Psi}, t)_d|    
\end{equation}
\end{small}
Intuitively, $W$ is the $L_\infty$ norm of derivative difference, and $W=0$ if and only if $\bm{\dot X}(t)$ strictly satisfies the ODE structure, which is equivalent to constraining $\bm{X}(t)$ on the manifold of the ODE solutions. The advantage of $L_\infty$ norm is further discussed in Supplementary Material Section \ref{sec:infty-reason}. In the ideal situation where $W\equiv 0$, the posterior distribution of $\bm{\Theta}(t)$, $\bm{\Psi}$, and $\bm{X}(t)$ shall be formulated as $P_{\bm{\Theta}(t),\bm{\Psi},\bm{X}(t)| W,\bm{Y}(\bm{\tau})}(\bm{\theta}(t), \bm{\psi}, \bm{x}(t)|W=0,\bm{Y}(\bm{\tau})=\bm{y}(\bm{\tau}))$.
However, such ideal posterior is not computable in practice. Therefore, we approximate $W$ by finite discretization on time points $\bm{I} = \{t_1, t_2, \ldots, t_n\}$, such that $\bm{\tau} \subset \bm{I} \subset[0,T]$. We similarly define $W_{\bm{I}}$ on the discretization set $\bm{I}$ as the $L_\infty$ distance of the derivative from GP and that from ODE:
\begin{small}
\begin{equation}\label{eq:norm_inf_w_i}
W_{\bm{I}}=\mathop{\sup}\limits_{t \in \bm{I},d \in \{1,...,D\}}|\dot X_d(t)-\mathbf{f}(\bm{X}(t),\bm{\Theta}(t), \bm{\Psi}, t)_d|
\end{equation}
\end{small}
Here $W_{\bm{I}}$ is the maximum on a finite set, and $W_{\bm{I}} \rightarrow W$ monotonically as $\bm{I}$ becomes dense. The associated computable likelihood of the discretized manifold constraint $W_{\bm{I}} = 0$ is
\begin{small}
\begin{align}\label{eq:tvmagi_mc_likelihood}
&P(W_{\bm{I}}=0 | \bm{X}(\bm I) = \bm{x}(\bm I), {\bm{\Theta}(\bm{I})} = \bm{\theta}(\bm{I}), \bm{\Psi} = \bm{\psi}) \notag\\
=&P(\bm{\dot X}(\bm{I}) - \mathbf{f}(\bm{X}(\bm{I}), \bm{\Theta}(\bm{I}), \bm{\Psi}, t_{\bm{I}}) = \bm{0} |\bm{X}(\bm{I})=\bm{x}(\bm{I}),\bm{\Theta}(\bm{I})=\bm{\theta}(\bm{I}),\bm{\Psi}=\bm{\psi}) \notag\\
=& P(\bm{\dot X}(\bm{I})=\mathbf{f}(\bm{x}(\bm{I}), \bm{\theta}(\bm{I}), \bm{\psi}, t_{\bm{I}})|\bm{X}(\bm{I})=\bm{x}(\bm{I}))
\end{align}
\end{small}
which is a multivariate Gaussian distribution since the time derivative $\dot{X}(t)$ of GP is also a GP with specific mean and covariance kernel.

\paragraph{The posterior.} Therefore, a computable discretized posterior for TVMAGI inference of $X(t)$, $\Theta(t)$, and $\Psi$ is:
\begin{small}
\begin{equation}\label{eq:posterior-practical}
P_{\bm{\Theta}(\bm{I}),\bm{\Psi}, \bm{X}(\bm{I})| W_{\bm{I}},\bm{Y}(\bm{\tau})}(\bm{\theta}(\bm{I}),\bm{\psi},\bm{x}(\bm{I})|W_{\bm{I}}=0,\bm{Y}(\bm{\tau})=\bm{y}(\bm{\tau})) 
\end{equation}
\end{small}
Equation \eqref{eq:posterior-practical} is the computable discretized posterior of TVMAGI inference. In this paper, we consider the Maximum A Posteriori (MAP) as the fast point estimate from TVMAGI, while the Posterior Mean and the Posterior Interval are the formal Bayesian inference results that further quantify the uncertainty. 


\paragraph{The closed-form derivation.} The posterior distribution of $X(t)$, $\Theta(t)$, and $\Psi$ in Eq.\eqref{eq:posterior-practical} can be further derived as
\begin{small}
\begin{align}
&p_{\bm{\Theta}(\bm{I}), \bm{\Psi},\bm{X}(\bm{I})|W_{\bm{I}}, \bm{Y}(\bm{\tau})}(\bm{\theta}(\bm{I}), \bm{\psi}, \bm{x}(\bm{I})|W_{\bm{I}}=0, \bm{Y}(\bm{\tau})=\bm{y}(\bm{\tau})) \notag\\
\propto\ &  P(\bm{\Theta}(\bm{I})=\bm{\theta}(\bm{I}), \bm{\Psi} = \bm{\psi},\bm{X}(\bm{I})=\bm{x}(\bm{I}), W_{\bm{I}}=0, \bm{Y}(\bm{\tau})=\bm{y}(\bm{\tau})) \label{eq:full-posterior} \\
\propto& \pi_{\bm{\Psi}}(\bm{\psi}) \times  \underbrace{ P(\bm{\Theta}({\bm{I}}) = \bm{\theta}({\bm{I}}) | \bm{\Psi} = \bm{\psi}) }_{\text{1st Part Eq.\eqref{eq:tvmagi-theta-prior}}} \times \underbrace{ P(\bm{X}(\bm{I})=\bm{x}(\bm{I})|\bm{\Theta}(\bm{I})=\bm{\theta}(\bm{I}),\bm{\Psi}=\bm{\psi}) }_{\text{2nd Part Eq.\eqref{eq:gp_prior}}} \notag \\
& \times \underbrace{ P(\bm{Y}(\bm{\tau})=\bm{y}(\bm{\tau})|\bm{X}(\bm{I})=\bm{x}(\bm{I}), \bm{\Theta}(\bm{I})=\bm{\theta}(\bm{I}),\bm{\Psi}=\bm{\psi}) }_{\text{3rd Part Eq.\eqref{eq:obs_likelihood}}} \notag \\
&\times \underbrace{ P(W_{\bm{I}}=0|\bm{Y}(\bm{\tau})=\bm{y}(\bm{\tau}),\bm{X}(\bm{I})=\bm{x}(\bm{I}),\bm{\Theta}(\bm{I})=\bm{\theta}(\bm{I}),\bm{\Psi}=\bm{\psi}) }_{\text{4th Part Eq.\eqref{eq:tvmagi_mc_likelihood}}} \label{eq:decomposition} \\
& = \pi_{\bm{\Psi}}(\bm{\psi}) \times \exp \Big\{-\frac{1}{2} \Big( \underbrace{ \sum^{P}_{p=1} \Big[|\bm{I}|\log(2\pi)+\log|\mathcal{K}^\Theta_p(\bm{I})|+|| \theta_p(\bm{I})-\mu_p^{\Theta}({\bm{I}})||^2_{\mathcal{K}^\Theta_p(\bm{I})^{-1}} \Big]}_{\text{1st Part Eq.\eqref{eq:tvmagi-theta-prior}}} \notag\\
& + \sum^D_{d=1}\Big[ \underbrace{|\bm{I}|\log(2\pi)+\log|\mathcal{K}^X_d(\bm{I})|+||x_d(\bm{I})-\mu_d^X(\bm{I})||^2_{\mathcal{K}^X_d(\bm{I})^{-1}}}_{\text{2nd Part Eq.\eqref{eq:gp_prior}}} +\underbrace{N_d \log(2\pi\sigma_d^2)+||x_d(\bm{\tau}_d)-y_d(\bm{\tau}_d)||^2_{\sigma_d^{-2}}}_{\text{3rd Part Eq.\eqref{eq:obs_likelihood}}} \notag\\
&+ \underbrace{|\bm{I}|\log(2\pi)+\log|C_d|+||\mathbf{f}_{d,\bm{I}}^{\bm{x}, \bm{\theta}, \bm{\psi}} - \dot \mu_d^X(\bm{I})- \mathcal{'K}^X_d(\bm{I})\mathcal{K}^X_d(\bm{I})^{-1} \{x_d(\bm{I})-\mu_d^X(\bm{I})\}||^2_{C_d^{-1}}\Big] \Big) \Big\}}_{\text{4th Part Eq.\eqref{eq:tvmagi_mc_likelihood}}} \label{eq:posterior-calculation}
\end{align}
\end{small}
where  $||v||_A^2=v^T Av$, $|\bm{I}|$ is the cardinality of $\bm{I}$, and $\mathbf{f}_{d, \bm{I}}^{\bm{x}, \bm{\theta}, \bm{\psi}}$ is short for the $d$-th component of $\mathbf{f}(\bm{x}(\bm{I}), \bm{\theta}(\bm{I}), \bm{\psi}, t_{\bm{I}})$, and $C_d = \mathcal{K''}^X_d(\bm{I}) - \mathcal{'K}^X_d(\bm{I})\mathcal{K}^X_d(\bm{I})^{-1}\mathcal{K'}^X_d(\bm{I})$ is the conditional covariance matrix of $\dot X_d(\bm{I})$ given $X_d(\bm{I})$.

A deeper look into the above equation reveals that Eq.\eqref{eq:full-posterior} is the joint probability in Bayesian statistics, and Eq.\eqref{eq:decomposition} further decomposes it into parts. The 1st Part Eq.\eqref{eq:tvmagi-theta-prior} corresponds to independent GP prior distribution of $\bm{\Theta}(\bm{I})$, as the prior of $\bm{\Theta}(t)$  and $\bm{\Psi}$ are independent. The 2nd Part Eq.\eqref{eq:gp_prior} is the prior of GP on $\bm{X}(\bm{I})$, because the prior of $\bm{X}(\bm{I})$ is independent from $\bm{\Theta}(t)$ and $\bm{\Psi}$. The 3rd Part Eq.\eqref{eq:obs_likelihood} is the level of observation noise, and given the value of underlying true components $\bm{X}(\bm{\tau})$, the additive Gaussian observation noise $\bm{\epsilon}(\bm{\tau})$ is independent from everything else. The 4th Part Eq.\eqref{eq:tvmagi_mc_likelihood} can be simplified to be the conditional probability of $\bm{\dot X}(\bm{I})$ given $\bm{X}(\bm{I})$ evaluated at $\mathbf{f}(\bm{x}(\bm{I}), \bm{\theta}(\bm{I}), \bm{\psi}, t_{\bm{I}})$. All four parts are multivariate Gaussian distributed. Especially, The 4th Part Eq.\eqref{eq:tvmagi_mc_likelihood} is Gaussian because conditional $\bm{\dot X}(\bm{I})$ given $\bm{X}(\bm{I})$ has a multivariate Gaussian distribution, provided that the GP kernel $\mathcal{K}^X$ is twice differentiable.

We choose Matern kernel with degree of freedom $\nu=2.01$ for both $\bm{\Theta}(t)$ and $\bm{X}(t)$ to guarantee a differentiable GP that allows more flexible patterns: 
\begin{equation}\label{eq:kernel}
\mathcal{K}_\nu(l)=\phi_1^2 \frac{2^{1-\nu}}{\Gamma(\nu)} (\sqrt{2\nu} \frac{l} {\phi_2})^\nu K_\nu(\sqrt{2\nu} \frac l \phi_2), \quad l=|s-t|
\end{equation}
where $K_\nu$ denotes the modified Bessel function of the second kind. In this case, $'\mathcal{K}=\frac{\partial}{\partial s}\mathcal{K}(s,t)$, $\mathcal{K}'=\frac{\partial}{\partial t}\mathcal{K}(s,t)$, and $\mathcal{K}''=\frac{\partial^2}{\partial s \partial t}\mathcal{K}(s,t)$ are all well-defined.

\section{Algorithm}\label{sec:algorithm}
This section provides a detailed computational scheme of TVMAGI, including the hyper-parameter settings. The implementation is available on GitHub\footnote{https://github.com/Canadasunyan/TVMAGI}. Overall, the Maximum A Posteriori (MAP) of $\bm{X}(\bm{I})$, $\bm{\Theta}(\bm{I})$, and $\bm{\Psi}$ is obtained by optimization, while the posterior mean/interval is obtained by Hamiltonian Monte Carlo. To set the hyper-parameters and initiate the optimizer, we introduce a multi-stage approach in the algorithm. The advantages of multi-stage algorithm are discussed towards the end of this section.

\subsection{Initialization and inference of the mean}\label{sec:algorithm-step1}
At the first stage, we impose a GP only on $\bm{X}(t)$ and substitute the time-varying $\bm{\theta}(t)$ with its unknown mean $\bm{\mu}^\Theta$ in the entire model. This formulation ignores the time-varying property of $\bm{\theta}(t)$ and treats it as time-constant, which fits in the time-constant parameters inference framework of \cite{Yange2020397118}. As such, we can use MAGI package (\cite{Yange2020397118}) to obtain point estimates for the parameters and system components, denoted as $\bm{\hat{\mu}}^\Theta$, $\bm{\psi}^{(0)}$, and $\bm{x}(\bm{I})^{(0)}$. The $\bm{\hat{\mu}}^\Theta$ is subsequently used as the prior mean value for the time-varying $\bm{\theta}(t)$ in an empirical Bayes fashion, and will be plugged in Eq.\eqref{eq:posterior-calculation}. The $\bm{\psi}^{(0)}$ and $\bm{x}(\bm{I})^{(0)}$ will be used as the initial values for $\bm{\psi}$ and $\bm{x}(\bm{I})$ in the later MAP optimization.

The hyper-parameters $(\phi_{1,d}^{X}, \phi_{2,d}^X)$ for kernel $\mathcal{K}_d^X$ in Eq.\eqref{eq:kernel} and the noise level $\sigma$ for each system component $X_d$, $d=1,...,D$, are also estimated in MAGI package, using the Gaussian process smoothing marginal likelihood (\cite{Yange2020397118}). The MAGI estimated noise level $\sigma^{(0)}$ will serve as initial value for later joint MAP optimization. 

\subsection{Point-wise inference of the time-varying parameters}\label{sec:algorithm-pointwise}
At the second stage of TVMAGI, we obtain an initial estimate for time-varying $\bm{\theta}(\bm{I})$ by removing the smoothing GP prior. That is, we maximize the partial posterior Eq.\eqref{eq:pointwise-calculation} conditioning on $\bm{\Theta}(\bm{I})$, without considering Part 1 Eq.\eqref{eq:tvmagi-theta-prior}:

\begin{small}
\begin{align}\label{eq:pointwise-calculation}
&\bm{\tilde x}(\bm{I}), \bm{\tilde \theta}(\bm{I}), \bm{\tilde \psi}, \bm{\tilde \sigma}= \argmax \limits_{\bm{x}, \bm{\theta},\bm{\psi},\sigma} p_{ \bm{\Psi},\bm{X}(\bm{I})|W_{\bm{I}}, \bm{Y}(\bm{\tau}), \bm{\Theta}(\bm{I})}(\bm{\psi}, \bm{x}(\bm{I})|W_{\bm{I}}=0, \bm{Y}(\bm{\tau})=\bm{y}(\bm{\tau}), \bm{\Theta}(\bm{I}) = \bm{\theta}(\bm{I})) \notag \\
\propto& \pi_{\bm{\Psi}}(\bm{\psi}) \times \underbrace{P(\bm{X}(\bm{I})=\bm{x}(\bm{I})  ) }_{\text{2nd Part Eq.\eqref{eq:gp_prior}}} \times \underbrace{P(\bm{Y}(\bm{\tau})=\bm{y}(\bm{\tau})|\bm{X}(\bm{I})=\bm{x}(\bm{I}) )}_{\text{3rd Part Eq.\eqref{eq:obs_likelihood}}} \notag \\ 
& \quad \times\underbrace{P(\bm{\dot X}(\bm{I})=\mathbf{f}(\bm{x}(\bm{I}), \bm{\theta}(\bm{I}), \bm{\psi}, t_{\bm{I}})|\bm{X}(\bm{I})=\bm{x}(\bm{I}))}_{\text{4th Part Eq.\eqref{eq:tvmagi_mc_likelihood}}} 
\end{align}
\end{small}

The optimization is initialized at $\bm{x}(\bm{I})^{(0)}$, $\bm{\psi}^{(0)}$, $\bm{\sigma}^{(0)}$, and the $\bm{\theta}(\bm{I})$ is initialized at $\bm{\hat{\mu}}^\Theta$. We denote the optimized $\bm{{\theta}}(\bm{I})$ as $\bm{\tilde \theta}(\bm{I})$, and $\bm{x}(\bm{I})^{(0)}$, $\bm{\psi}^{(0)}$, $\sigma^{(0)}$ are updated to a new optimum $\bm{\tilde x}(\bm{I})$, $\bm{\tilde \psi}$ and $\tilde \sigma$. We call $\bm{\tilde \theta}(\bm{I})$ the point-wise estimate since there is no requirement on the smoothness or continuity of $\bm{\tilde \theta}(t)$ on $\bm{I}$. Although wiggling and possibly overfitting the data, the point-wise estimate $\bm{\tilde \theta}(\bm{I})$ captures the trend of parameter changes, which provides information to set the hyper-parameters of GP kernels $\mathcal{K}_p^\Theta$ for $\bm{\Theta}(t)$.

\subsection{GP hyper-parameters for time-varying ODE parameters} \label{sec:theta-hyper-parameters}

The length-scale parameter $\phi^{\Theta}$ controls how fast $\bm\theta(t)$ could change. Provided the point-wise estimate $\bm{\tilde \theta}(\bm{I})$, we use Gaussian Process smoothing method to set the hyper-parameters $\phi_{1,p}^\Theta, \phi_{2,p}^\Theta$ of GP kernels $\mathcal{K}_p^\Theta$ in Eq.\eqref{eq:kernel}. We shall treat $\bm{\tilde \theta}(\bm{I})$ as observations of $\bm{\Theta}(\bm{I})$, and operate on each dimension of time-varying ODE parameters separately. 

Recall the prior $\Theta_p(\bm{I}) \sim \mathcal{GP}(\hat{\mu}^\Theta_p, \mathcal{K}_p^\Theta(\bm{I},\bm{I}))$, where the mean $\hat{\mu}^\Theta_p$ is obtained in Section \ref{sec:algorithm-step1}. We use the empirical Bayes approach again to set $\phi_{1,p}^\Theta, \phi_{2,p}^\Theta$ by maximizing its posterior density at $\tilde{\theta}_p(\bm{I})$:
\begin{equation}
\hat{\phi}_{1,p}^\Theta, \hat{\phi}_{2,p}^\Theta =\argmax_{\phi_1,\phi_2} \pi_{\bm{\Phi}_p}(\bm{\phi})p({\tilde \theta_p}(\bm{I})| \bm{\phi}) \\
\end{equation}
where ${\tilde \theta_p}(\bm{I})| \bm{\phi} \sim \mathcal{N}(\hat{\mu}^\Theta_p, \mathcal{K}(\bm{\phi}) + \mathrm{diag}(\delta^2))$, the $\delta$ is the nuisance parameter governing the induced noise in point-wise estimate ${\tilde \theta_p}(\bm{I})$, and the $\pi_{\bm{\Phi}_p}(\cdot)$ is the hyper-prior. In practice, the hyper-prior $\pi_{\bm{\Phi}_p}(\cdot)$ is often set to be uniform on a reasonable interval depending on the context to ensure desired level of smoothness for the time-varying ODE parameter $\theta_p(t)$.

\subsection{Maximum A Posteriori (MAP) optimization}\label{sec:map-optim}
All the hyper-parameters are now set and will be held as constant when optimizing Eq.\eqref{eq:posterior-calculation} to get the MAP, with initial values $\bm{ \theta}(\bm{I})^{(0)} = \bm{\tilde \theta}(\bm{I})$, $\bm{x}(\bm{I})^{(0)}=\bm{\tilde x}(\bm{I})$, $\bm{\psi}^{(0)}=\bm{\tilde \psi}$ and $\bm{\sigma}^{(0)}=\bm{\tilde \sigma}$, all from Section \ref{sec:algorithm-pointwise}. The joint posterior function Eq.\eqref{eq:posterior-calculation} of $\bm{x}(\bm{I})$, $\bm{\theta}(\bm{I})$, $\bm{\psi}$ and $\bm{\sigma}$ is optimized with Adam optimizer (\cite{kingma2014adam}) in PyTorch to get the MAP estimation of TVMAGI. Finally, to mitigate the potential issue of Adam optimizer converging to local optimum, we suggest trying multiple initial values, including starting $\bm{x}(\bm{I})$ at linear interpolations from the observations $\bm{y}(\bm{\tau})$.

\subsection{Interval estimation of parameters}
In addition to the MAP point estimate, we also quantify of the parameter uncertainty in TVMAGI using posterior samples. In particular, we sample the posterior function Eq.\eqref{eq:posterior-calculation} using Hamilton Monte Carlo (HMC) \citep{neal2011mcmc}, while holding all the hyper-parameters at the same constant value as in Section \ref{sec:map-optim}. Details about the HMC algorithm can we found in Supplementary Material Section \ref{sec:hmc-alg}. Specifically in all illustration examples of this paper, we set step size $\epsilon=10^{-5}$, number of leap-frog steps $L=100$, sample size 8000, burn-in ratio 0.5, and the HMC is initialized at the MAP estimate. 

\subsection{Advantages of our multi-stage algorithm}\label{sec:advantage}
Compared with joint optimization of hyperparameters and parameters together, the multi-stage optimization method enjoys several advantages. First, the GP hyperparameters $\bm \Phi^X$ for the system components are set at the first stage and held as constant in the rest of the optimization so that the inverse of kernel matrix only needs to be computed once, following the recommendation of \cite{golightly2010markov}. Second, GP hyperparameters $\bm \Phi^\Theta$ for the time-varying parameters could not be set without any information about $\bm \Theta(I)$. Therefore, a multi-stage procedure is necessary, where a point-wise $\bm {\tilde \theta}(I)$ is obtained in one stage without GP, and then GP hyperparameters $\bm \Phi^\Theta$ is estimated in the following stage based on $\bm {\tilde \theta}(I)$. Lastly, The multi-stage optimization ensures that each step of the optimization starts with sensible initial value obtained from previous modularized optimization, thus drastically decreasing the chance of Adam optimizer stuck in local mode. Experiments have shown that our carefully designed multi-stage optimization is faster and achieves better results than joint optimization with randomized starting values.


\section{Benchmark methods and evaluation metrics}
\subsection{Benchmark methods}
In this section we provide a  brief introduction of two common approaches for time-varying parameter inference in ODE: numerical integration methods, represented by Runge-Kutta method, and Bayesian filtering methods, represented by Ensemble Adjustment Kalman Filter (EAKF). Supplementary Material Section \ref{sec:additional-benchmark-discussion} has some additional theoretical discussion about the limitations and the statistical rigor of the benchmark methods for ODE inference when time-varying parameters and time-constant parameters co-exist.

\subsubsection*{Runge-Kutta method}
Runge-Kutta method (\cite{lapidus1971numerical}) is a brute-force way for parameter inference in ODE systems. As a non-linear least square method, Runge-Kutta method minimizes the MSE of observations and reconstructed trajectory using numerical integration from the proposed initial conditions and parameters. The objective function is given by: \\
\begin{equation*}
\min \limits_{\bm{x_0}, \bm{\theta}(\bm{I}), \bm{\psi}} \sum_{\tau \in \bm{\tau}_d}\sum^D_{d=1}(\bm{y}_d(\tau)-\mathbb{X}^{RK4}_\tau(\bm{x_0}, \bm{\theta}(t), \bm{\psi})_d)^2
\end{equation*}
where $\mathbb{X}^{RK4}$ denotes the reconstructed trajectory using the 4th Order Runge-Kutta method. 

\subsubsection*{Ensemble Adjustment Kalman Filter}
Ensemble Adjustment Kalman Filter (EAKF) (\cite{shaman2012forecasting}) is a variation of Kalman Filter that is popular for parameter calibration of ODE systems in practice. It is a specially designed fast Bayesian filtering method. As a data assimilation technique, EAKF represents filtered distribution using Monte Carlo samples, and replaces the covariance matrix with sample covariance. The Kalman update assumes all probability distributions involved are Gaussian. As a major difference with Ensemble Kalman Filter (EnKF), EAKF uses a deterministic update instead of stochastic update. EAKF has been used in estimating the influenza disease spread SIRS model parameter (\cite{shaman2012forecasting}). In COVID-19 disease spread modeling, EAKF has also been used to study time-varying fatality rate (\cite{yang2020estimating}). 

Supplementary Material Section \ref{sec:others} includes a few more Bayesian filtering benchmark methods of Extended Kalman Filter (EKF), Unscented Kalman Filter (UKF), and EnKF. All the Bayesian filtering methods have the inherent limitation that all parameters must be assumed time-varying, and thus cannot accommodate time-constant parameters. 

\subsection{Evaluation Metrics} \label{sec:evaluation}
To assess the quality of the parameter estimates and the system recovery, we consider two metrics based on root mean squared error (RMSE). First, we examine the accuracy of the parameter estimates, using \emph{parameter RMSE}. For the time-constant parameters, we directly calculate the RMSE of the parameter estimates to the true parameter value across simulations. For the time-varying parameters, we additionally average over discretization set $\bm{I}$ for the RMSE. Second, we examine the system recovery, using \emph{trajectory RMSE}. Due to the potential identifiability issue that different parameters can give similar system observations, we measure how well the system components are recovered as another independent evaluation. To calculate the trajectory RMSE, we use numerical integration to reconstruct the trajectory based on the TVMAGI inferred parameters and initial conditions. The RMSE of the reconstructed trajectory to the true system is then calculated at observation time points. 

We emphasize that the numerical integration is only used for evaluation purpose, and throughout our TVMAGI approach, no numerical integration is ever needed. For better distinction, we refer to the MAP of $\bm{x}(\bm{I})$ directly from TVMAGI as the \emph{inferred trajectory}, and refer to the numerically integrated $\bm{x}(t)$ based on the TVMAGI inferred parameters and initial conditions as the \emph{reconstructed trajectory}.

To assess the quality of the interval estimates, we consider the Frequentist coverage of our posterior intervals. For the time-constant parameters, we directly calculate the proportion of repeated simulations where our posteiror interval covers the truth. For the time-varying parameters, we additionally average over discretization set $\bm{I}$ for the coverage. The coverage of the inferred trajectory can be similarly calculated, averaging over discretization set $\bm{I}$. We do not compute the coverage of the reconstructed trajectory as it will require numerical solver for each posterior sample of the parameters and initial conditions.

\section{Results}\label{sec:results}
We illustrate the accuracy and efficiency of TVMAGI through three realistic simulation studies of ODE models in epidemiology, ecology, and system biology. We begin with a disease compartmental model that demonstrates the effectiveness of TVMAGI for problems with partially observed system component(s). We then use an ecology example to show how TVMAGI can mitigate the identifiability issue through the informative GP prior that favors smoother time-varying parameters. Lastly, we apply TVMAGI on a system biology example with non-stationary rapid-changing time-varying parameters, and presents TVMAGI's competitive performance with one additional tailor-made benchmark method for such ODE.

\subsection{SEIRD model}  
Consider a COVID-19 cases/deaths modeling using an infectious disease Susceptible-Exposed-Infectious-Recovered-Deceased (SEIRD) compartmental ODE model \cite{hethcote2000mathematics, hao2020reconstruction}, where the entire population is classified into S, E, I, R, D components, and any transitions from one state to another state (i.e., the disease spreading dynamics) are modeled as ODE:
\begin{equation}
\frac{dS}{dt}=-\frac{\beta IS}{N}, \quad
\frac{dE}{dt}=\frac{\beta IS}{N} - v^e E,  \quad
\frac{dI}{dt}=v^e E - v^i I, \quad
\frac{dD}{dt}=v^i I \cdot p^d 
\end{equation}
$N$ is the total population, and the cumulative recovered population is $R = N - S -E -I-D$. The $S, E, I$ and $D$ denote the susceptible, exposed, infected population and cumulative death respectively. The 4 parameters of interest are investigated: rate of contact by an infectious individual ($\beta$), rate of transferring from state of exposed to infectious ($v^e$), rate of leaving infectious period ($v^i$) and fatality rate ($p^d$). During a pandemic, parameters in the SEIRD model can evolve over time due to pharmaceutical and non-pharmaceutical interventions. We assume that $\beta$ is time-varying due to the mutation of disease and policy interventions during a specific time; $p^d$ is time-varying depending on the sufficiency of medical treatments; $v^e$ is time-varying due to the different levels of public awareness or complacency, and $v^i$ is assumed to be unknown time-constant parameter to avoid identifiability issues.

\begin{figure}[htp]
     \centering
         \includegraphics[width=5in]{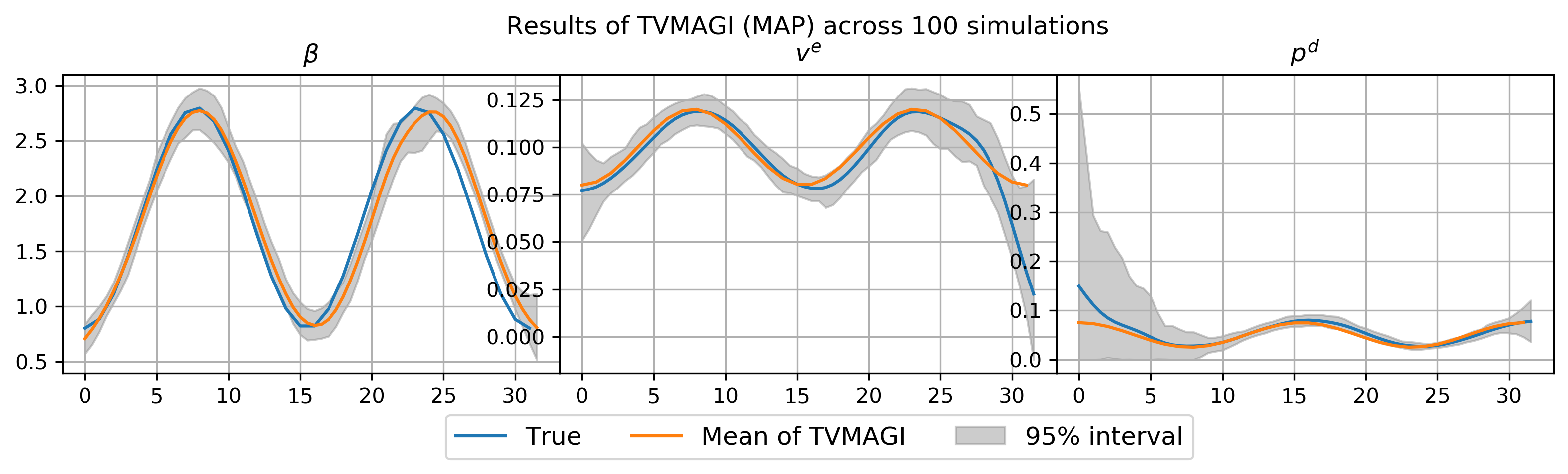}
         \includegraphics[width=5in]{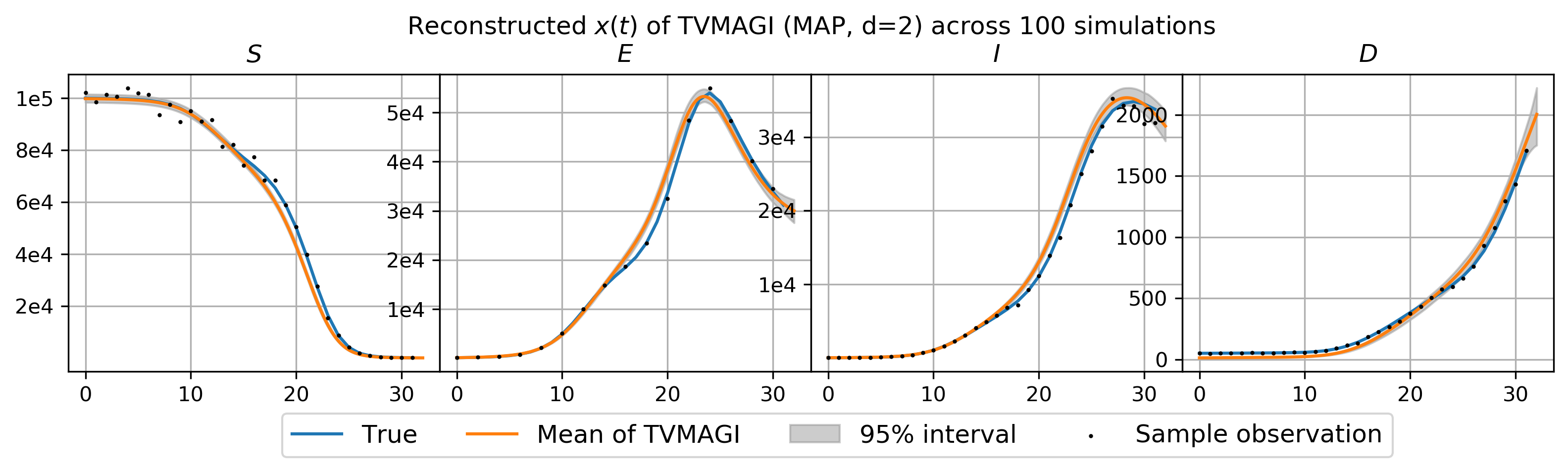}
\caption{Results of parameter inference (upper) and reconstructed trajectory (lower) of TVMAGI in 100 simulated datasets for SEIRD model. The mean and the 95\% interval here refer to the point estimates across 100 simulated datasets. One sample simulation dataset is also presented to visualize the noise level and observation schedule.}
\label{fig:TVMAGI-seird}
\end{figure}

\begin{table}
\begin{center}
\caption{Accuracy comparison for the SEIRD model based on 100 simulation datasets. The mean of RMSE is reported first with the standard deviation across 100 replications followed after $\pm$ for the parameters and the reconstructed trajectories. The last column is the coverage of interval estimates for the parameters and the inferred trajectories. The last row shows the computing time (in seconds) needed to obtain point estimates from all methods.}
\label{tab:seird-1}
\scalebox{0.7}{
\begin{tabular}{
  cc
  >{}r<{}@{$\pm$}>{}r<{}>{}r<{}@{$\pm$}>{}r<{}>{}r<{}@{$\pm$}>{}r<{} | >{}r|cc
}
\toprule
 &  & \multicolumn{6}{c|}{Point Estimate} & \multicolumn{2}{c}{TVMAGI Posterior Samples} \\
 \midrule
 & RMSE & \multicolumn{2}{c}{TVMAGI-MAP} & \multicolumn{2}{c}{Runge-Kutta} & \multicolumn{2}{c|}{EAKF} & Posterior Mean RMSE & Interval Coverage \\
\midrule
\parbox[t]{2mm}{\multirow{4}{*}{\rotatebox[origin=c]{90}{Parameter}}} &    $\beta$ & $\bm{0.114}$ & $\bm{0.039}$ & 0.178 & 0.094 & 0.706 & 0.010 & 0.110 $\pm$ 0.043 & 98.2\% \\
  &      $v^e$ & $\bm{0.009}$ & $\bm{0.010}$ & 0.051 & 0.030 & 0.057 & 0.003 & 0.007 $\pm$ 0.005 &  97.4\% \\
  &      $v^i$ & 0.005 & 0.003 & $\bm{0.004}$ & $\bm{0.003}$ & 0.151 & 0.008 & 0.007 $\pm$  0.004 &  91.0\% \\
& $p^d$ & $\bm{0.019}$ & $\bm{0.029}$ & 0.083 & 0.073 & 0.039 & 0.003  & 0.011  $\pm$  0.008 & 98.0\% \\
\midrule
\parbox[t]{2mm}{\multirow{4}{*}{\rotatebox[origin=c]{90}{Trajectory}}} &      $S$ & $\bm{581.7}$ & $\bm{272.1}$ & 1084.8 & 195.3 & 3868.3 & 132.2 & 615.3 $\pm$ 294.5& 98.8\% \\
 &      $E$ & $\bm{704.7}$ & $\bm{218.3}$ & 951.7 & 142.3 & 5376.1 & 496.6 & 660.7 $\pm$ 202.3& 96.6\%  \\
 &      $I$ & $\bm{439.0}$ & $\bm{140.4}$ & 556.2 & 90.6 & 3167.0 & 404.9  & 415.0 $\pm$ 158.7 & 96.4\% \\
&      $D$ & 38.3 & 4.9 & $\bm{33.3}$ & $\bm{5.0}$ & 907.3 & 48.6   & 14.0 $\pm$ \ \ \ \ 5.2 & 94.2\%  \\
 \bottomrule
  \multicolumn{2}{c}{Computing Time (s)}  & 1006.7 & 115.54 & 2904.4 & 195.5 & 7.3 & 0.4 & - \\
\bottomrule
\end{tabular}
}
\end{center}
\end{table}

In the experiment we set $v^i=0.1$, $\beta_t = 1.8 - \cos(\pi t / 8)$, $v^e_t=0.1 - 0.02\cos(\pi t/8)$, $p^d_t = 0.05 + 0.025 \cos(\pi t/8)$, and focus on a time horizon of 32 days. The initial values of four components are set as $(100000, 100, 50, 50)$ for $(S,E,I,D)$. We assume $S$, $I$, $D$ are observed on daily frequency with log-normal multiplicative observation noise at 3\% level. The exposed population $E$ is assumed to be only sparsely observable at 3\% noise level, with one observation per two days, due to the high cost of data acquisition from sampling test. Such pandemic settings (\cite{dong2020interactive, mwalili2020seir}) capture the periodic fluctuation of parameters often observed in the real world. 

We apply TVMAGI on a log-transformed system (by taking the log of populations in each of the $S, E, I, D$ state) over 100 simulation datasets, with 2 discretizations per day. Figure \ref{fig:TVMAGI-seird} shows the results of parameter inference and the TVMAGI reconstructed trajectory $\bm{X}(\bm{I})$ of the ODE system. The parameter RMSE and trajectory RMSE introduced in Section \ref{sec:evaluation} are presented in Table \ref{tab:seird-1}, where TVMAGI is shown to be more accurate than Runge-Kutta or EAKF.

For point estimates, Figure \ref{fig:TVMAGI-seird} and Table \ref{tab:seird-1} shows that, even when the exposed population is sparsely observed, TVMAGI is still capable of providing good results of inference. As the most important parameter when assessing the spread of disease, $\beta_t$ can be accurately and robustly inferred. $v^i$ can also be accurately inferred as constant. $p^d_t$ has larger variability at the start, as initial deaths are too few to provide enough information. In comparison, the variability of $v^e_t$ inference increases at the end of the period, because susceptible population has decreased to nearly zero while infectious population reaches plateau. Despite variations in the inferred parameters, the inferred system trajectories are all very close to the truth, confirming the intuition that the system is possibly less sensitive to $p^d$ in earlier state and $v^e$ in later stage. Supplementary Material Figure \ref{fig:compare-seird-param} and Figure \ref{fig:compare-seird-x} has the visual illustration for Runge-Kutta or EAKF, and their accuracy is far from satisfactory: Runge-Kutta method will overfit the observation noise, and EAKF reconstructed trajectory completely misses the truth.

\begin{figure}[htp]
     \centering
         \includegraphics[width=2in]{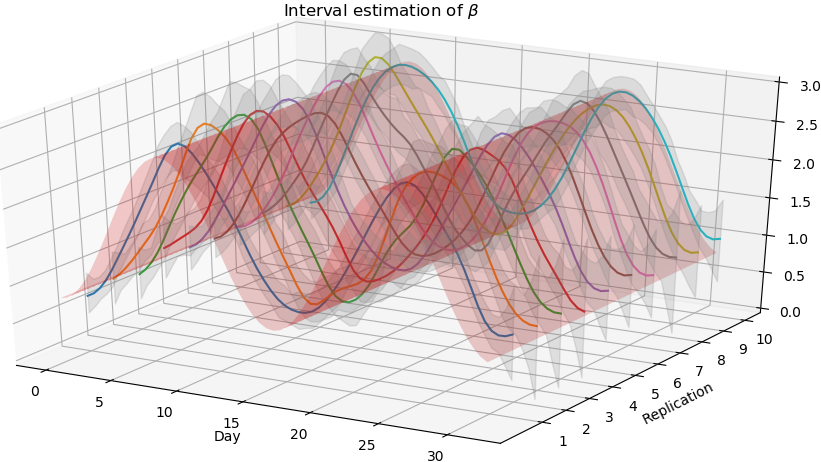}
         \includegraphics[width=2in]{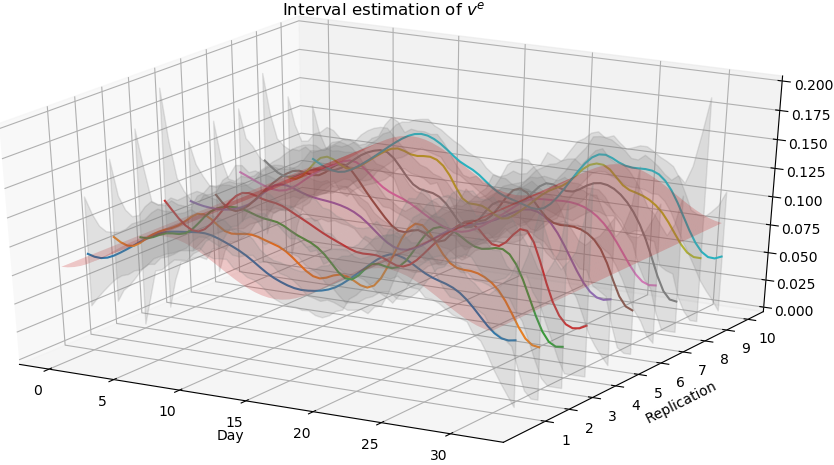}
         \includegraphics[width=2in]{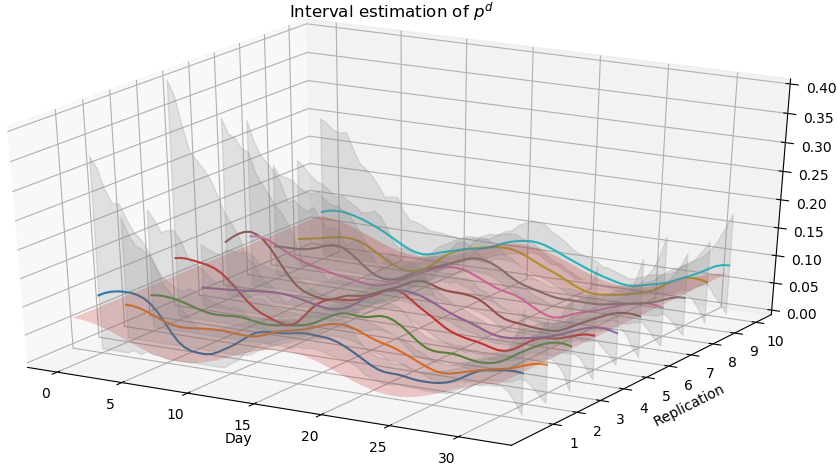}
        \caption{Illustration of interval estimation of SEIRD model of 10 sample datasets for $\beta$ (left), $v^e$ (middle), $p^d$ (right). The shadow indicates the 95\% posterior interval from HMC samples, the solid lines indicate the posterior mean, and the red surface indicates the true value.}
        \label{fig:seird_sampling}
\end{figure}

For Interval estimates, Figure \ref{fig:seird_sampling} gives a visual illustration for 10 sample datasets. The coverage of Posterior Interval across 100 simulated datasets is included in Table \ref{tab:seird-1}. The emperical coverage of the interval is reasonable around the 95\% nominal value. The intervals are wider for $p^d$ at the starting time, and wider for $v^e$ at the ending time, which are consistent with the intuition about their sensitivity discussed above. More interval estimation results are available in Supplementary Material Section \ref{sec:seird-hmc}.

On the computational cost, Table \ref{tab:seird-1} also shows that TVMAGI is much faster than the Runge-Kutta numerical integration methods. EAKF is fast, but gives unreliable results (see Supplementary Material Section \ref{sec:bayesian-problems} for more discussion on the reliability of EAKF).

\subsection{Lotka-Volterra Model}

Lotka-Volterra (LV) model (a.k.a. predator-prey model) is widely used to describe population fluctuation of predators and preys and their interactions in the ecosystem (\cite{goel1971volterra}). With the introduction of time-varying parameters, the system becomes weakly identifiable during certain time range, which creates a challenge in the inference. Specifically, the ODE system is characterized as:
\begin{small}
\begin{equation}
\frac{dx}{dt}=\alpha_t x - \beta xy, \quad
\frac{dy}{dt}=\delta xy - \gamma_t y 
\end{equation}
\end{small}
where $x$ and $y$ denote the population of preys and predators. $\alpha_t$ indicates the birth rate of the prey and $\gamma_t$ denotes the death rate of the predator, both of which are assumed to fluctuate according to seasonality. $\beta$ and $\delta$ describe the interaction relationships between predators and preys, and are assumed constant. We set the parameters $\beta=0.75$, $\delta=1$, $\alpha_t=0.6 + 0.3 \cos(\pi t/5)$, and $\gamma_t=1 + 0.1 \sin(\pi t /5)$. The time is measured on a yearly basis, and data for 20 years are generated with monthly observations contaminated by 3\% multiplicative log-normal noise. The initial values of predators and preys are 1 and 3, as an ideal ratio in real ecology systems (\cite{donald2003resistance}).

\begin{figure}[htp]
     \centering
     \begin{subfigure}[b]{1\textwidth}
         \centering
         \includegraphics[width=5in]{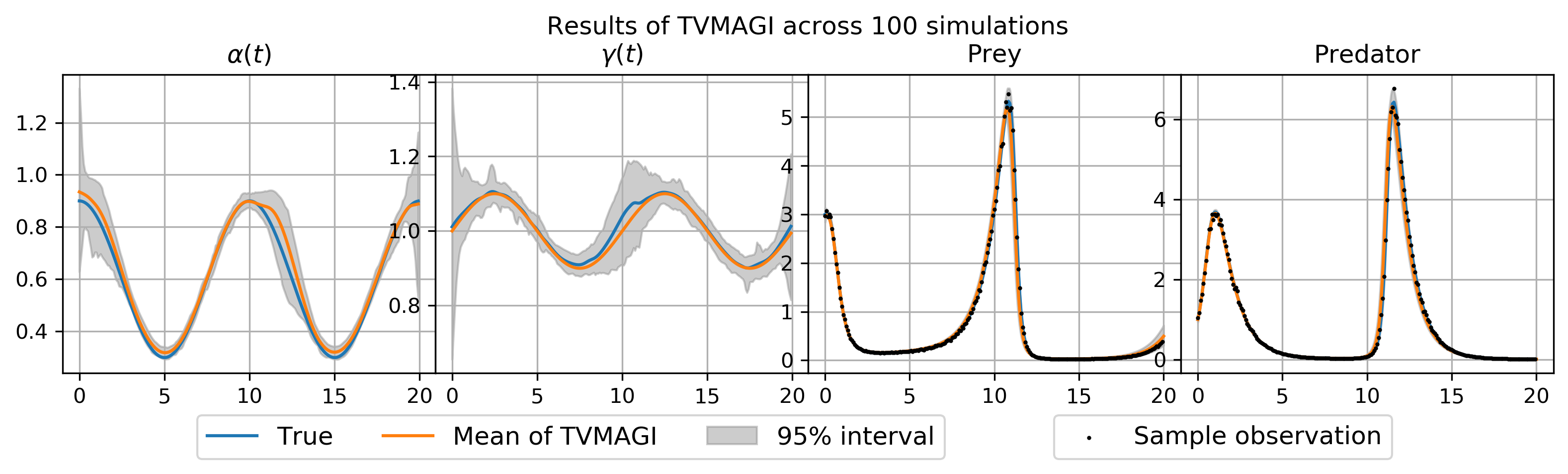}
     \end{subfigure}
\caption{Comparison of inferred $\bm{\theta}(t)$ and reconstructed $\bm{X}(t)$ of LV model. The mean and the 95\% interval here refer to the point estimates across 100 simulated datasets. One sample simulation dataset is also plotted to visualize the noise level.}
\label{fig:lv}
\end{figure}




\begin{table}[htp]
\begin{center}
\caption{Accuracy comparison of estimated parameters and reconstructed trajectory in the LV model based on 100 simulation datasets. See legend of Table \ref{tab:seird-1} for detailed description.}
\label{tab:res_lv}
\scalebox{0.7}{
\begin{tabular}{
  cc
  >{}r<{}@{$\pm$}>{}r<{}>{}r<{}@{$\pm$}>{}r<{}>{}r<{}@{$\pm$}>{}r<{} | c|c
}
\toprule+

&  & \multicolumn{6}{c|}{Point Estimate} & \multicolumn{2}{c}{TVMAGI Posterior Samples} \\
 \midrule
 & RMSE & \multicolumn{2}{c}{TVMAGI} & \multicolumn{2}{c}{Runge-Kutta} & \multicolumn{2}{c|}{EAKF} & Posterior mean RMSE & Interval Coverage \\
\midrule
\parbox[t]{2mm}{\multirow{4}{*}{\rotatebox[origin=c]{90}{Parameter}}} & $\alpha(t)$ & $\bm{0.0395}$ & $\bm{0.0302}$ & 0.1155 & 0.0920 & 0.2330 & 0.1589 & 0.0450 $\pm$ 0.0346 & 40.5\% \\
 & $\beta$ & $\bm{0.0154}$ & $\bm{0.0098}$ & 0.0480 & 0.0069 & 0.1041 & 0.0621 & 0.0187 $\pm$ 0.0122 & 62.0\% \\
 & $\delta$ & $\bm{0.0156}$ & $\bm{0.0111}$ & 0.0166 & 0.0058 & 0.1605 & 0.0831 & 0.0160 $\pm$ 0.0119 & 58.0\% \\
& $\gamma(t)$ & $\bm{0.0304}$ & $\bm{0.0226}$ & 0.0863 & 0.0756 & 0.0974 & 0.1847 & 0.0341 $\pm$ 0.0201 & 46.0\% \\

\midrule
\parbox[t]{2mm}{\multirow{2}{*}{\rotatebox[origin=c]{90}{Traj.}}} & $x$ (prey) & 0.0606 & 0.0550 & $\bm{0.0314}$ & $\bm{0.0085}$ &  0.4701 & 0.0903 & 0.0838  $\pm$ 0.0568 & 69.1\% \\
 & $y$ (predator) & 0.0813 & 0.0622 & $\bm{0.0384}$ & $\bm{0.0131}$ & 0.2773 & 0.0646 & 0.0989  $\pm$ 0.0714 & 62.2\%  \\
 \bottomrule
  \multicolumn{2}{c}{Computing Time (s)}  & 910.8 & 113.7 & 2042.1 & 85.2 & 22.7 & 1.1 & - \\
\bottomrule
\end{tabular}
}
\end{center}
\end{table}

Figure~\ref{fig:lv} shows the estimated time-varying parameters and the reconstructed trajectory $\bm{X}(\bm{I})$, with parameter RMSE and trajectory RMSE presented in Table~\ref{tab:res_lv}. Our recovered system components $x$ and $y$ are very close to the truth, despite the weak identifiability of the parameter $\alpha_t$ and $\gamma_t$ when the $x$ and $y$ are at peak (year=12). Most notably, $\alpha_t$ could deviate from the truth in the attempt to best fit the observed noisy data at the peak of $x_t$, resulting in a biased inference of the time-varying parameters at the weakly identifiable time points, although all deviations are still within the range of smoothness constraints on $\alpha_t$. Nevertheless, both TVMAGI inferred system components $x$ and $y$ are still accurate. 

Comparing with benchmark models in Table~\ref{tab:res_lv}, TVMAGI gives the most accurate parameter inference thanks to the GP smoothing prior that mitigates the identifiability issue. The numerical method of Runge-Kutta gives better trajectory inference, but it cannot handle the identifiability issue in the parameters (see also Supplementary Material Figure \ref{fig:compare-lv}). The coverage from TVMAGI is not ideal, possibly due to the bias in $\alpha(t)$ estimate and the variance in $\gamma(t)$ estimate -- if the GP smoothing prior is too strong, the point estimates will be biased, and if the GP smoothing prior is too weak, the point estimates will have large variance (see Supplementary Material Figure \ref{fig:LV-hmc}). The comparison on computational cost again demonstrates the expected advantage of TVMAGI over Runge-Kutta, while EAKF is the fastest method with the worst accuracy.

This example illustrates the performance of TVMAGI in the presence of weak identifiability -- the inferred time-varying parameters at the weakly identified time points could subject to deviation from the truth, although the parameters are smooth and still fit the observed data well.

\subsection{HIV model}
In this example, we compare TVMAGI with a state-of-the-art two-stage Efficient Local Estimation (ELE) method proposed by \cite{HLW} in an HIV dynamic model that they studied. This is a challenging case for GP modeling as the true time-varying parameter has non-periodic non-stationary trends with rapid changes. The ODE model that characterizes the response of anti-viral regimens during HIV infection is given by:
\begin{align}\label{eq:ode-hiv}
\left\{
\begin{aligned}
& T'(t) \equiv \frac{d}{dt} T(t)=\lambda-\rho T(t)-k[1-r(t)]T(t)X(t) \\
& T^{*\prime}(t) \equiv \frac{d}{dt} T^{*}(t) =k[1-r(t)]T(t)X(t)-\delta T^*(t) \\
& X'(t) \equiv \frac{d}{dt} X(t) =N\delta T^*(t) -c X(t) 
\end{aligned}\right.
\end{align}
$T(t)$ denotes the concentration of uninfected CD4+ T cells, which can be accurately measured clinically; $T^*(t)$ denotes the unknown unobservable concentration of infected T cells; $X(t)$ is the HIV-1 viral load in plasma, and can be observed with noise. $\lambda$ is the rate of new T cell generation; $\rho$ is the death rate of T cells; $k$ is the infection rate of T cells by HIV virus; $\delta$ is the death rate of infected cells; $N$ is the total production of new virions by an infected T cell; $c$ denotes the known constant rate of free virion clearance (\cite{perelson1996hiv}); $r(t)$ is the time-varying antiviral drug efficacy coefficient, which may decay through time due to drug resistance. Our simulation settings are based on \cite{HLW, huang2003modeling} as  $\lambda=36$, $\rho=0.108$, $k=5 \times 10^{-4}$, $\delta=0.1$, $N=1000$, $c=3.5$, $X(0)=1000$, $T(0)=350$, $T^*(0)=20$, and $r (t ) = \cos(\pi t/500)$. Time horizon is set as 100 days, with observation noise level at 5\%. To use the ELE method of \cite{HLW}, the ODE system must fit in the linear form of Eq.\eqref{simplified_hiv}:
\begin{equation}\label{simplified_hiv}
X'(t) = \sum_{i=1}^d a_i(t)Z_i(t) -c X(t)
\end{equation}
where $Z_i(t)$ is the known covariate, and $a_i(t)$ is the unknown time-varying coefficient. \cite{HLW} transformed the system Eq.\eqref{eq:ode-hiv} into Eq.\eqref{simplified_hiv} by taking $d=2$, $a_1(t)=-NT^{*\prime}(t)$, $a_2(t)= N k[1-r(t)]X(t)$, $Z_1(t)=1$ and $Z_2(t)=T(t)$, and then used their ELE method estimate time-varying coefficients $a_i(t)$. For benchmark comparison, we treat $a_1(t)$ and $a_2(t)$ in Eq.\eqref{simplified_hiv} as unknown time-varying parameters for TVMAGI. 

\begin{figure}[htp]
     \centering
     \begin{subfigure}[b]{1\textwidth}
         \centering
         \includegraphics[width=6in]{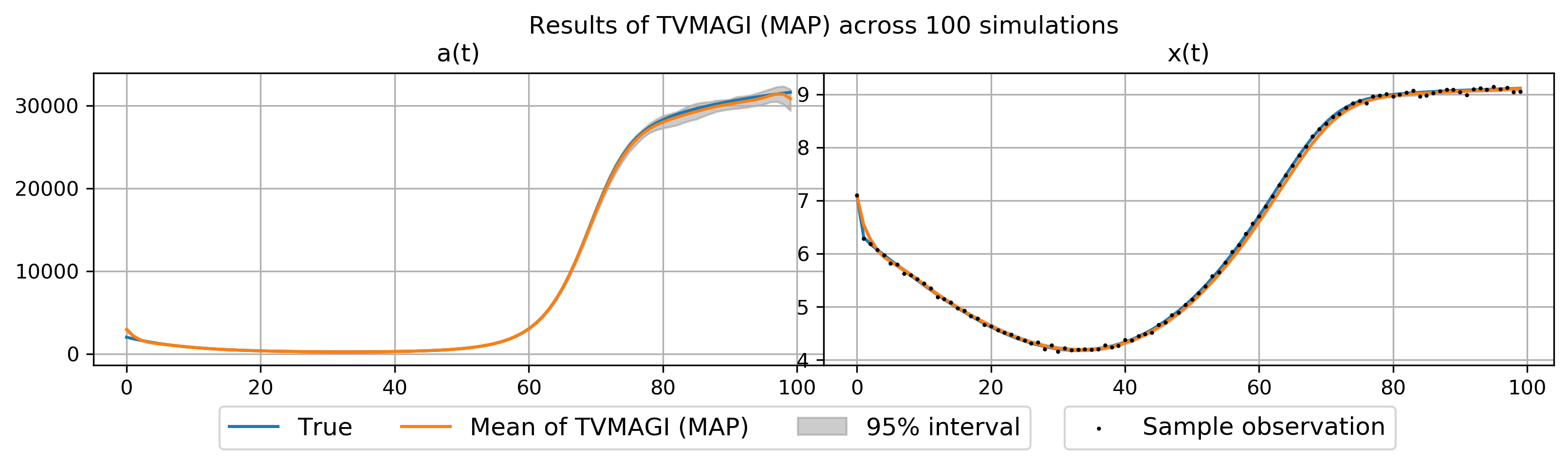}
     \end{subfigure}
\caption{TVMAGI inferred $a(t)$ and reconstructed $\bm{X}(t)$ of HIV model.}
\label{fig:hiv_reconstruct}
\end{figure}

Figure~\ref{fig:hiv_reconstruct} shows the TVMAGI inferred time-varying parameter $a(t) = \sum_{i=1}^d a_i(t)Z_i(t) $ and the reconstructed trajectory ${X}(t)$. The parameter/trajectory RMSEs of TVMAGI and the benchmark methods are reported in Table \ref{tab:hiv}. TVMAGI has a small advantage over the state-of-the-art method on HIV model inference of $X(t)$. Further visual comparison to benchmark methods (Supplementary Material Figure \ref{fig:compare-hiv}) shows that TVMAGI is slightly more accurate at the beginning phase of the system, which is in fact the most challenging phase for HIV inference as viral load drops sharply due to the drug effect. TVMAGI also achieves competitive inference result on $a(t)$, which is of clinical importance for the generation rate of HIV virus (\cite{cao2012penalized}). The TVMAGI posterior interval coverage is less ideal because of the decreased accuracy in $a(t)$ towards the ending period (Supplementary Material Figure \ref{fig:hiv-hmc}). Most importantly, while the benchmark method requires a highly restricted form of ODE formulation, TVMAGI assumes no specific form of ODE equations, and is thus applicable for general ODE systems, albeit with longer computing time. 


\begin{table}[htp]
\begin{center}
\caption{Accuracy comparison of estimated parameters and trajectories in the HIV model based on 100 simulation datasets. See legend of Table \ref{tab:seird-1} for detailed description.}
\label{tab:hiv}
\scalebox{0.7}{
\begin{tabular}{
  c
  >{}r<{}@{$\pm$}>{}r<{} >{}r<{}@{$\pm$}>{}r<{} >{}r<{}@{$\pm$}>{}r<{} >{}r<{}@{$\pm$}>{}r<{} | c|c
}
\toprule
  & \multicolumn{8}{c|}{Point Estimate} & \multicolumn{2}{c}{TVMAGI Posterior Samples}\\
 \midrule
  RMSE & \multicolumn{2}{c}{TVMAGI} & \multicolumn{2}{c}{Runge-Kutta} & \multicolumn{2}{c}{EAKF} & \multicolumn{2}{c|}{ ELE } & Posterior mean RMSE & Interval Coverage\\
\midrule
  $a(t)$ & $\bm{281.9}$ & $\bm{71.8}$ & 695.7 & 50.9 & 818.6 & 54.9 & 291.5 & 48.4  & 359.0 $\pm$ 127.8 & 74.1\% \\
\midrule
  $x(t)$  & 0.057 & 0.002 & $\bm{0.038}$ & $\bm{0.003}$ &  0.181 & 0.004 & 0.075 & 0.003 & 0.069 $\pm$ 0.004 & 65.9\% \\
  \bottomrule
  Computing Time (s) & 897.6 & 72.1 & 1940.2 & 79.7 & 5.4 &  0.3 & 10.7 & 0.1 & - & - \\
\bottomrule
\end{tabular}
}
\end{center}
\end{table}

Overall in this example, we compare TVMAGI with an additional benchmark method that can only be applied to the ODEs with a specific form, where TVMAGI is shown to provide competitive inference accuracy while having much more general applicability. The application in HIV model also illustrates that TVMAGI could work well with non-stationary trends in the time-varying parameters, where the time-varying parameters is not periodic and have rapid changes in part of the time horizon. 

\section{Sensitivity analysis}

In this section we conduct two sensitivity analysis: the number of discretization and the selection of GP kernel. For the number of discretization, our theoretical derivation ensures that the inference result will converge as the discretization increase, and we recommend gradually increasing the number of discretization points until the result is stabilized. Here we empirically demonstrate such convergence by presenting results with various discretization level. For the GP kernel selection, we relax the GP kernel of time-varying parameter to be Matern kernel with different degrees of freedom $\nu=2.5$ and $\nu=1.5$. When $\nu=1.5$, $\bm \theta(t)$ is only required to be continuous, but not necessarily differentiable. Both experiments are conducted under the SEIRD model, with table summaries presented here in this section. More visualizations are available in Supplementary Material Section \ref{sec:SI-sensitivity}.

\subsection{Number of discretization}

In this section we explore the sensitivity of TVMAGI to the number of discretization. In the paper we used discretization level of 2 in the SEIRD model, that is, with 32 observation points, we have a total of 64 discretization points (2 discretization per observation). For comparison, we use the same observation data with discretization level of 1 and level of 4, corresponding to 32 discretization points (1 discretization per observation) and 128 discretization points (4 discretization per observation), respectively. Table \ref{tab:methodcompare} shows that the inference accuracy indeed converges. However, the computational time scales up linearly with the number of discretization points. In practice, we recommend gradually increasing the number of discretization points until the result is stabilized, trying to balance the inference accuracy with the computing cost.

\begin{table}

\caption{Parameter and trajectory RMSE of TVMAGI in SEIRD model under different discretization level. The computing time (in seconds) is reported in the last row.}  
\label{tab:methodcompare}
\begin{center}
\scalebox{0.7}{
\begin{tabular}{c r|r|r|r}
\toprule
& \multirow{2}*{RMSE} & \multicolumn{3}{c}{Discretization Level}  \\ 
\cline{3-5}
&         & 1 & 2 & 4 \\
\midrule 
\parbox[t]{2mm}{\multirow{4}{*}{\rotatebox[origin=c]{90}{Parameter}}} & $\beta$ & 0.156 $\pm$ 0.083 & 0.114 $\pm$ 0.039 & 0.102 $\pm$ 0.036 \\
& $v^e$ & 0.007 $\pm$ 0.007 & 0.009 $\pm$ 0.010 & 0.010 $\pm$ 0.010 \\
& $v^i$ & 0.006 $\pm$ 0.004 & 0.005 $\pm$ 0.003 & 0.005 $\pm$ 0.003 \\
& $p^d$  & 0.018 $\pm$ 0.027 & 0.019 $\pm$ 0.029 & 0.021 $\pm$ 0.033 \\
\cdashline{1-5}[0.8pt/2pt]
\parbox[t]{2mm}{\multirow{4}{*}{\rotatebox[origin=c]{90}{Trajectory}}} & $S$  & 1253.4 $\pm$\  332.0 & 581.7 $\pm$ 272.1  & 603.9 $\pm$ 355.2 \\
& $E$  & 1010.6 $\pm$ 231.9 & 704.7 $\pm$ 218.3 & 679.5 $\pm$ 209.1 \\
& $I$ & 584.6 $\pm$\ 195.6 & 439.0 $\pm$ 140.4 & 401.0 $\pm$ 132.8 \\
& $D$ & \ \ 44.7 $\pm$\ \  \ \ 8.8 & \ \ 38.3  $\pm$  \ \ \ 4.9 & \ \ 39.0 $\pm$ \ \ \ 3.4 \\
\midrule
\multicolumn{2}{c|}{computing time (s)} & \ \ 622.1  $\pm$ \ 49.6 & 1013.3  $\pm$  109.1 & 1773.9 $\pm$  161.3 \\
\bottomrule
\end{tabular}
}
\end{center}
\vspace{-1.5em}
\end{table}

\subsection{Selection of kernel}
In this section we discuss how the kernel selection will affect the performance of TVMAGI. In the paper we recommend modeling $\bm{\theta}(t)$ as Gaussian process with Matern kernel $\nu=2.01$ to guarantee a continuous and differentiable time-varying parameters while maintaining high flexibility. We can also use other GP kernels or hyperparameters to control the smoothness. For example, Matern kernel with $\nu=2.5$ can be used for even more smooth GP with a simple closed-form kernel. The condition of differentiability can also be further relaxed if we substitute the kernel with $\nu=1.5$, and then the parameters are only assumed with continuity without differentiability, allowing more flexible patterns for the time-varying parameters $\bm{\theta}(t)$. Table \ref{tab:kernel-seird} shows the result under both kernels, where the the performance is similar to the recommended $\nu = 2.01$ in SEIRD model, indicating that TVMAGI is not sensitive to the choice of kernels.

\begin{table}

\caption{Parameter and trajectory RMSE of TVMAGI in SEIRD using different kernels.}  
\label{tab:kernel-seird}
\begin{center}
\scalebox{0.7}{
\begin{tabular}{c r|r|r|r}
\toprule
& \multirow{2}*{RMSE} & \multicolumn{3}{c}{$\nu$ for $\bm{\theta}(t)$}  \\ 
\cline{3-5}
&         & 1.5 & 2.5 & 2.01 \\
\midrule 
\parbox[t]{2mm}{\multirow{4}{*}{\rotatebox[origin=c]{90}{Parameter}}} & $\beta$ & 0.085 $\pm$ 0.005 & 0.139 $\pm$ 0.021 & 0.114 $\pm$ 0.039  \\
& $v^e$ & 0.012 $\pm$ 0.004  &  0.015 $\pm$ 0.005 & 0.009 $\pm$ 0.010 \\
& $v^i$ &  0.004 $\pm$ 0.005 &  0.003 $\pm$ 0.004 & 0.005 $\pm$ 0.003 \\
& $p^d$  & 0.058 $\pm$ 0.004 & 0.044 $\pm$ 0.012 & 0.019 $\pm$ 0.029 \\
\cdashline{1-5}[0.8pt/2pt]
\parbox[t]{2mm}{\multirow{4}{*}{\rotatebox[origin=c]{90}{Trajectory}}} & $S$  & 423.0 $\pm$ 130.2 & 853.3 $\pm$ 288.5 & 581.7 $\pm$ 272.1  \\
& $E$  &  412.3 $\pm$ 111.7 & 836.1 $\pm$ 186.5 & 704.7 $\pm$ 218.3\\
& $I$ &  356.3 $\pm$ \ \ 94.1 & 569.4 $\pm$ 171.8 &  439.0 $\pm$ 140.4 \\
& $D$ &  69.1 $\pm$ 21.5 & 20.4 $\pm$ \ 8.6 & 38.3 $\pm$\ \ 4.9 \\
\bottomrule
\end{tabular}
}
\end{center}
\vspace{-1.5em}
\end{table}

\section{Discussion}\label{sec:discussion}

In this paper, we introduce a Bayesian approach, TVMAGI, for time-varying parameters inference in ODE dynamic systems. TVMAGI models time-varying parameters and system components as Gaussian process, and is constrained to have the derivative processes satisfy the ODE dynamics. We show that TVMAGI is statistically principled and illustrate its general applicability through three simulation examples. Results have shown that TVMAGI yields accurate and robust parameter inference from noisy observations, with reasonable interval estimates as well. Moreover, TVMAGI can mitigate the identifiability issue and the overfitting issue in the time-varying parameters using the informative GP smoothing prior. TVMAGI is also generally applicable in the presence of missing observations. 

TVMAGI is more accurate than the benchmark methods because TVMAGI addresses the challenges of the numerical integration method and the Bayesian filtering method for ODE time-constant and time-varying parameter inference. Numerical integration methods are the gold standard for the ODE parameter inference when all parameters are time-constant. However, with the presence of time-varying parameters, due to the lack of smoothness structure on $\theta(t)$, the inferred time-varying parameters from Runge-Kutta will overfit the noisy observation data, resulting in volatile $\theta(t)$ with little information about the true trends. The Bayesian filtering approach, on the other hand, cannot infer time-constant parameter $\bm\psi$ because the update in $\bm\psi$ is not permissible in a state-space model fashion. We can nevertheless enforcing an update on the time-constant parameter, but there will be no guarantee on the accuracy of the reconstructed trajectory. The ELE two-stage approach relies on a regression technique that can only be used if the ODE has linear dependency on the time-varying parameters. Therefore, TVMAGI is the only approach that is theoretically sound, practically accurate, and generally applicable for the ODE inference problem when time-constant and time-varying parameters co-exist.

On the computational time comparison, TVMAGI has the notable advantage of reduced computation cost compared to numerical integration method, while the inference is more accurate compared to the fast-yet-unreliable Bayesian filtering methods. Even for the three small-sized illustration problems in this paper, TVMAGI is more than twice as fast than the numerical integration method of Runge-Kutta with better accuracy. When dealing with large-scale system, the gain in computational time is likely to be even larger, as TVMAGI computational time would scale linearly as the dimension of system components grow, while inference with numerical integration method such as Runge-Kutta typically scales super linearly. Therefore, TVMAGI has strong potential in large-scale systems, where numerical integration is expensive. 

There are two settings that may require tuning in TVMAGI. First, the number of discretization can affect the inference results. When observed components are sparse, the number of discretization should increase until the results are stabilized. However, over-densed discretization will lead to higher computation cost. For example, in SEIRD model, we set discretization as 2 data points per day for optimized performance, as further increasing the discretization will not improve the result accuracy. Second, the inference results on TVMAGI can be affected by hyper-parameter settings of the GP kernel for $\bm{\theta}(t)$. To achieve the desired variability level of time-varying parameters, we find it helpful to use informative hyper-prior that specifies the range of length-scale (a.k.a. bandwidth) parameter of the GP kernel for $\bm{\theta}(t)$ to prevent obvious over-smoothing or over-fitting.

The Gaussian process modeling of $\bm\theta(t)$ with Matern kernel $\nu=2.01$ ensures continuously differentiable time-varying parameters, which prevents overfitting the parameter to the observation noise. The variability in time-varying parameter $\bm\theta (t)$ can be further controlled by the length-scale GP hyperparameter $\phi_2$ through its hyper-prior. The Matern kernel together with the hyper-prior on the length-scale hyperparameter ensures the smoothness and the degree of variability in $\bm \theta(t)$, which in turn prevents over-fitting and mitigates identifiability issues. If a more flexible $ \bm\theta (t)$ is desired, Matern kernel $\nu=1.5$ with hyper-prior favoring smaller GP hyperparameter $\phi_2$ can be used to allow rapid non-differentiable changes in $\bm\theta (t)$.

One limitation of TVMAGI is its inherent bias. Just like any other Bayesian approaches, TVMAGI could be biased towards smoother curves due to the GP prior. The results of inference would be less accurate when the true time-varying parameters have rapid changes, and the posterior interval coverage could suffer. But as shown in the examples, the magnitude of such bias is small in practice, and our accuracy is still comparable with state-of-the-art approaches while TVMAGI having much better universal applicability. 

Although we carefully designed the multi-stage optimization and sampling schedule, occasionally the Adam optimizer or the HMC sampler could still get stuck. Among the total of 300 simulated datasets across three examples, the algorithm got stuck in one particular dataset of the SEIRD model. In the stuck case, some manual tuning of the hyper-parameters or jittering of the sampled parameters might be needed. We will continue to improve the robustness of our proposed algorithm and our software implementation.

TVMAGI is also not suitable if the underlying time-varying parameter is a jump process. In this case, methods in change point detection literature might be more applicable \cite{cuenod2011parameter}. Alternatively, we can place the prior of continuous-time Markov chain or Poisson process on $\theta(t)$, instead of Gaussian process, to model the jump process.

There are also many other interesting future directions for TVMAGI. We currently focus on empirical performance of TVMAGI through simulation examples. More theoretical study on the convergence property, identifiability issue, and asymptotic behavior of the time-varying parameter estimate are all natural directions of future research. The deterministic systems specified by ODEs are studied in this paper, and it would be of future interest to extend TVMAGI for partial differential equation of spatial-temporal dynamics \cite{xun2013parameter}, or stochastic differential equation of inherent noise modeling \cite{kou2004generalized}.

\bibliographystyle{chicago}

\bibliography{References}

\begin{thebibliography}{}

\bibitem[\protect\citeauthoryear{Abbott, Hellewell, Thompson, Sherratt, Gibbs,
  Bosse, Munday, Meakin, Doughty, Chun, et~al.}{Abbott
  et~al.}{2020}]{abbott2020estimating}
Abbott, S., J.~Hellewell, R.~N. Thompson, K.~Sherratt, H.~P. Gibbs, N.~I.
  Bosse, J.~D. Munday, S.~Meakin, E.~L. Doughty, J.~Y. Chun, et~al. (2020).
\newblock Estimating the time-varying reproduction number of sars-cov-2 using
  national and subnational case counts.
\newblock {\em Wellcome Open Research\/}~{\em 5\/}(112), 112.

\bibitem[\protect\citeauthoryear{Cao, Huang, and Wu}{Cao
  et~al.}{2012}]{cao2012penalized}
Cao, J., J.~Z. Huang, and H.~Wu (2012).
\newblock Penalized nonlinear least squares estimation of time-varying
  parameters in ordinary differential equations.
\newblock {\em Journal of computational and graphical statistics\/}~{\em
  21\/}(1), 42--56.

\bibitem[\protect\citeauthoryear{Chen and Wu}{Chen and Wu}{2008}]{HLW}
Chen, J. and H.~Wu (2008).
\newblock Efficient local estimation for time-varying coefficients in
  deterministic dynamic models with applications to {HIV-1} dynamics.
\newblock {\em Journal of the American Statistical Association, 103:481,
  369-384\/}.

\bibitem[\protect\citeauthoryear{Cuenod, Favetto, Genon-Catalot, Rozenholc, and
  Samson}{Cuenod et~al.}{2011}]{cuenod2011parameter}
Cuenod, C.-A., B.~Favetto, V.~Genon-Catalot, Y.~Rozenholc, and A.~Samson
  (2011).
\newblock Parameter estimation and change-point detection from dynamic contrast
  enhanced mri data using stochastic differential equations.
\newblock {\em Mathematical biosciences\/}~{\em 233\/}(1), 68--76.

\bibitem[\protect\citeauthoryear{Donald and Stewart~Anderson}{Donald and
  Stewart~Anderson}{2003}]{donald2003resistance}
Donald, D.~B. and R.~Stewart~Anderson (2003).
\newblock Resistance of the prey-to-predator ratio to environmental gradients
  and to biomanipulations.
\newblock {\em Ecology\/}~{\em 84\/}(9), 2387--2394.

\bibitem[\protect\citeauthoryear{Dondelinger, Husmeier, Rogers, and
  Filippone}{Dondelinger et~al.}{2013}]{dondelinger2013ode}
Dondelinger, F., D.~Husmeier, S.~Rogers, and M.~Filippone (2013).
\newblock Ode parameter inference using adaptive gradient matching with
  gaussian processes.
\newblock In {\em Artificial intelligence and statistics}, pp.\  216--228.
  PMLR.

\bibitem[\protect\citeauthoryear{Dong, Du, and Gardner}{Dong
  et~al.}{2020}]{dong2020interactive}
Dong, E., H.~Du, and L.~Gardner (2020).
\newblock An interactive web-based dashboard to track covid-19 in real time.
\newblock {\em The Lancet infectious diseases\/}~{\em 20\/}(5), 533--534.

\bibitem[\protect\citeauthoryear{Donnet and Samson}{Donnet and
  Samson}{2013}]{donnet2013review}
Donnet, S. and A.~Samson (2013).
\newblock A review on estimation of stochastic differential equations for
  pharmacokinetic/pharmacodynamic models.
\newblock {\em Advanced drug delivery reviews\/}~{\em 65\/}(7), 929--939.

\bibitem[\protect\citeauthoryear{Goel, Maitra, and Montroll}{Goel
  et~al.}{1971}]{goel1971volterra}
Goel, N.~S., S.~C. Maitra, and E.~W. Montroll (1971).
\newblock On the volterra and other nonlinear models of interacting
  populations.
\newblock {\em Reviews of modern physics\/}~{\em 43\/}(2), 231.

\bibitem[\protect\citeauthoryear{Golightly and Wilkinson}{Golightly and
  Wilkinson}{2010}]{golightly2010markov}
Golightly, A. and D.~J. Wilkinson (2010).
\newblock Markov chain monte carlo algorithms for sde parameter estimation.
\newblock {\em Learning and Inference for Computational Systems Biology\/},
  253--276.

\bibitem[\protect\citeauthoryear{Hao, Cheng, Wu, Wu, Lin, and Wang}{Hao
  et~al.}{2020}]{hao2020reconstruction}
Hao, X., S.~Cheng, D.~Wu, T.~Wu, X.~Lin, and C.~Wang (2020).
\newblock Reconstruction of the full transmission dynamics of covid-19 in
  wuhan.
\newblock {\em Nature\/}~{\em 584\/}(7821), 420--424.

\bibitem[\protect\citeauthoryear{Hethcote}{Hethcote}{2000}]{hethcote2000mathematics}
Hethcote, H.~W. (2000).
\newblock The mathematics of infectious diseases.
\newblock {\em SIAM review\/}~{\em 42\/}(4), 599--653.

\bibitem[\protect\citeauthoryear{Huang and Liu}{Huang and
  Liu}{2006}]{huang2006}
Huang, Y. and H.~Liu, D.and~Wu (2006).
\newblock Hierarchical bayesian methods for estimation of parameters in a
  longitudinal hiv dynamic system.
\newblock {\em Biometrics\/}~{\em 62\/}(2), 413–423.

\bibitem[\protect\citeauthoryear{Huang, Rosenkranz, and Wu}{Huang
  et~al.}{2003}]{huang2003modeling}
Huang, Y., S.~L. Rosenkranz, and H.~Wu (2003).
\newblock Modeling hiv dynamics and antiviral response with consideration of
  time-varying drug exposures, adherence and phenotypic sensitivity.
\newblock {\em Mathematical biosciences\/}~{\em 184\/}(2), 165--186.

\bibitem[\protect\citeauthoryear{Kingma and Ba}{Kingma and
  Ba}{2014}]{kingma2014adam}
Kingma, D.~P. and J.~Ba (2014).
\newblock Adam: A method for stochastic optimization.
\newblock {\em arXiv preprint arXiv:1412.6980\/}.

\bibitem[\protect\citeauthoryear{Kou and Xie}{Kou and
  Xie}{2004}]{kou2004generalized}
Kou, S.~C. and X.~S. Xie (2004).
\newblock Generalized langevin equation with fractional gaussian noise:
  subdiffusion within a single protein molecule.
\newblock {\em Physical review letters\/}~{\em 93\/}(18), 180603.

\bibitem[\protect\citeauthoryear{Lapidus and Seinfeld}{Lapidus and
  Seinfeld}{1971}]{lapidus1971numerical}
Lapidus, L. and J.~H. Seinfeld (1971).
\newblock {\em Numerical solution of ordinary differential equations}.
\newblock Academic press.

\bibitem[\protect\citeauthoryear{Li, Brown, Lee, and Gupta}{Li
  et~al.}{2002}]{Li2002}
Li, L., M.~Brown, K.~Lee, and S.~Gupta (2002).
\newblock Estimation and inference for a spline-enhanced population
  pharmacokinetic model.
\newblock {\em Biometrics\/}~{\em 58\/}(3), 601--11.

\bibitem[\protect\citeauthoryear{Li and Muldowney}{Li and
  Muldowney}{1995}]{seir}
Li, M.~Y. and J.~S. Muldowney (1995).
\newblock Global stability for the seir model in epidemiology.
\newblock {\em Mathematical Biosciences\/}~{\em 125\/}(2), 155--164.

\bibitem[\protect\citeauthoryear{Miao, Xia, Perelson, and Wu}{Miao
  et~al.}{2011}]{miao2011identifiability}
Miao, H., X.~Xia, A.~S. Perelson, and H.~Wu (2011).
\newblock On identifiability of nonlinear ode models and applications in viral
  dynamics.
\newblock {\em SIAM review\/}~{\em 53\/}(1), 3--39.

\bibitem[\protect\citeauthoryear{Mwalili, Kimathi, Ojiambo, Gathungu, and
  Mbogo}{Mwalili et~al.}{2020}]{mwalili2020seir}
Mwalili, S., M.~Kimathi, V.~Ojiambo, D.~Gathungu, and R.~Mbogo (2020).
\newblock Seir model for covid-19 dynamics incorporating the environment and
  social distancing.
\newblock {\em BMC Research Notes\/}~{\em 13\/}(1), 1--5.

\bibitem[\protect\citeauthoryear{Neal}{Neal}{2011}]{neal2011mcmc}
Neal, R.~M. (2011).
\newblock {MCMC Using Hamiltonian Dynamics}.
\newblock In S.~Brooks, A.~Gelman, G.~Jones, and X.~Meng (Eds.), {\em {Handbook
  of Markov Chain Monte Carlo}}, Chapman \& Hall/CRC Handbooks of Modern
  Statistical Methods, Chapter~5, pp.\  113--162. CRC Press.

\bibitem[\protect\citeauthoryear{Pei and Shaman}{Pei and
  Shaman}{2020}]{pei2020initial}
Pei, S. and J.~Shaman (2020).
\newblock Initial simulation of sars-cov2 spread and intervention effects in
  the continental us.
\newblock {\em MedRxiv\/}.

\bibitem[\protect\citeauthoryear{Perelson, Neumann, Markowitz, Leonard, and
  Ho}{Perelson et~al.}{1996}]{perelson1996hiv}
Perelson, A.~S., A.~U. Neumann, M.~Markowitz, J.~M. Leonard, and D.~D. Ho
  (1996).
\newblock Hiv-1 dynamics in vivo: virion clearance rate, infected cell
  life-span, and viral generation time.
\newblock {\em Science\/}~{\em 271\/}(5255), 1582--1586.

\bibitem[\protect\citeauthoryear{Schmidt, Kr{\"a}mer, and Hennig}{Schmidt
  et~al.}{2021}]{schmidt2021probabilistic}
Schmidt, J., N.~Kr{\"a}mer, and P.~Hennig (2021).
\newblock A probabilistic state space model for joint inference from
  differential equations and data.
\newblock {\em arXiv preprint arXiv:2103.10153\/}.

\bibitem[\protect\citeauthoryear{Shaman and Karspeck}{Shaman and
  Karspeck}{2012}]{shaman2012forecasting}
Shaman, J. and A.~Karspeck (2012).
\newblock Forecasting seasonal outbreaks of influenza.
\newblock {\em Proceedings of the National Academy of Sciences\/}~{\em
  109\/}(50), 20425--20430.

\bibitem[\protect\citeauthoryear{Takeuchi, Du, Hieu, and Sato}{Takeuchi
  et~al.}{2006}]{LVmodel}
Takeuchi, Y., N.~Du, N.~Hieu, and K.~Sato (2006).
\newblock Evolution of predator–prey systems described by a lotka–volterra
  equation under random environment.
\newblock {\em Journal of Mathematical Analysis and Applications\/}~{\em
  323\/}(2), 938--957.

\bibitem[\protect\citeauthoryear{Wenk, Gotovos, Bauer, Gorbach, Krause, and
  Buhmann}{Wenk et~al.}{2019}]{wenk2019gpode}
Wenk, P., A.~Gotovos, S.~Bauer, N.~S. Gorbach, A.~Krause, and J.~M. Buhmann
  (2019).
\newblock Fast gaussian process based gradient matching for parameter
  identification in systems of nonlinear odes.
\newblock In K.~Chaudhuri and M.~Sugiyama (Eds.), {\em Proceedings of the
  Twenty-Second International Conference on Artificial Intelligence and
  Statistics}, Volume~89 of {\em Proceedings of Machine Learning Research},
  pp.\  1351--1360. PMLR.

\bibitem[\protect\citeauthoryear{Wu}{Wu}{2005}]{wu2005}
Wu, H. (2005).
\newblock Statistical methods for hiv dynamic studies in aids clinical trials.
\newblock {\em Statistical methods in medical research\/}~{\em 14\/}(2),
  118--134.

\bibitem[\protect\citeauthoryear{Xun, Cao, Mallick, Maity, and Carroll}{Xun
  et~al.}{2013}]{xun2013parameter}
Xun, X., J.~Cao, B.~Mallick, A.~Maity, and R.~J. Carroll (2013).
\newblock Parameter estimation of partial differential equation models.
\newblock {\em Journal of the American Statistical Association\/}~{\em
  108\/}(503), 1009--1020.

\bibitem[\protect\citeauthoryear{Yang, Wong, and Kou}{Yang
  et~al.}{2021}]{Yange2020397118}
Yang, S., S.~W.~K. Wong, and S.~C. Kou (2021).
\newblock Inference of dynamic systems from noisy and sparse data via
  manifold-constrained gaussian processes.
\newblock {\em Proceedings of the National Academy of Sciences\/}~{\em
  118\/}(15).

\bibitem[\protect\citeauthoryear{Yang, Kandula, Huynh, Greene, Van~Wye, Li,
  Chan, McGibbon, Yeung, Olson, et~al.}{Yang et~al.}{2020}]{yang2020estimating}
Yang, W., S.~Kandula, M.~Huynh, S.~K. Greene, G.~Van~Wye, W.~Li, H.~T. Chan,
  E.~McGibbon, A.~Yeung, D.~Olson, et~al. (2020).
\newblock Estimating the infection fatality risk of covid-19 in new york city,
  march 1-may 16, 2020.
\newblock {\em MedRxiv\/}.

\end{thebibliography}

\clearpage
\renewcommand{\appendixtocname}{Supplementary Material}
\renewcommand\appendixname{Supplementary Material}
\renewcommand\appendixpagename{Supplementary Material}

\begin{appendices}
\beginsupplement

\section{Advantage of the $L_\infty$ norm in Eq.\eqref{eq:norm_inf} of main text}\label{sec:infty-reason}
In this section we illustrate how $L_\infty$ norm in Eq.\eqref{eq:norm_inf} of main text facilitates theoretical construction, compared to $L_2$ norm. First, with $L_\infty$ norm in $W$, it is clear that on the discretization subset $I$, the corresponding $W_I$ will simply be the maximum over $I$. However, with the $L_2$-norm of $\int_0^T (X(t)  - f(X(t), \Theta(t), \Psi, t))^2 dt$, the formulation of the corresponding $W_I$ is not as clear. Second, using $L_\infty$ makes the theoretical justification easier. To mathematically study the properties of TVMAGI while avoiding Borel paradox, one can use the fact that $\{W_I < \epsilon\} \equiv \cap_{i \in I} \{W_i < \epsilon\}$, thanks to $W_I$ being the $L_\infty$ norm over the set $I$. Third, the $L_\infty$ norm in Eq.\eqref{eq:norm_inf} and Eq.\eqref{eq:norm_inf_w_i} automatically transforms into $L_2$ loss for likelihood calculation in Eq.\eqref{eq:tvmagi_mc_likelihood} and Eq.\eqref{eq:posterior-calculation} through a simple mathematical derivation, which facilitates computation while maintaining the theoretical rigor. This is because when a Gaussian distributed vector is constrained to have zero deviation with some fixed value (i.e., vector $L_\infty$ distance to the fixed value is zero), the fixed value will be plugged into the Gaussian probability density function, inducing an $L_2$ loss in the target function Eq.\eqref{eq:posterior-calculation}.

\section{HMC algorithm}\label{sec:hmc-alg}

We outline the HMC procedure for sampling from a target probability distribution.  The interested reader may refer to  \cite{neal2011mcmc} for more thorough introduction to HMC. Algorithm \ref{alg:HMC} provides the details of our HMC implementation.

\begin{algorithm}
\caption{HMC sampling in TVMAGI}\label{alg:HMC}
\textbf{Input}: \\
$U$: Log likelihood function in Eq.(\ref{eq:posterior-calculation})   \\
$\epsilon$: step size of HMC \\
$L$: number of leaf frog steps \\
N: number of samples  \\
\textbf{Initialize}: $\bm{x}(\bm{I}), \bm{\theta}(\bm{I}), \bm{\psi}, \bm{\sigma}$
\begin{algorithmic}

\For{$i$ in 1:N}
    \State $\bm{q_{\text{current}}}= \textbf{vector}(\bm{x}(\bm{I})$, $\bm{\theta}(\bm{I})$, $\bm{\psi}$, $\bm \sigma$)
    \State $\bm{q}=\bm{q_{\text{current}}}$
    \State $\bm{p}= \textbf{rnorm}(\textbf{length}(\bm{q}), 0, 1)$
    \State $\bm{p_{\text{current}}} = \bm{p}$
    \State $\bm{p}=\bm{p}-\epsilon \nabla U(\bm{q})/2$
    \For{$j$ in 1:L}
        \State $\bm{q}=\bm{q}+\epsilon \bm{p}$
        \State $\bm{p}=\bm{p}-\epsilon \nabla U(\bm{q})$
    \EndFor
    \State $\bm{p}=\bm{p}-\epsilon \nabla U(\bm{q})/2$
    \If {$\textbf{runif}(1) < \textbf{exp}(U(\bm{q_{\text{current}}})-U(\bm{q})+\textbf{sum}(\bm{p_{\text{current}}^2}-\bm{p^2})/2)$}
        \State \textbf{return} $\bm{q}$  \Comment{(Accept)}
    \Else
        \State \textbf{return} $\bm{q_\text{current}}$ \Comment{(Reject)}
    \EndIf
\EndFor

\end{algorithmic}
\end{algorithm}

\section{Additional benchmark methods of Bayesian filtering approaches}\label{sec:others}
\begin{figure}[htp]
     \centering
     \begin{subfigure}[b]{1\textwidth}
         \centering
         \includegraphics[width=5in]{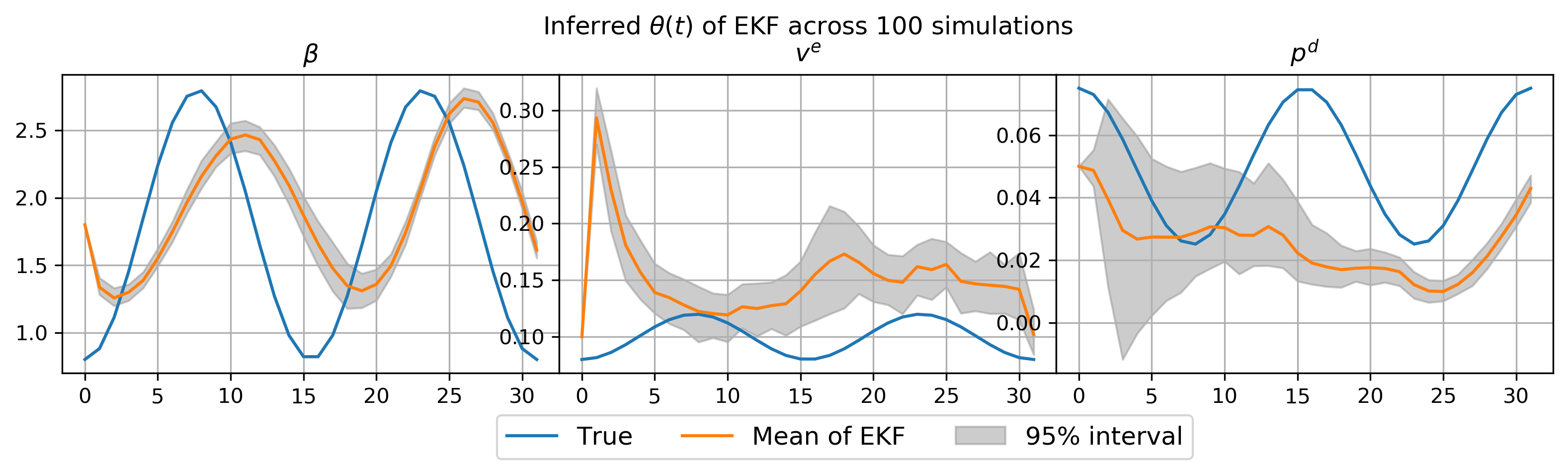}
         \caption{EKF results}
     \end{subfigure}
     \begin{subfigure}[b]{1\textwidth}
         \centering
         \includegraphics[width=5in]{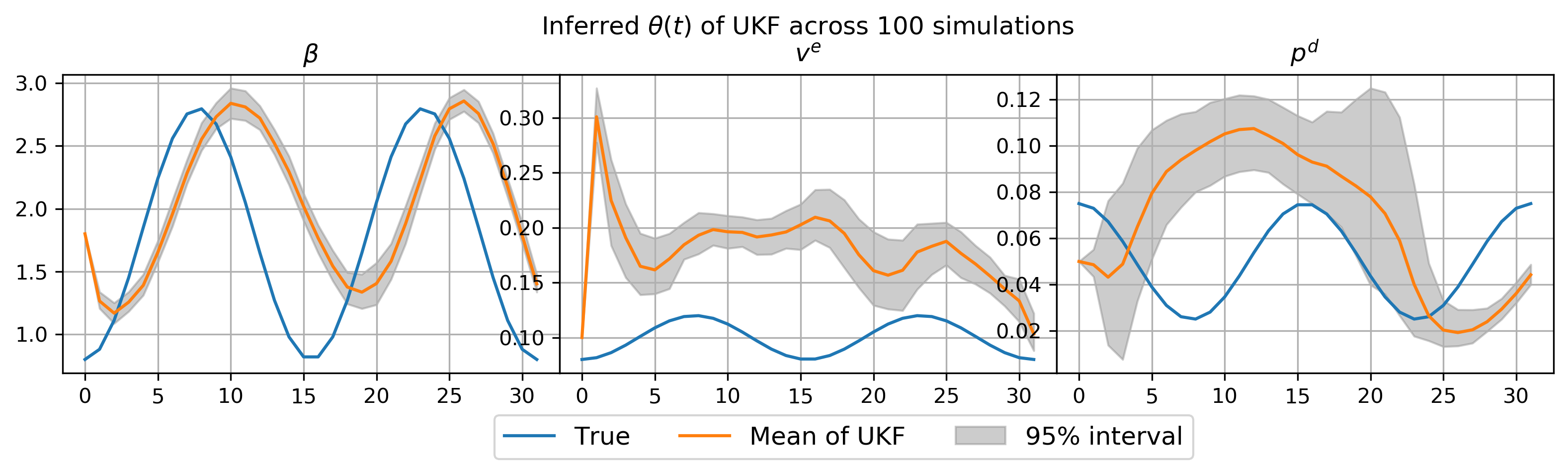}
         \caption{UKF results}
     \end{subfigure}
     \begin{subfigure}[b]{1\textwidth}
         \centering
         \includegraphics[width=5in]{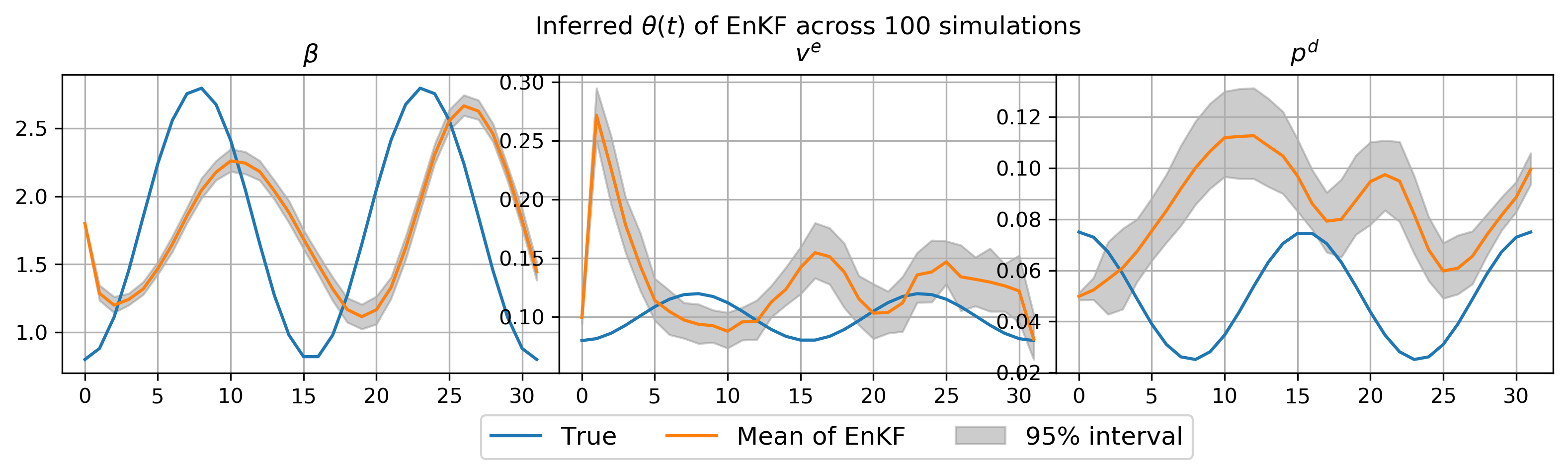}
         \caption{EnKF results}
     \end{subfigure}
\caption{Inferred $\theta(t)$ of EKF, UKF and EnKF approach for SEIRD model. The inferred parameters cannot capture the ODE structure, thus yielding inaccurate reconstructed trajectory.}
\label{fig:otherbf-seird-param}
\end{figure}

\begin{figure}[htp]
     \centering
     \begin{subfigure}[b]{1\textwidth}
         \centering
         \includegraphics[width=5in]{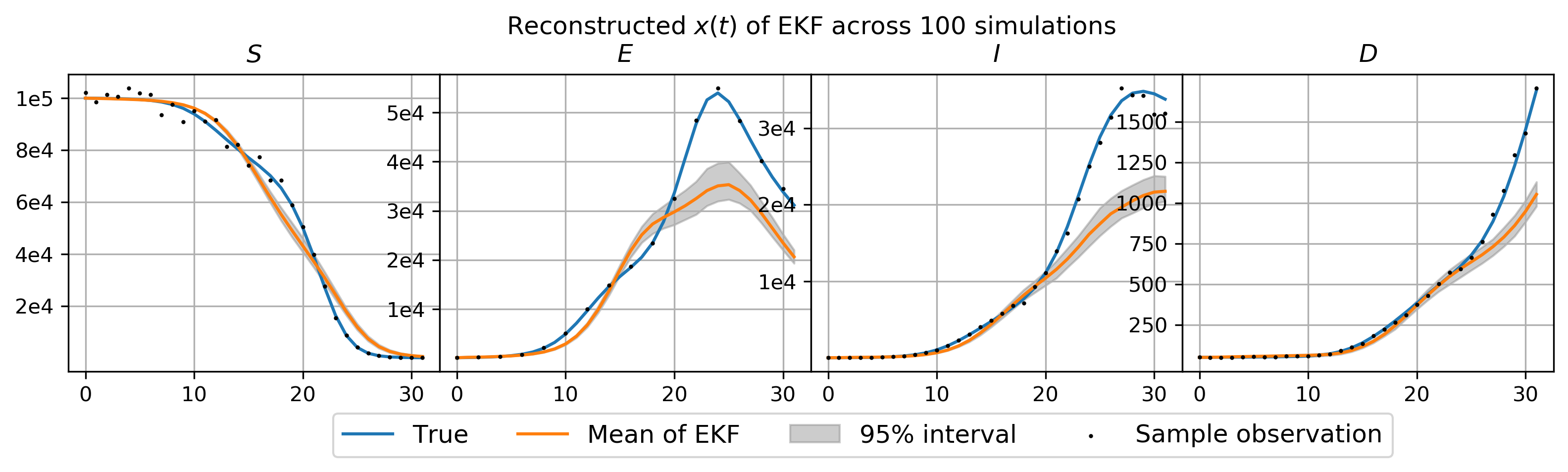}
         \caption{EKF results}
     \end{subfigure}
     \begin{subfigure}[b]{1\textwidth}
         \centering
         \includegraphics[width=5in]{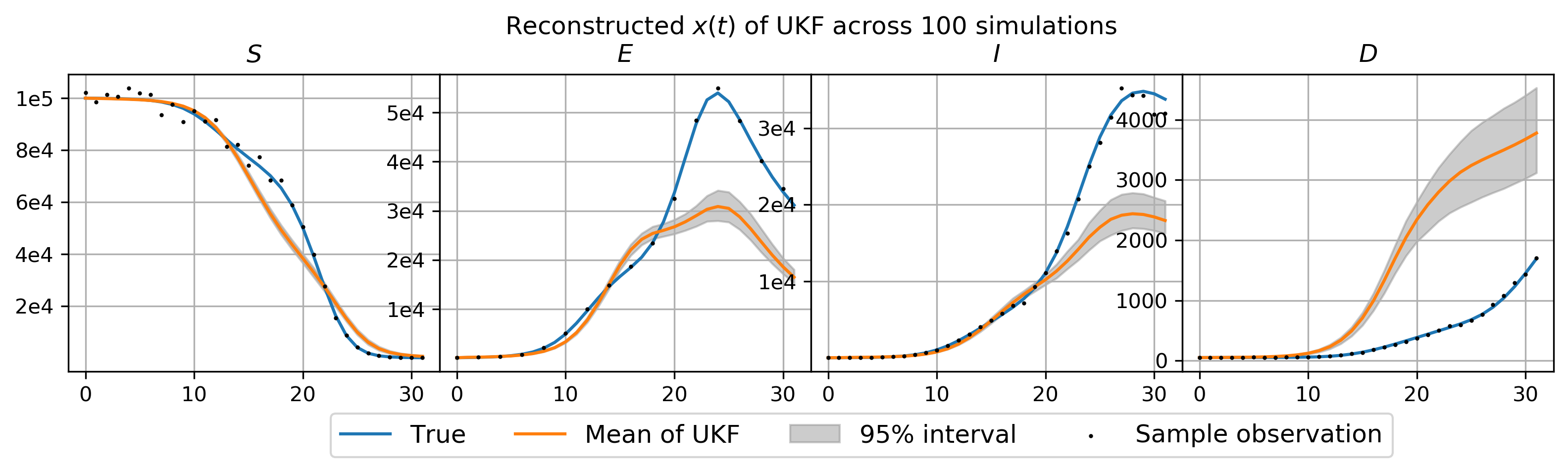}
         \caption{UKF results}
     \end{subfigure}
     \begin{subfigure}[b]{1\textwidth}
         \centering
         \includegraphics[width=5in]{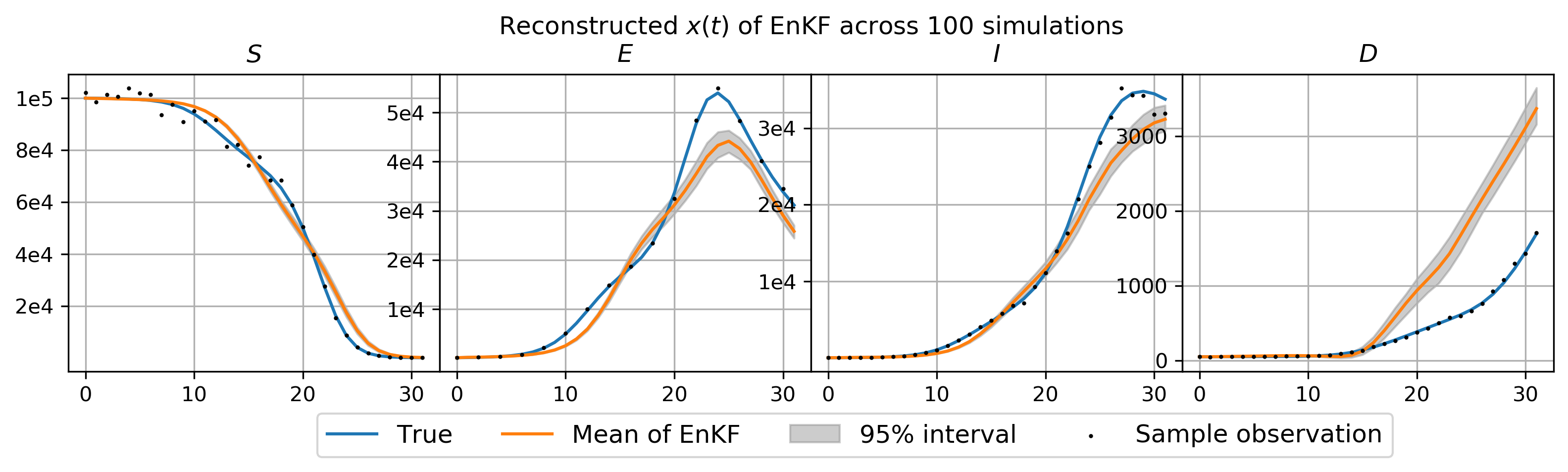}
         \caption{EnKF results}
     \end{subfigure}
\caption{Reconstructed trajectory of EKF, UKF and EnKF approaches for SEIRD model. The accuracy is far from satisfactory.}
\label{fig:otherbf-seird-x}
\end{figure}

\begin{figure}[htp]
     \centering
     \begin{subfigure}[b]{1\textwidth}
         \centering
         \includegraphics[width=5in]{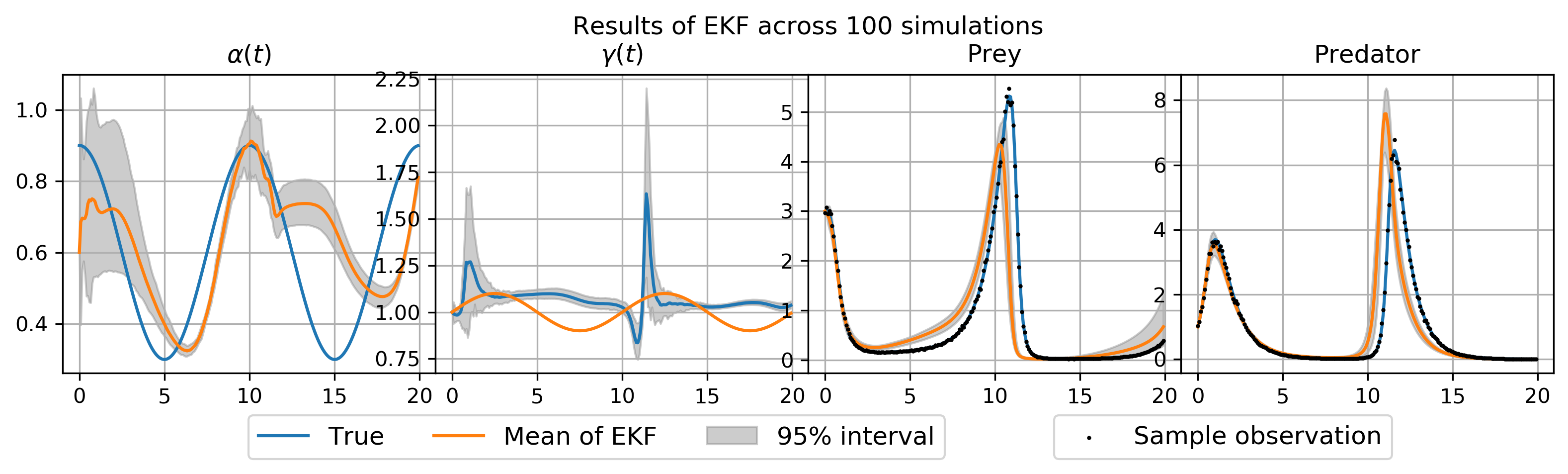}
         \caption{EKF results}
     \end{subfigure}
     \begin{subfigure}[b]{1\textwidth}
         \centering
         \includegraphics[width=5in]{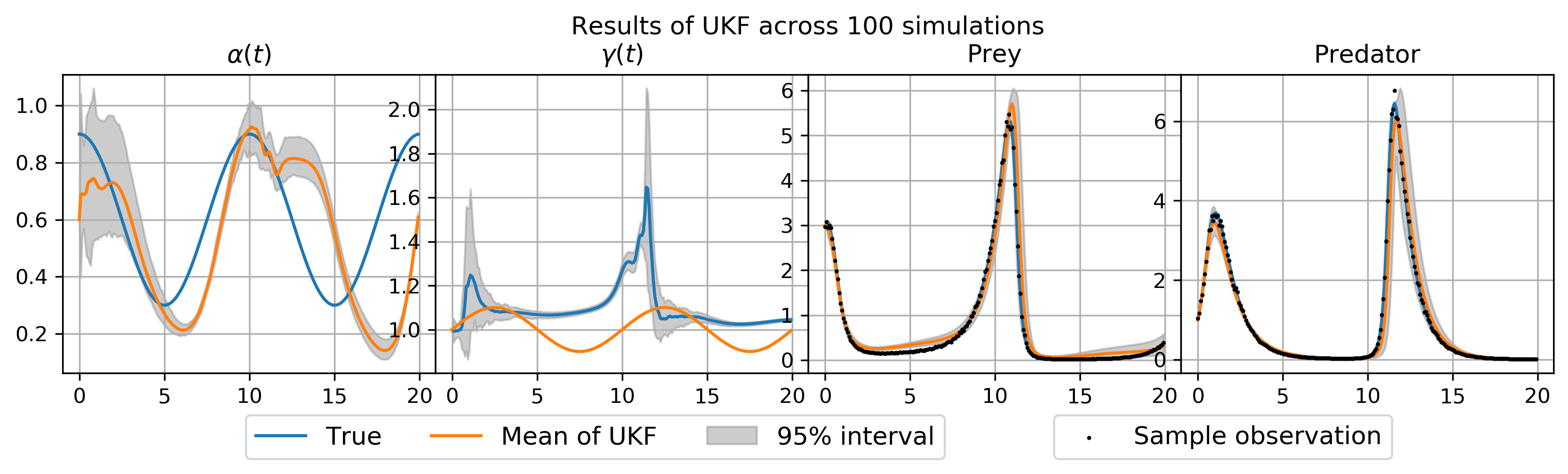}
         \caption{UKF results}
     \end{subfigure}
     \begin{subfigure}[b]{1\textwidth}
         \centering
         \includegraphics[width=5in]{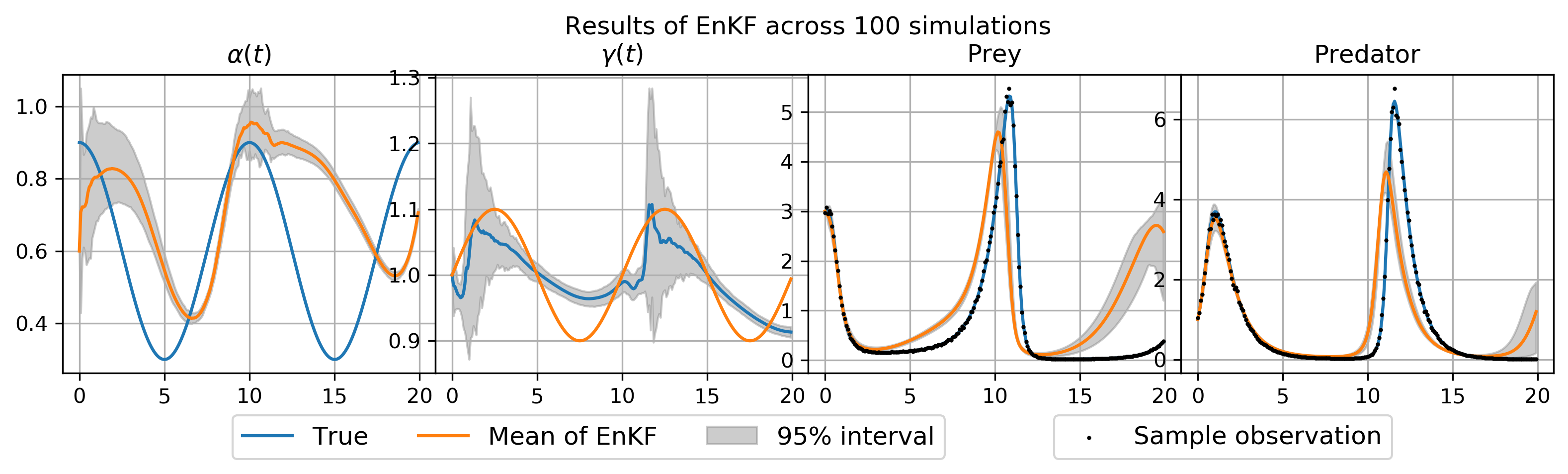}
         \caption{EnKF results}
     \end{subfigure}
\caption{Inferred parameter and reconstructed trajectory of EKF, UKF and EnKF approaches for LV model. The accuracy is far from satisfactory.}
\label{fig:otherbf-lv}
\end{figure}

\begin{figure}[htp]
     \centering
     \begin{subfigure}[b]{1\textwidth}
         \centering
         \includegraphics[width=5in]{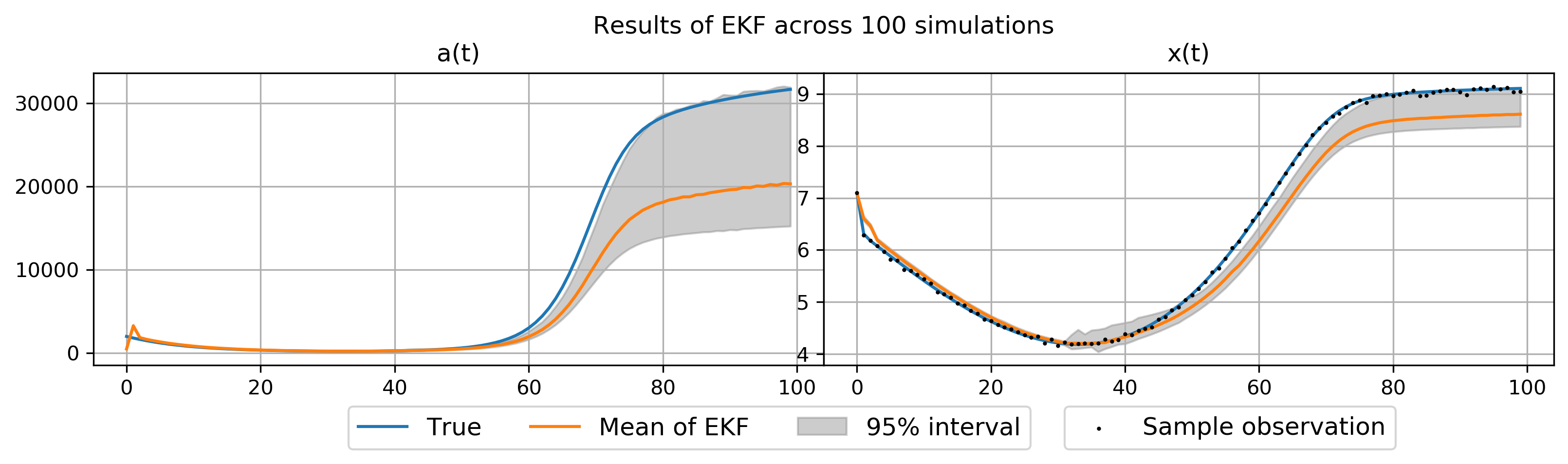}
         \caption{EKF results}
     \end{subfigure}
     \begin{subfigure}[b]{1\textwidth}
         \centering
         \includegraphics[width=5in]{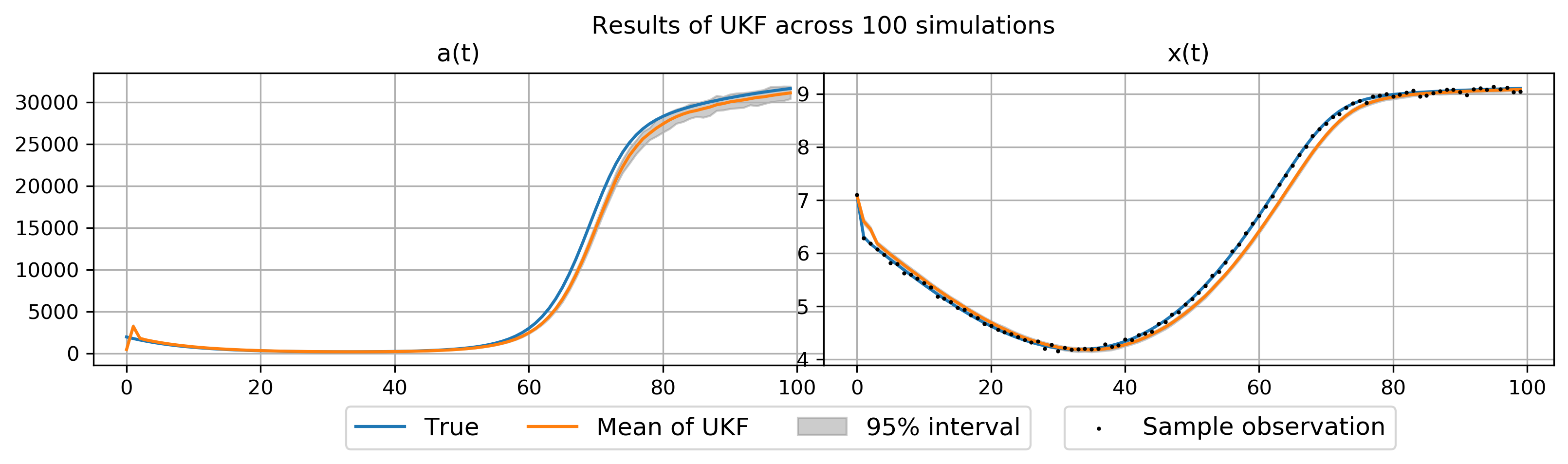}
         \caption{UKF results}
     \end{subfigure}
     \begin{subfigure}[b]{1\textwidth}
         \centering
         \includegraphics[width=5in]{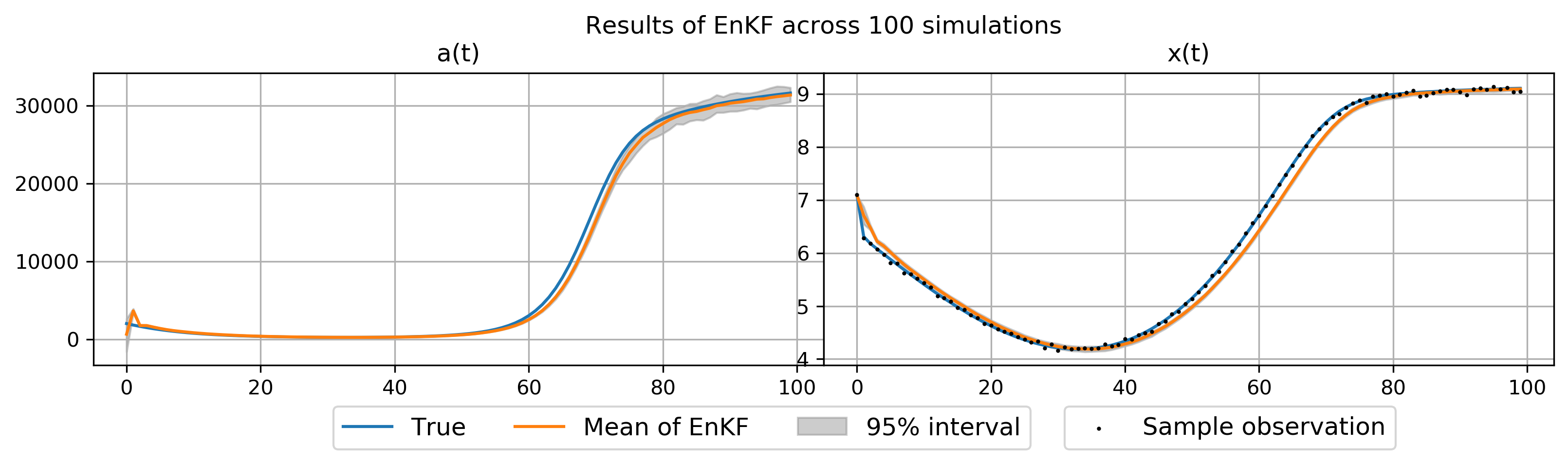}
         \caption{EnKF results}
     \end{subfigure}
\caption{Inferred parameter and reconstructed trajectory trajectory of EKF, UKF and EnKF approaches for HIV model. The accuracy is worse than TVMAGI, Runge-Kutta, or ELE.}
\label{fig:otherbf-hiv}
\end{figure}

Compared with Ensemble Adjustment Kalman Filter (EAKF), other Bayesian filtering methods, such as Extended Kalman Filter (EKF), Unscented Kalman Filter (UKF) and Ensemble Kalman Filter (EnKF) are less discussed in ODE parameter inference applications. To further illustrate the Bayesian filtering approaches, we also provide inference results using the other 3 methods. Fig.\ref{fig:otherbf-seird-param} and Fig.\ref{fig:otherbf-seird-x} illustrate the results of SEIRD model. Fig.\ref{fig:otherbf-lv} shows the results of LV model and Fig.\ref{fig:otherbf-hiv} shows the results of HIV model.

\clearpage

\clearpage


\clearpage

\section{Additional results of the main benchmark methods}

\begin{figure}[htp]
     \centering
     \begin{subfigure}[b]{1\textwidth}
         \centering
         \includegraphics[width=5in]{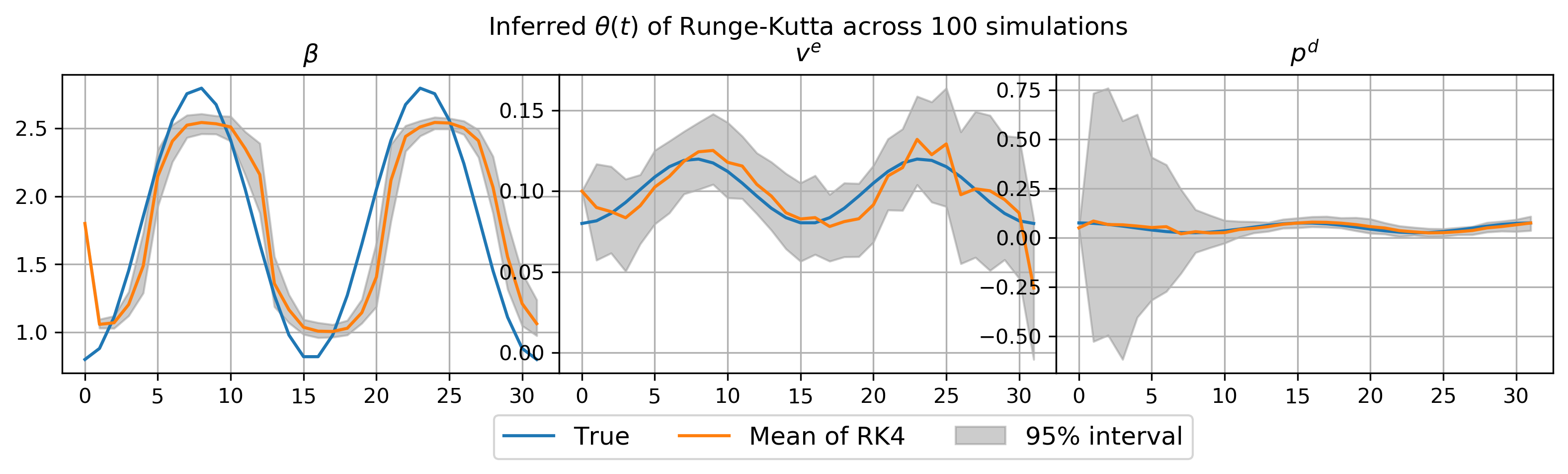}
         \caption{Inferred parameters using Runge-Kutta}
     \end{subfigure}
     \begin{subfigure}[b]{1\textwidth}
         \centering
         \includegraphics[width=5in]{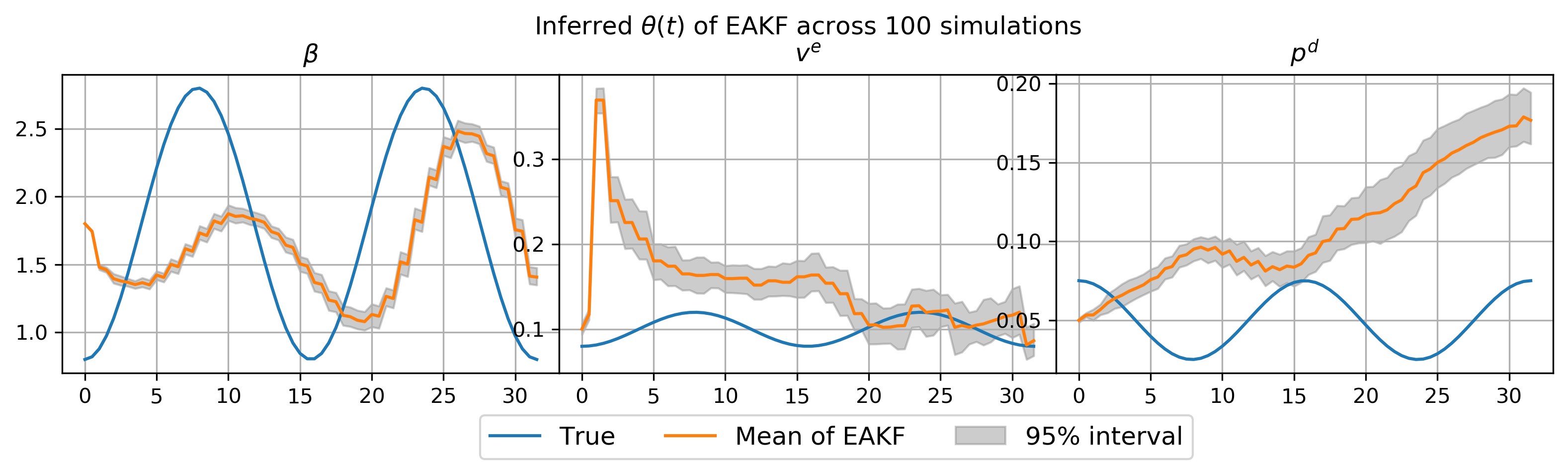}
         \caption{Inferred parameters using EAKF}
     \end{subfigure}
\caption{Results of parameter inference in 100 simulated datasets using Runge-Kutta and EAKF for SEIRD model.}
\label{fig:compare-seird-param}
\end{figure}

\begin{figure}[htp]
     \centering
     \begin{subfigure}[b]{1\textwidth}
         \centering
         \includegraphics[width=5in]{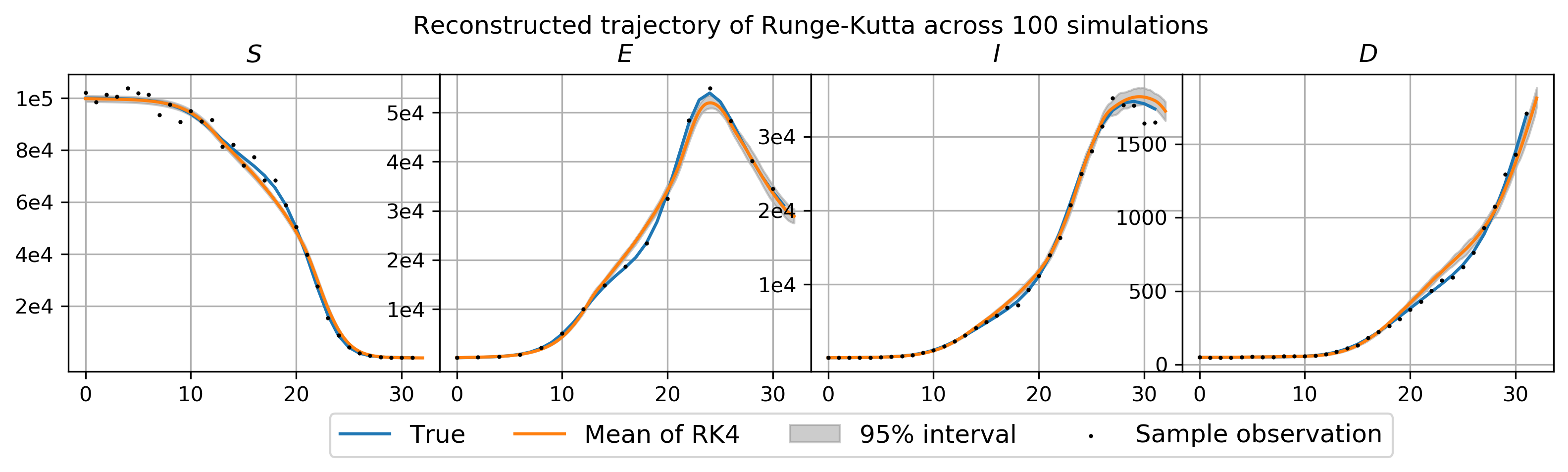}
         \caption{Reconstructed trajectory using Runge-Kutta}
     \end{subfigure}
     \begin{subfigure}[b]{1\textwidth}
         \centering
         \includegraphics[width=5in]{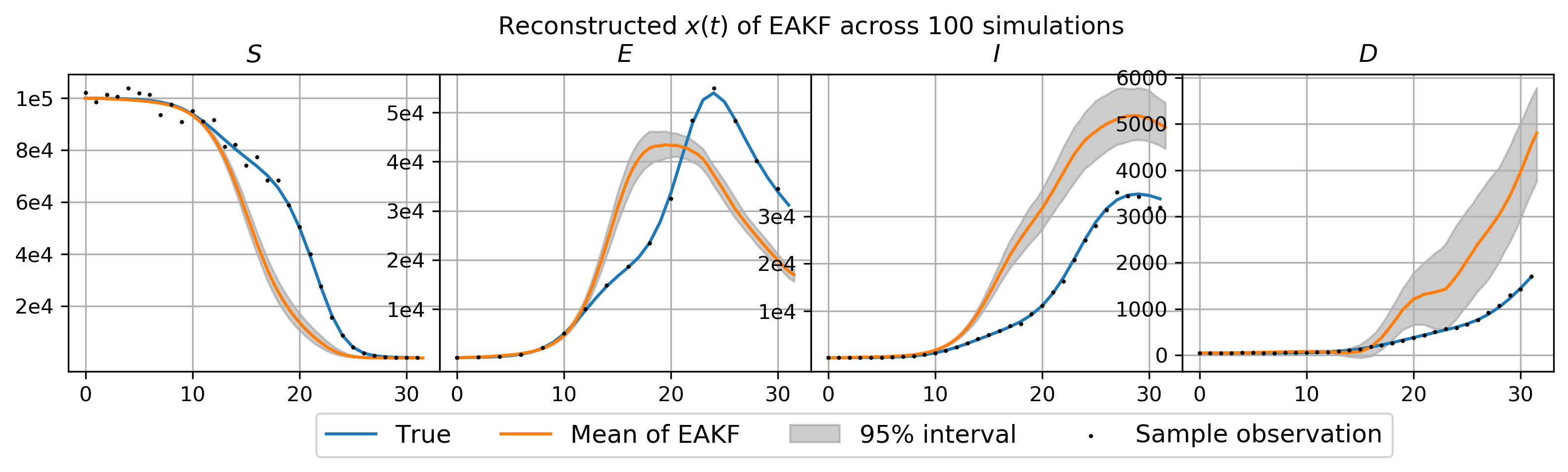}
         \caption{Reconstructed trajectory using EAKF}
     \end{subfigure}
\caption{Reconstructed trajectory using inferred parameters of Runge-Kutta and EAKF methods for SEIRD model. One sample simulation dataset is also presented to visualize the noise level and observation schedule.}
\label{fig:compare-seird-x}
\end{figure}

\begin{figure}[htp]
     \centering
     \begin{subfigure}[b]{1\textwidth}
         \centering
         \includegraphics[width=5in]{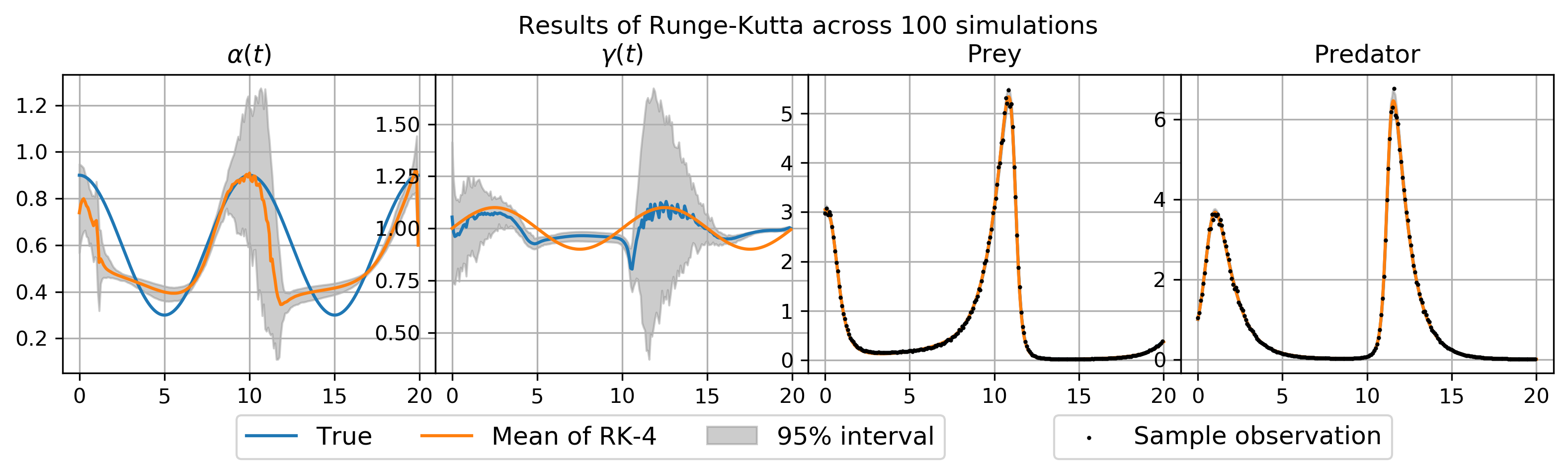}
         \caption{Runge-Kutta results}
     \end{subfigure}
     \begin{subfigure}[b]{1\textwidth}
         \centering
         \includegraphics[width=5in]{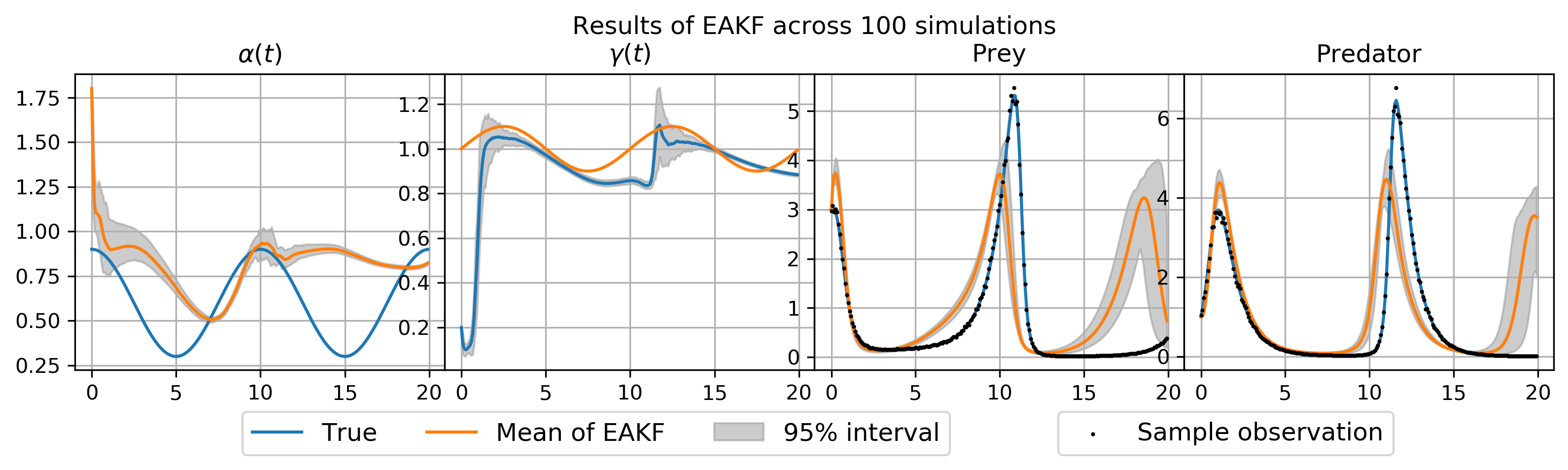}
         \caption{EAKF results}
     \end{subfigure}
\caption{Comparison of inferred $\bm{\theta}(t)$ and reconstructed $\bm{X}(t)$ of LV model. We also plot one sample simulation dataset to visualize the noise level.}
\label{fig:compare-lv}
\end{figure}

\begin{figure}[htp]
     \centering
     \begin{subfigure}[b]{1\textwidth}
         \centering
         \includegraphics[width=5in]{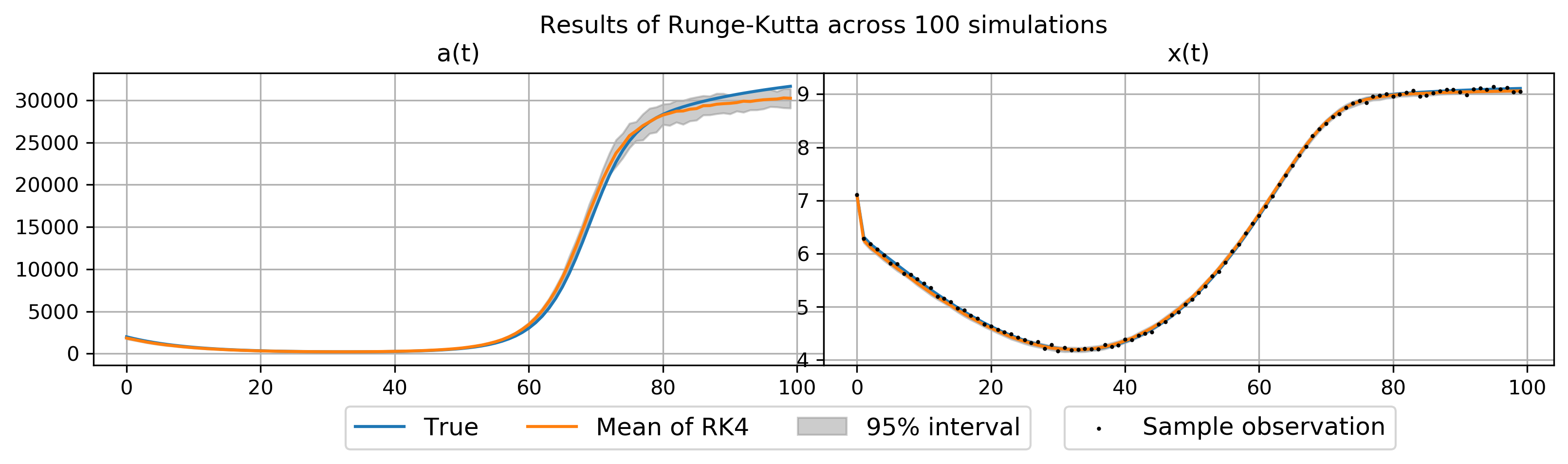}
         \caption{Runge-Kutta results}
     \end{subfigure}
     \begin{subfigure}[b]{1\textwidth}
         \centering
         \includegraphics[width=5in]{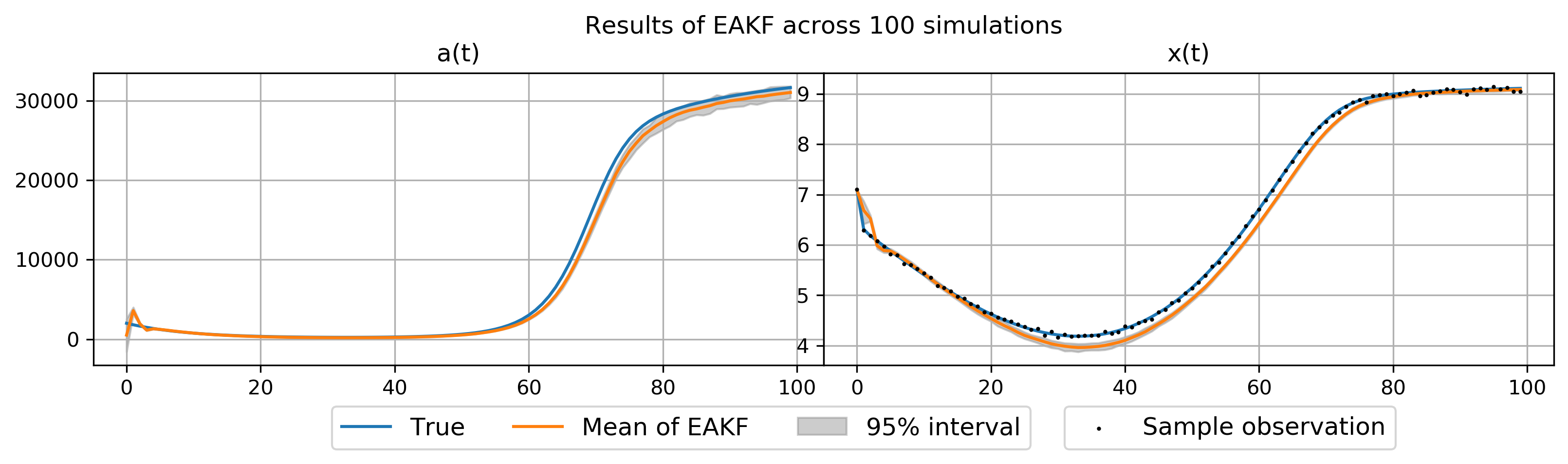}
         \caption{EAKF results}
     \end{subfigure}
     \begin{subfigure}[b]{1\textwidth}
         \centering
         \includegraphics[width=5in]{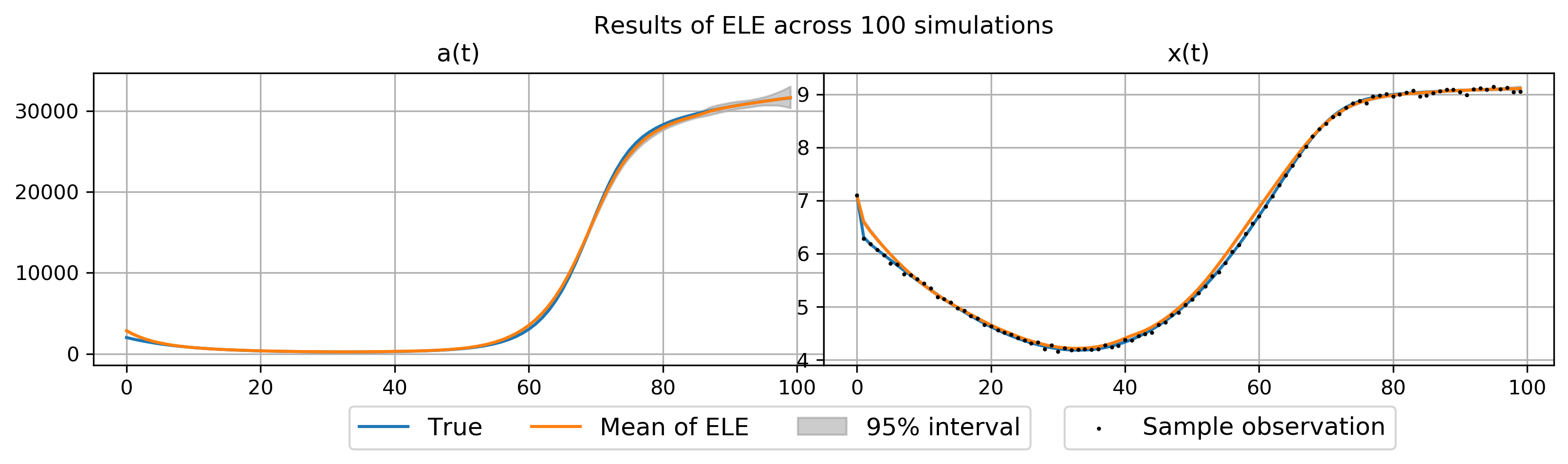}
         \caption{Results of ELE \citep{HLW}}
     \end{subfigure}
\caption{Comparison of inferred $\bm{\theta}(t)$ and reconstructed $\bm{X}(t)$ of HIV model. We also plot one sample simulation dataset to visualize the noise level.}
\label{fig:compare-hiv}
\end{figure}

Due to the limited space of the main text, we include visualizations of the main benchmark methods of Runge-Kutta method and EAKF method for the three examples in the main text here. Fig.\ref{fig:compare-seird-param} is for SEIRD model parameter inference, and Fig.\ref{fig:compare-seird-x} is for SEIRD model reconstructed trajectory. Fig.\ref{fig:compare-lv} is for LV model. Fig.\ref{fig:compare-hiv} is for HIV model.

\clearpage

\section{Additional discussion of the main benchmark methods}\label{sec:additional-benchmark-discussion}

In this section, we have a detailed discussion on the numerical integration methods, represented by Runge-Kutta method, and the Bayesian filtering methods, represented by EAKF.

\subsection{Runge-Kutta method}

Numerical integration methods are the gold standard for the ODE parameter inference when all parameters are time-constant. However, with the presence of time-varying parameters, there are several inherent disadvantages of numerical integration methods. First and foremost, without any structure on the time-varying parameters, the numerical integration method will give time-varying parameter estimate that perfectly fits the observation data, resulting in overfitting issues in the time-varying parameter. Second, with the increase of time points and size of the system, the objective function becomes expensive to evaluate, resulting in high computation cost. Third, the numerical methods are sensitive to the initial value of the optimization, while searching for a good initial point can be challenging in the high-dimensional scenarios, as optimization of objective functions using algorithms such as Adam can be easily stuck at local minimum. We point out that using random initial values in our examples can lead to high level of error, making numerical integration methods completely fail. In this case, all the optimization for Runge-Kutta method are initialized at the TVMAGI initial points from Section \ref{sec:algorithm-step1} in the main text examples. 


\subsection{Bayesian filtering methods}\label{sec:bayesian-problems}

Although Bayesian filtering method is the fastest, examples have shown that applying Bayesian filtering methods to ODE parameter inference problems have failed to provide satisfactory results. Even though we included the Bayesian filtering methods as baseline comparison methods, we emphasize that state-space model (Bayesian filtering) and ODE parameter inference (TVMAGI) are fundamentally different problems. To illustrate the difference in a simplified framework from a theoretical perspective, consider the following example of time-constant parameter inference where all parameters are denoted as $\bm \theta$. 


The fundamental difference is the lack of randomness in the state transition given the model parameters, and thus the Bayesian update given the observation will have zero effect. With the ODE structure, there is no randomness in $\bm{x}_{t} | \bm{x}_{t-1}, \bm{\theta}$. The state transition distribution $p(\bm{x}_{t} | \bm{x}_{t-1}, \bm{\theta})$ essentially has shifted Dirac delta distribution. Therefore, regardless of emission probability $p(\bm{y}_t | \bm{x}_t)$, the hidden state $\bm{x}_t$ will not depend on $\bm{y}_t$. As such, all Bayesian updates will have zero effect to shift distribution of $p(\bm{x}_{t} | \bm{x}_{t-1}, \bm{\theta})$, which is still shifted Dirac delta distribution. The exact Bayesian filtering results will simply be the solution of ODE dynamics given the initial sample of $\bm{x}_0$ and the parameter $\bm{\theta}$. In this case, the parameter estimation in the exact Bayesian filtering reduces to using numerical solver to generate the entire ODE curve given $\bm{x}_0, \bm{\theta}$, and then using a least square approach to compare the solved curve and the observations to find the best $\bm{x}_0, \bm{\theta}$. The exact Bayesian filtering in this case degenerates to a numerical integration method. From another particle filter perspective, each particle of sampled $\bm{x}_0, \bm{\theta}$ will evolve in time according to ODE without any randomness, and the parameter estimation becomes a brute-force search of the particle of sampled $\bm{x_0, \theta}$ that provide the smallest mean squared error to the observation. In light of this, the Bayesian filtering/smoothing is more suitable for inference of Stochastic Differential Equations (SDEs) parameters (see \cite{donnet2013review, golightly2010markov}). 

However, Bayesian filter methods still can be applied in ODE inference problem if we can forego some statistical rigor. We can treat \textbf{all} parameters as time-varying, and artificially introduce additional randomness in $\bm{\theta}_t | \bm{\theta}_{t-1}$ and $\bm{\psi}_t | \bm{\psi}_{t-1}$. In this case, the simultaneous estimation of system components $\bm{x}(t)$, time-constant parameter $\bm{\psi}$, time-varying parameter $\bm{\theta}(t)$ becomes an estimation of joint hidden state $(\bm{x}_t, \bm{\psi}_t, \bm{\theta}_t)$. Then Bayesian filtering methods such as EKF, UKF, EnKF and EAKF become applicable. However, this is not a statistically principled approach, because (1) time-constant parameter $\bm{\psi}$ is now changing with time, and (2) the inference on the distribution is not exact as Gaussian distributional approximation is used on system components $\bm{x}(t)$. Nevertheless, this method could work empirically, with the notable success of SIRS-EAKF model \cite{shaman2012forecasting, yang2020estimating}.

In our example, we found the Bayesian filter methods are highly sensitive to the initialization of the parameters. Randomized initialization often lead to insensible results. Therefore, all the parameters for Bayesian filter methods are initialized at the TVMAGI initial points from Section \ref{sec:algorithm-step1} in the main text examples,  which is the same as Runge-Kutta method. 

Even then, our numerical experiments suggest that Bayesian filtering methods are the fastest, but often yield unreliable inference results. Among EKF, UKF, EnKF and EAKF, we see that UKF, EnKF and EAKF yield similar results, all outperforming EKF in LV and HIV examples, while EKF has a slight advantage in SEIRD model inference compared with other Bayesian filtering methods, although not robust. All of them yield orders of magnitude worse trajectory RMSE compared to TVMAGI or Runge-Kutta. Two factors contribute to the unsatisfactory trajectory RMSE of Bayesian filtering methods. First, all filtering approaches fail to yield an accurate and robust parameter estimation, especially in SEIRD model when the observation points are limited. Second, the variance of time-varying parameters is large at weakly identifiable time points (Figure \ref{fig:otherbf-lv}), and the estimates for the later time points are no longer accurate due to the cascading effect. 

The failure of Bayesian filtering in our setting is not surprising. First, it violates the assumption of the model, as it is not principled to allow time-constant parameters $\bm{\psi}$ to change over time. Although we used the average $ \bm{\bar \psi}$ of the filtered parameter $\bm{\psi}_t$ to be the final estimate for the $\bm{\psi}$, the approximation error can still be large. The idea of changing a time-constant parameter to be time-varying during inference and later plug-in the average of the inferred values is not theoretically sound. The trajectory RMSE precisely evaluates how accurate the estimated parameters can be used to reconstruct the entire system given the ODE structure, of which $\bm{\bar \psi}$ would fail. Second, contrary to the typical setting of Bayesian filtering in machine learning where there is a long sequence of data, our experiments are designed to see how the method performs with short time series and sparse observations, as in most scientific experiment settings. The lack of long series of data poses a challenge to Bayesian filtering methods. Third, the ODE structure is no longer exactly followed in Bayesian filtering, which loosens the structure constraints and creates additional loss of information from the observation data that is already sparse.

\section{Additional results of TVMAGI interval estimates} \label{sec:seird-hmc}

In this section we present the visualization for the interval estimation results. We see that for long time series of observations, the estimated intervals tend to be narrow and may not contain the true values, which is a limitation of TVMAGI. 

\begin{figure}[htp]
     \centering
     \begin{subfigure}[b]{0.5\textwidth}
         \centering
         \includegraphics[width=3in]{plots/SEIRD-ve-HMC.png}
         \caption{Interval estimation of $v^e$}
     \end{subfigure}
     \begin{subfigure}[b]{0.49\textwidth}
         \centering
         \includegraphics[width=3in]{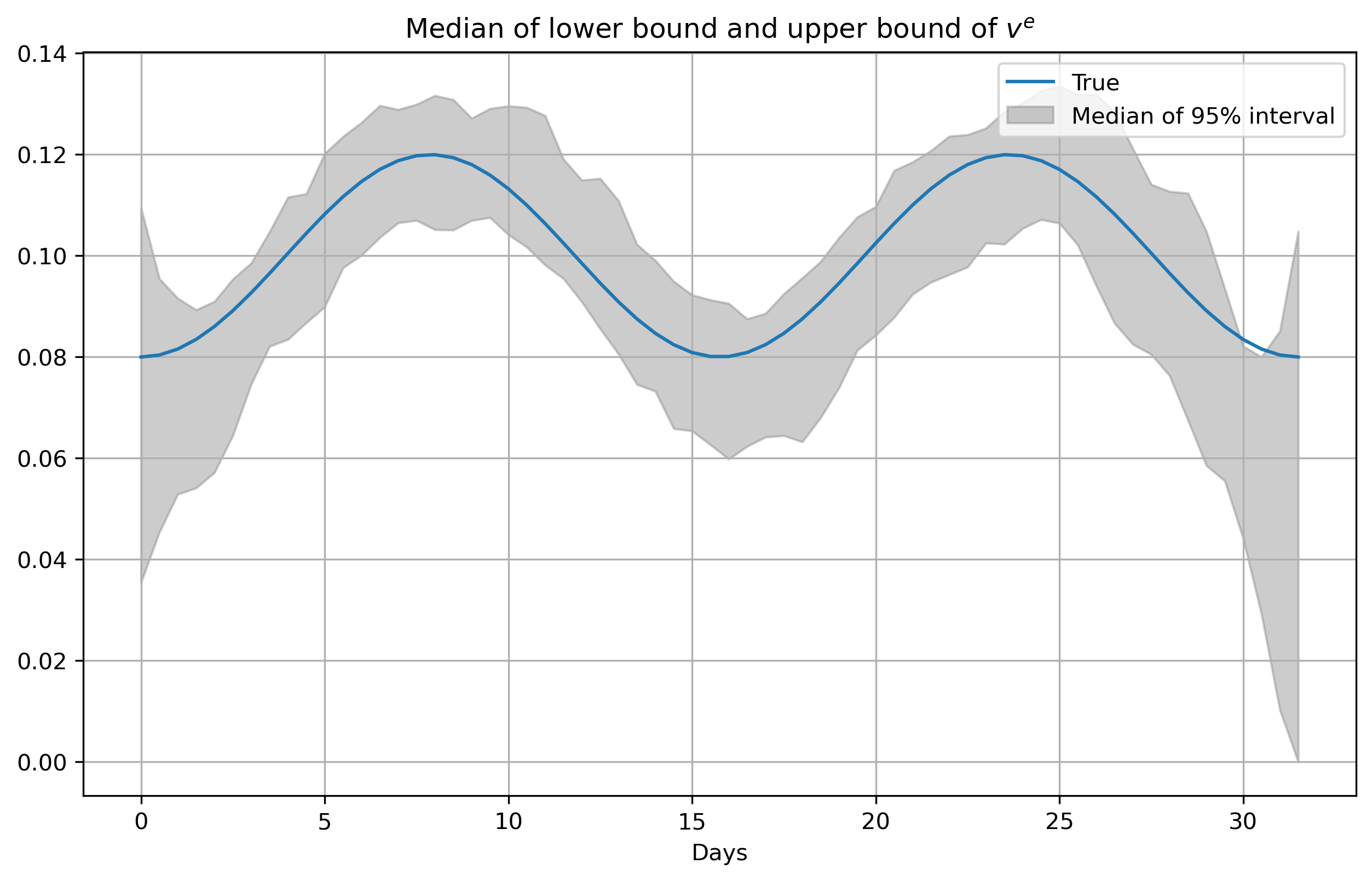}
         \caption{Median of upper \& lower bound of $v^e$ interval}
     \end{subfigure}
     \begin{subfigure}[b]{0.5\textwidth}
         \centering
         \includegraphics[width=3in]{plots/SEIRD-pd-HMC.png}
         \caption{Interval estimation of $p^d$}
     \end{subfigure}
     \begin{subfigure}[b]{0.49\textwidth}
         \centering
         \includegraphics[width=3in]{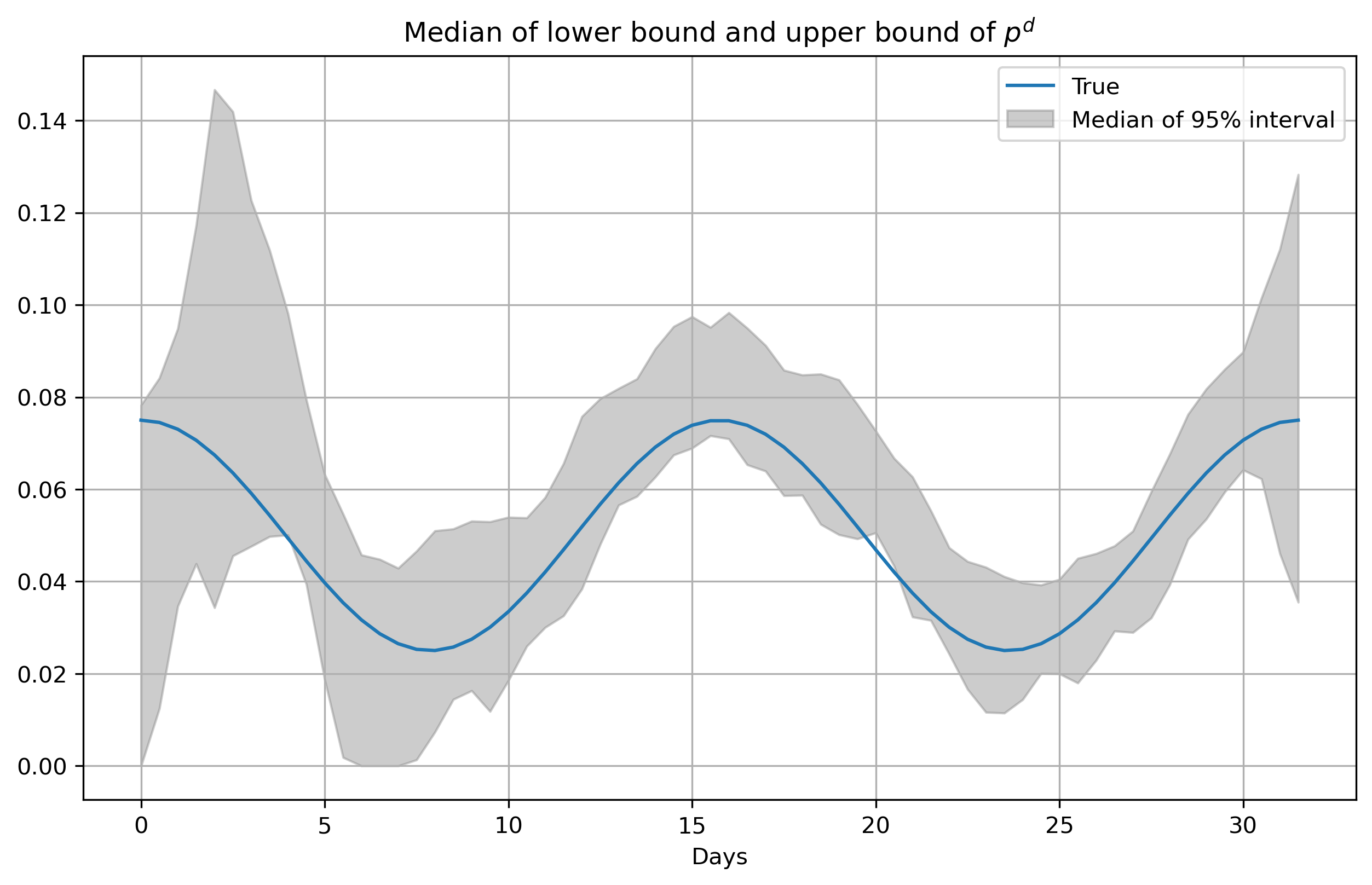}
         \caption{Median of upper \& lower bound of $p^d$ interval}
     \end{subfigure}
\caption{TVMAGI estimated interval of $v^e$ and $p^d$ in SEIRD model. We randomly plot 10 replications. }
\label{fig:seird-hmc-2}
\end{figure}

\begin{figure}[htp]
     \centering
     \begin{subfigure}[b]{0.5\textwidth}
         \centering
         \includegraphics[width=3in]{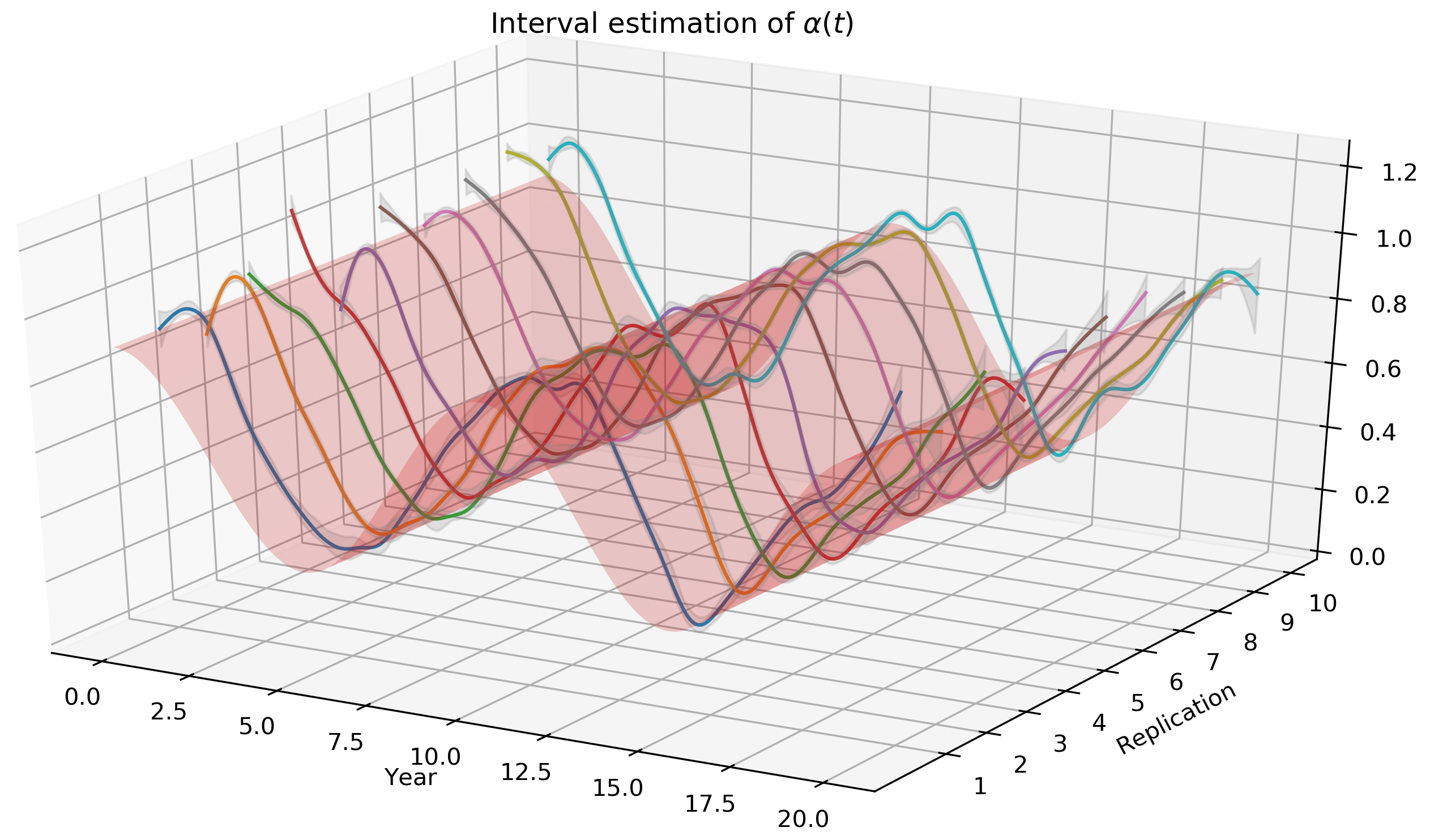}
         \caption{Interval estimation of $\alpha(t)$}
     \end{subfigure}
     \begin{subfigure}[b]{0.49\textwidth}
         \centering
         \includegraphics[width=3in]{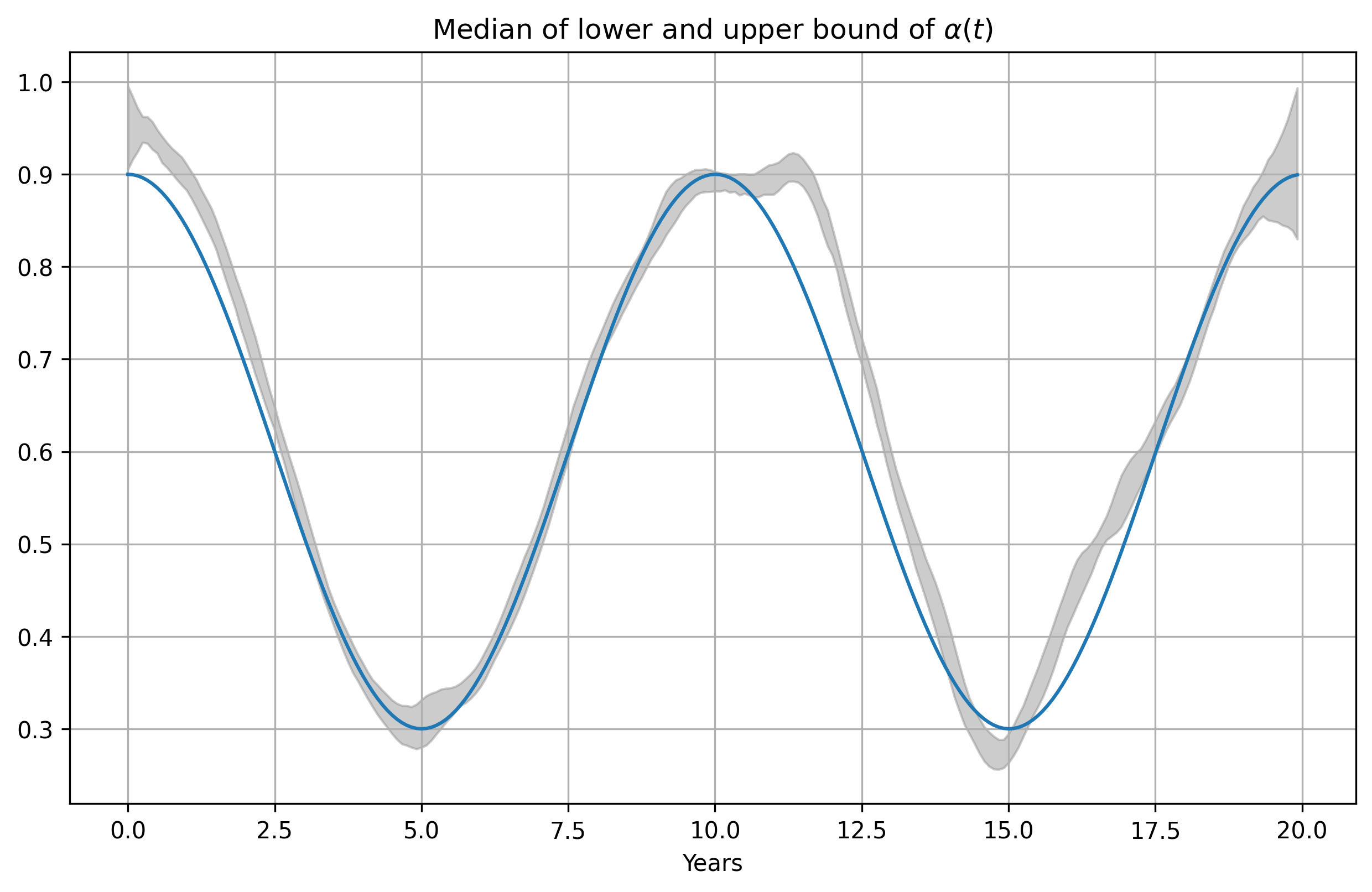}
         \caption{Median of upper \& lower bound of $\alpha(t)$ interval}
     \end{subfigure}
     \begin{subfigure}[b]{0.5\textwidth}
         \centering
         \includegraphics[width=3in]{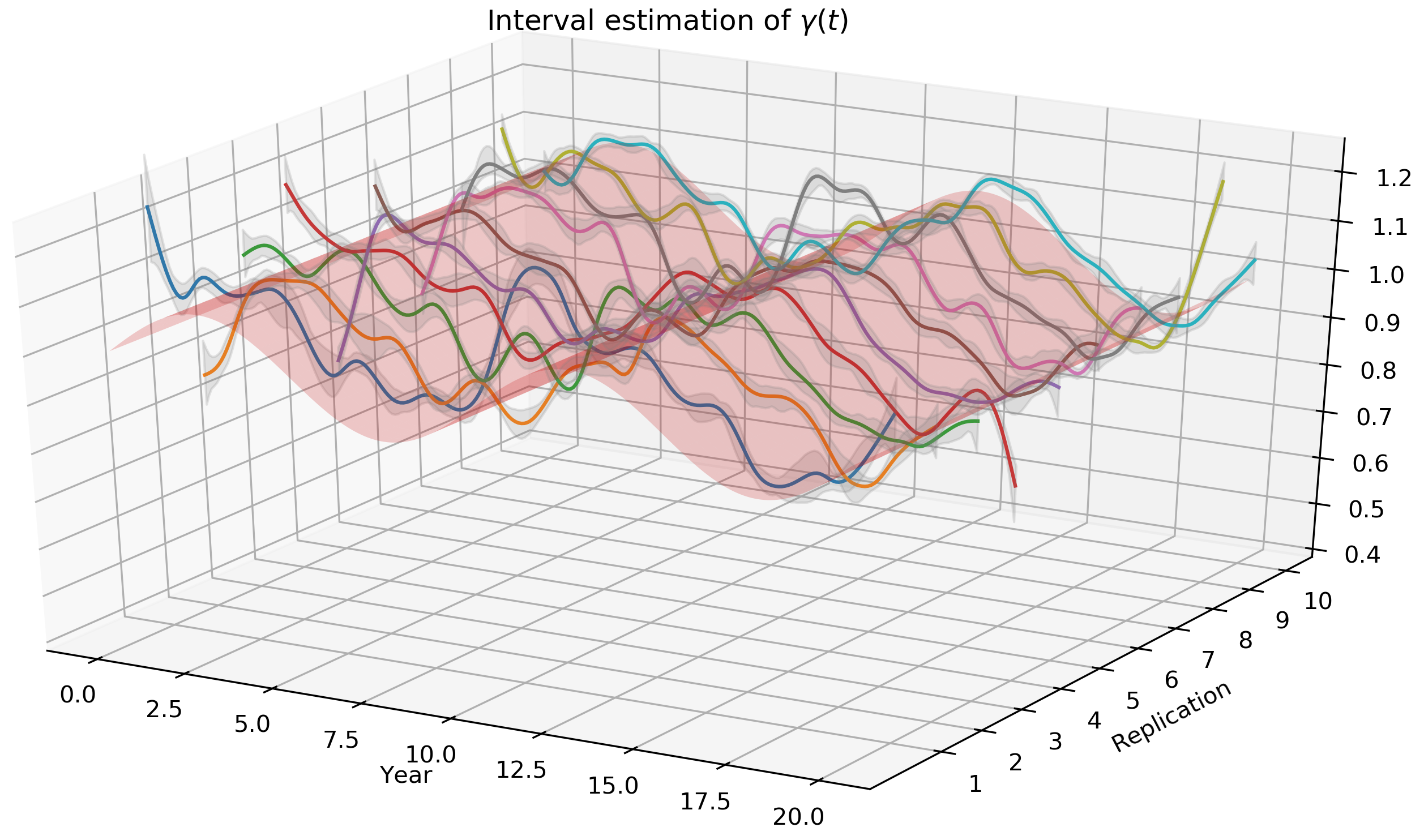}
         \caption{Interval estimation of $\gamma(t)$}
     \end{subfigure}
     \begin{subfigure}[b]{0.49\textwidth}
         \centering
         \includegraphics[width=3in]{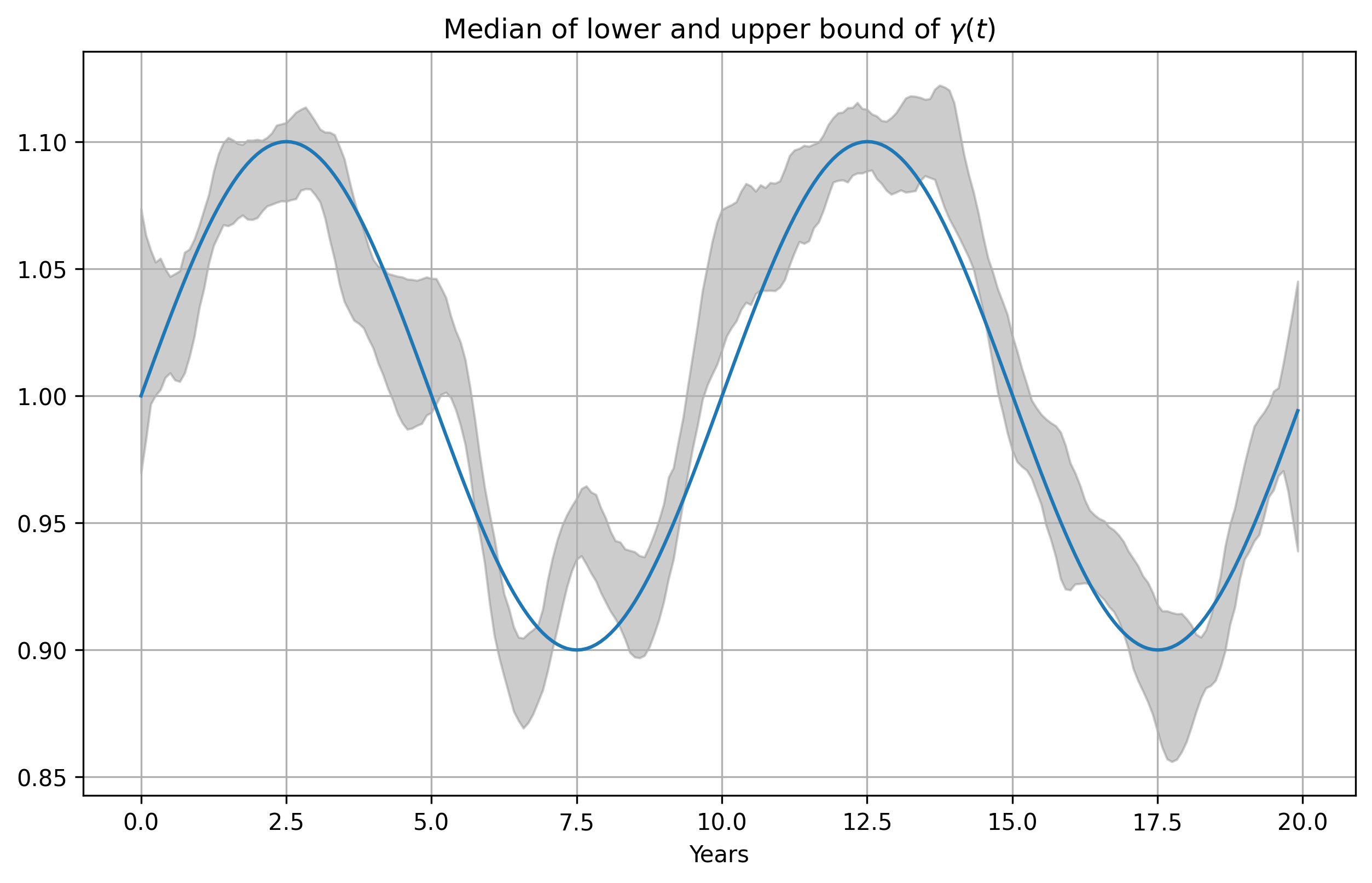}
         \caption{Median of upper \& lower bound of $\gamma(t)$ interval}
     \end{subfigure}
\caption{Estimated interval of $\alpha(t)$ and $\gamma(t)$ in LV model. We randomly plot 10 replications. }
\label{fig:LV-hmc}
\end{figure}

\clearpage
\begin{figure}[htp]
     \centering
     \begin{subfigure}[b]{0.5\textwidth}
         \centering
         \includegraphics[width=3in]{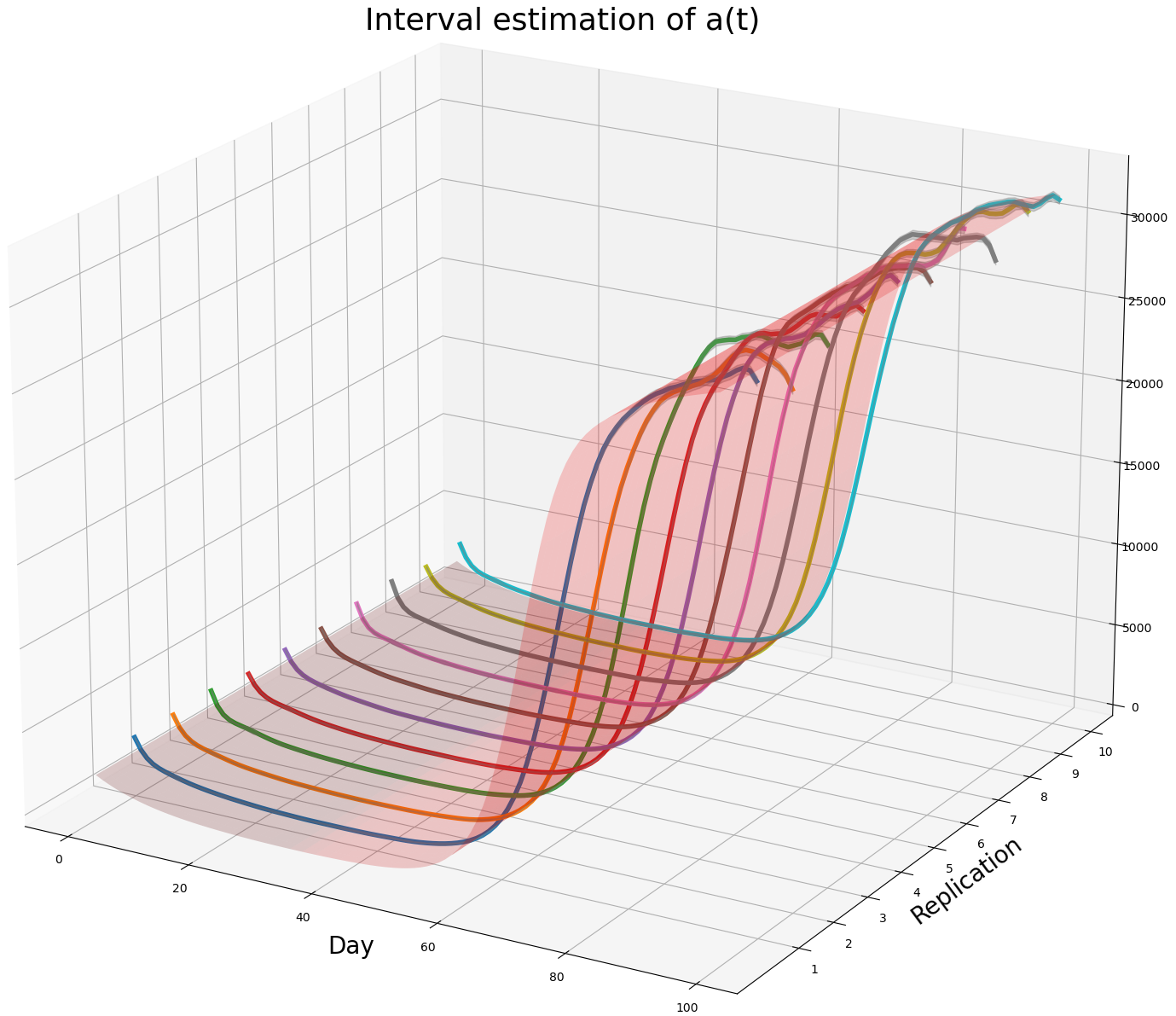}
         \caption{Interval estimation of $a(t)$}
     \end{subfigure}
     \begin{subfigure}[b]{0.49\textwidth}
         \centering
         \includegraphics[width=3in]{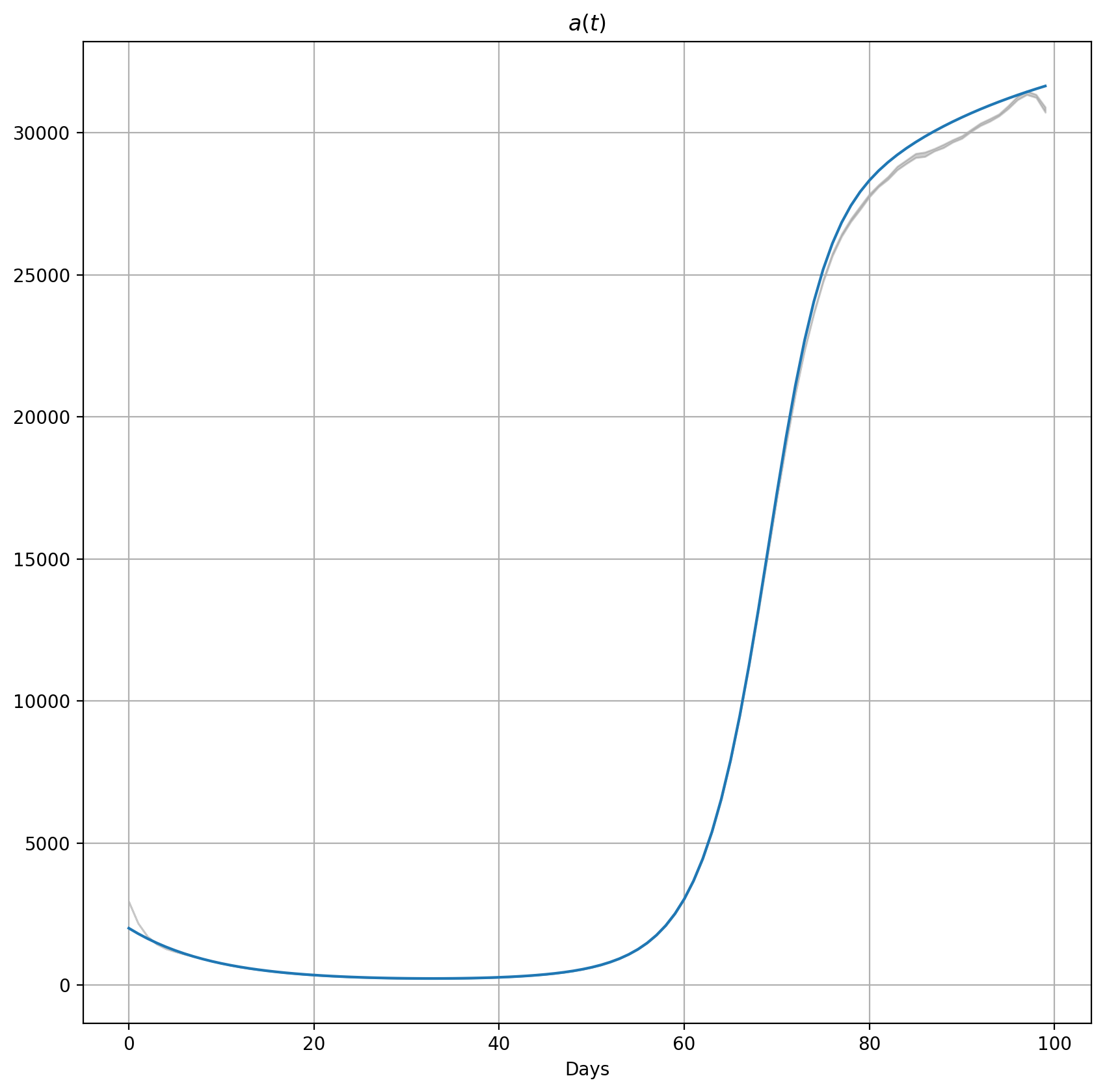}
         \caption{Median upper and lower bound of $a(t)$}
     \end{subfigure}
\caption{Estimated interval of $a(t)$ in HIV model. We randomly plot 10 replications. }
\label{fig:hiv-hmc}
\end{figure}

\clearpage

\section{Additional results on TVMAGI sensitivity study}\label{sec:SI-sensitivity}

In this section we give additional visualizations for TVMAGI sensitivity study when varying the discretization level, and Matern kernel degree of freedom $\nu$.

\begin{figure}[htp]
     \centering
     \begin{subfigure}[b]{1\textwidth}
         \centering
         \includegraphics[width=5in]{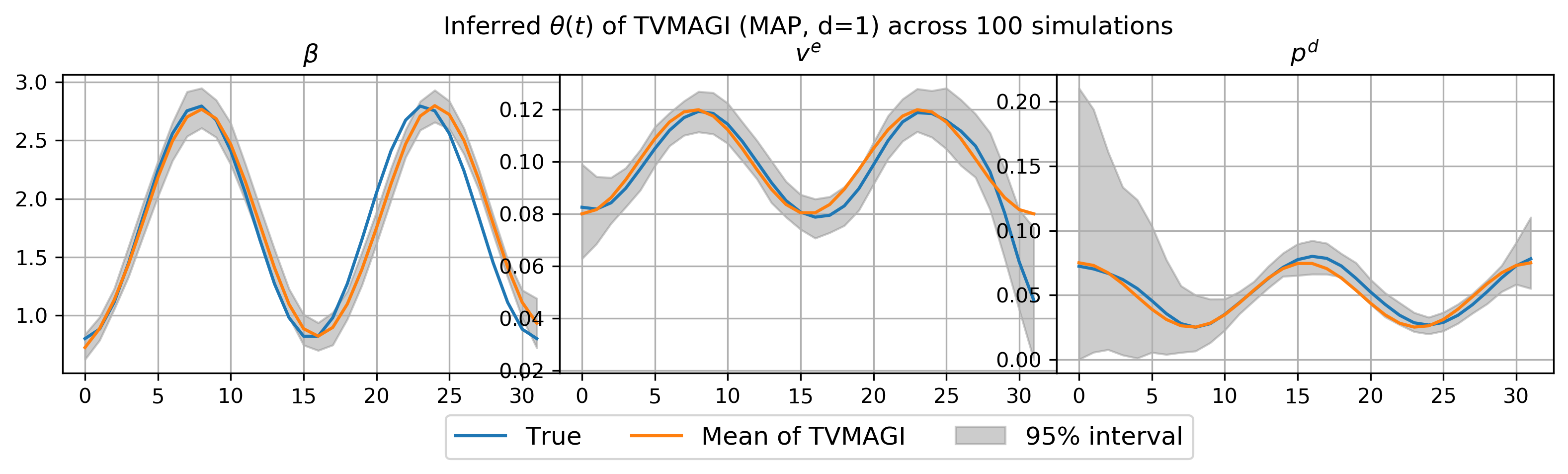}
         \caption{Discretization level = 1}
     \end{subfigure}
     \begin{subfigure}[b]{1\textwidth}
         \centering
         \includegraphics[width=5in]{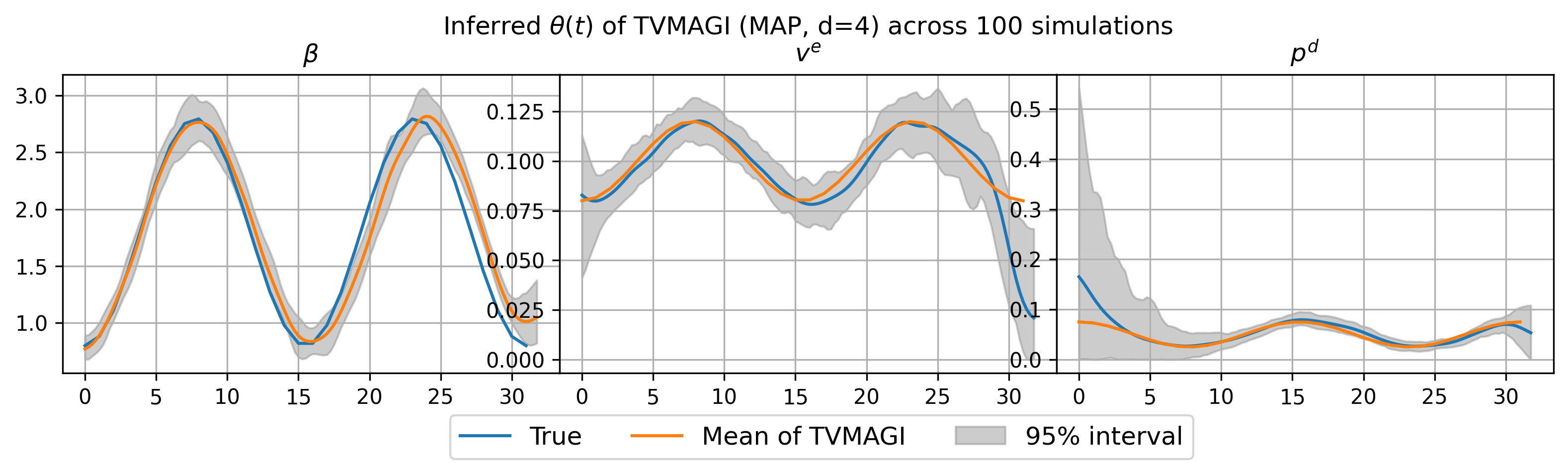}
         \caption{Discretization level = 4}
     \end{subfigure}
\caption{TVMAGI inferred parameters in SEIRD model with different discretization level.}
\label{fig:TVMAGI-dis-param}
\end{figure}

\begin{figure}[htp]
     \centering
     \begin{subfigure}[b]{1\textwidth}
         \centering
         \includegraphics[width=5in]{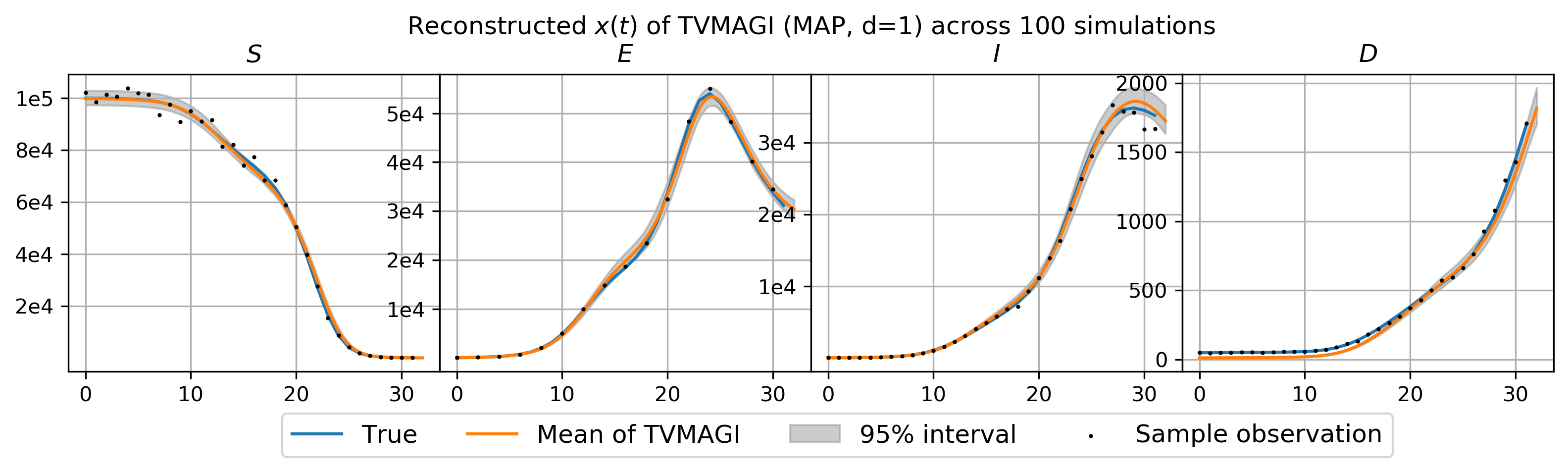}
         \caption{Discretization level = 1}
     \end{subfigure}
     \begin{subfigure}[b]{1\textwidth}
         \centering
         \includegraphics[width=5in]{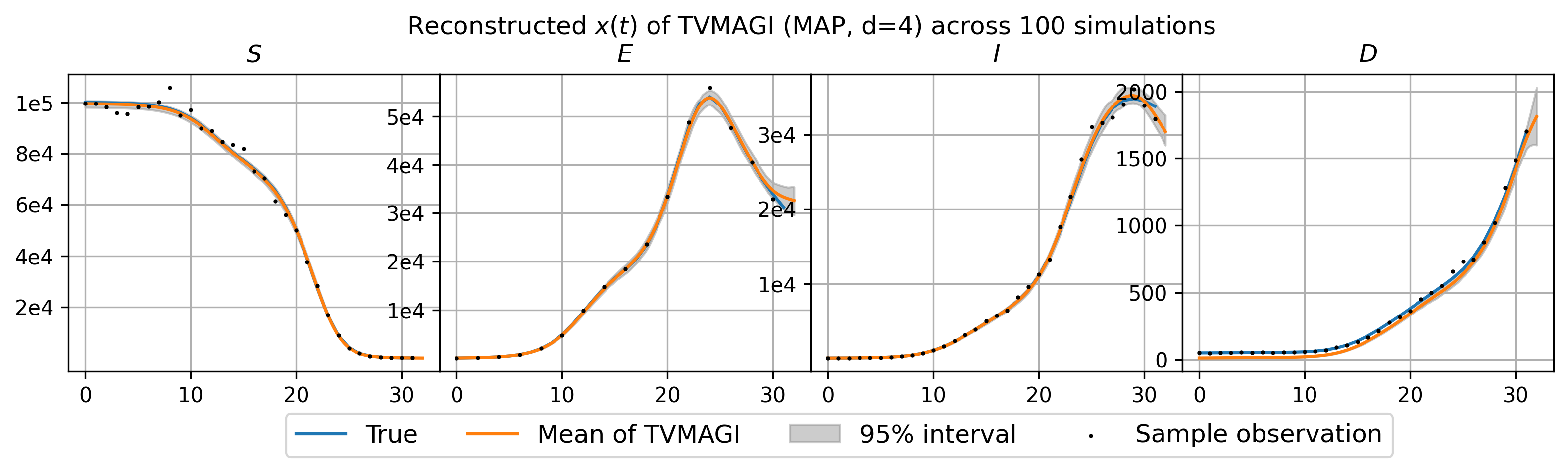}
         \caption{Discretization level = 4}
     \end{subfigure}
\caption{TVMAGI reconstructed trajectories in SEIRD model with different discretization level.}
\label{fig:TVMAGI-dis-x}
\end{figure}

\begin{figure}[htp]
     \centering
     \begin{subfigure}[b]{1\textwidth}
         \centering
         \includegraphics[width=5in]{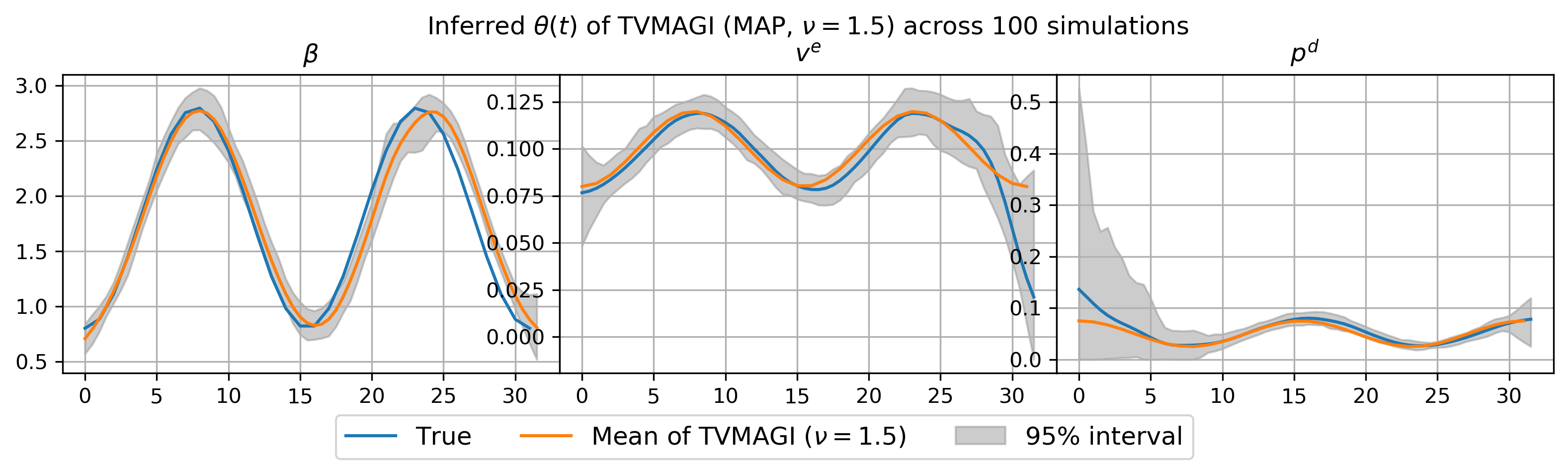}
         \caption{Inferred parameters of $\nu=1.5$}
     \end{subfigure}
     \begin{subfigure}[b]{1\textwidth}
         \centering
         \includegraphics[width=5in]{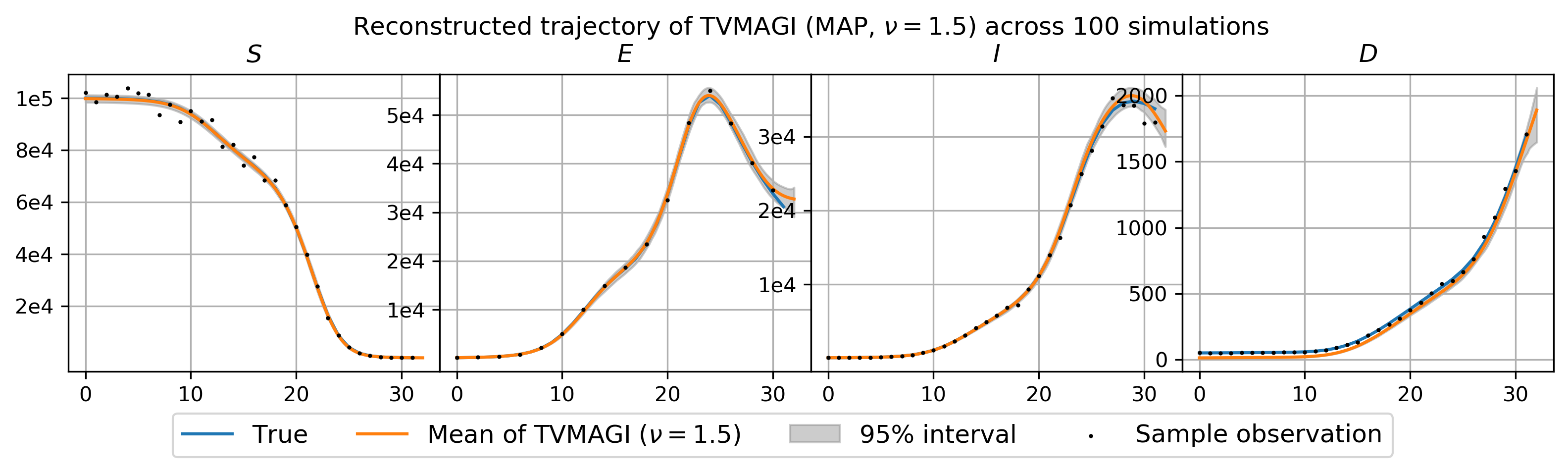}
         \caption{Reconstructed trajectory of $\nu=1.5$}
     \end{subfigure}
\caption{TVMAGI inferred parameters and reconstructed trajectories of SEIRD model with Matern kernel $\nu=1.5$}
\label{fig:nu-1.5}
\end{figure}

\begin{figure}[htp]
     \centering
     \begin{subfigure}[b]{1\textwidth}
         \centering
         \includegraphics[width=5in]{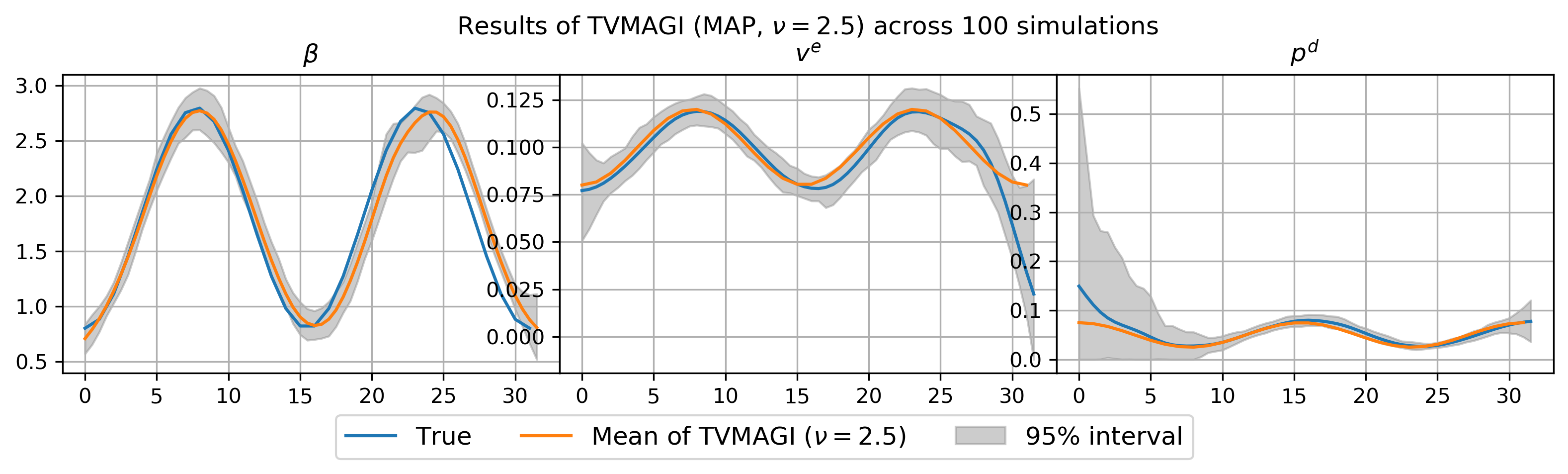}
         \caption{Inferred parameters of $\nu=2.5$}
     \end{subfigure}
     \begin{subfigure}[b]{1\textwidth}
         \centering
         \includegraphics[width=5in]{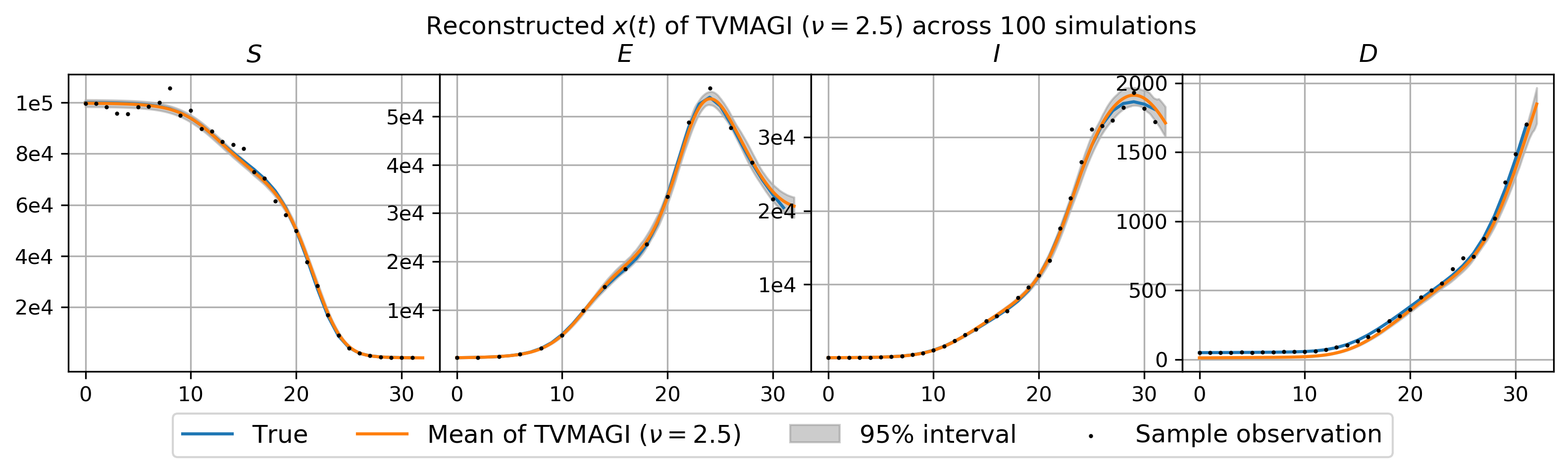}
         \caption{Reconstructed trajectory of $\nu=2.5$}
     \end{subfigure}
\caption{TVMAGI inferred parameters and reconstructed trajectories of SEIRD model with Matern kernel $\nu=2.5$}
\label{fig:nu-2.5}
\end{figure}

\end{appendices}

\end{document}